\documentclass[review]{elsarticle}

\usepackage[a4paper, total={6in, 8in}]{geometry}

\journal{Journal of Computational Physics}

\usepackage{listings}

\usepackage{amsmath}

\usepackage{epstopdf}

\usepackage{graphicx}
\usepackage[%
    colorlinks=true,
    pdfborder={0 0 0},
    linkcolor=red
]{hyperref}
\usepackage{multirow}
\usepackage[none]{hyphenat}
\usepackage{float}
\usepackage{subfig}
\usepackage{url}

\usepackage{epsfig}
\usepackage{algorithm}

\usepackage{adjustbox}
\usepackage{longtable}
\usepackage{color}
\usepackage{xcolor}
\usepackage{import}
\usepackage{bbold}
\usepackage{amssymb}

\usepackage{neuralnetwork}
\usepackage{tikz}
\usepackage{xstring}
\usepackage{xpatch}
\usepackage{comment}
\usepackage{adjustbox}
\usepackage{numprint}
\usepackage{siunitx}
\usetikzlibrary{fit,positioning}

\newcommand{\RGBcomposite}{\text{RGB composite}}

\newcommand{\numcol}{1}

\newcommand{\subfwp}{145px}
\newcommand{\SNRINPUT}{\text{SNR}_{\text{INPUT}}}

\newcommand{\myvector}[1]{\mathbf{#1}}

\newcommand{\mainfold}[2]{\mathbb{#1}^{#2}}
\newcommand{\mat}[1]{\mathbf{#1}}
\newcommand{\tensor}[1]{\mathcal{#1}}

\newcommand{\hide}[1]{}

\newcommand{\addFigures}[3]{\begin{figure}[ht]
	\centering
	\subfigure[$\RGBcomposite$s of $\tensor{X}$]{
    \label{sec:er:denoisepure} 
    \includegraphics[width=\subfwp]{Images/AVRIS_nonoise}}
    \subfigure[$\RGBcomposite$s of $\tensor{R}$ ($\SNRINPUT=20$dB)]{
    \label{sec:er:denoisenoiseadd} 
    \includegraphics[width=\subfwp]{Images/SNR20}}
    \caption{$\RGBcomposite$s of $\tensor{X}$ and $\tensor{R}$ (band 25, 65 and 80 for red, green and blue)}\label{sec:er:RGBdataset}
\end{figure}}

\newcommand{\onetwocol}[2]{\ifthenelse{\isodd{\numcol}}{#1}{#2}}

\newcommand{\W}[1]{\mat{W}_#1}

\newcommand{\wfTrain}[1]{\times_1 \W{1} \times_2 \W{2} \times_3 \W{3}}
\newcommand{\iwfTrain}[1]{\times_1 \W{1}^T \times_2 \W{2}^T \times_3 \W{3}^T}

\def\arA{{\mathbf b}}

\def\arC{{\mathbf c}}
\def\ara{{\mathbf a}}
\def\arr{{\mathbf r}}

\newcommand{\arD}[1]{{\mathbf d}_{\rm #1}({\rm iter})}

\newcommand{\xqK}[1]{{\bf x}_{\rm q}^{}({\rm #1})}

\newcommand{\xK}[2]{{\bf x}_{\rm #1}({\rm #2})}
\newcommand{\xlK}[1]{{\bf x}_{\rm l}^{}({\rm #1})}
\newcommand{\ylK}[1]{{\bf y}_{\rm l}^{}({\rm #1})}

\newcommand{\xqi}[2]{{x}_{{\rm q}}^{#1}({\rm #2})}

\newcommand{\xdimi}[1]{{x}_{\rm #1}^{i}}

\newcommand{\ydimi}[1]{{y}_{\rm #1}^{i}({\rm iter})}
\newcommand{\ddimi}[1]{{d}_{\rm #1}^{i}({\rm iter})}

\newcommand{\varphitSternaryAllParams}[5]{{((#1-#4)/#2)}^{#3} + #5}
\newcommand{\varphitGaussian}[2]{\frac{#2}{5}(exp(-\frac{{#1}^2}{2}) - exp(-\frac{{(#1-2)}^2}{2}))}

\newcommand{\varphiTernaryone}[6]{\varphi^{\mathcal{P}}(#1,#2,#3,#4,#5,#6)}

\newcommand\denselyConnectNodes[2]{
  \foreach \n [count=\lyrIdx, remember=\lyrIdx as \previdx, remember=\n as \prevn] in #2 {
    \foreach \y in {1,...,\n} {
      \ifnum \lyrIdx > 1
        \foreach \x in {1,...,\prevn}
          \draw[->] (#1-\previdx-\x) -- (#1-\lyrIdx-\y);
      \fi
    }
  }
}

\definecolor{fc}{HTML}{1E90FF}
\definecolor{h}{HTML}{228B22}
\definecolor{bias}{HTML}{87CEFA}
\definecolor{noise}{HTML}{8B008B}
\definecolor{conv}{HTML}{FFA500}
\definecolor{pool}{HTML}{B22222}
\definecolor{up}{HTML}{B22222}
\definecolor{view}{HTML}{FFFFFF}
\definecolor{bn}{HTML}{FFD700}
\tikzset{fc/.style={black,draw=black,fill=fc,rectangle,minimum height=1cm}}
\tikzset{h/.style={black,draw=black,fill=h,rectangle,minimum height=1cm}}
\tikzset{bias/.style={black,draw=black,fill=bias,rectangle,minimum height=1cm}}
\tikzset{noise/.style={black,draw=black,fill=noise,rectangle,minimum height=1cm}}
\tikzset{conv/.style={black,draw=black,fill=conv,rectangle,minimum height=1cm}}
\tikzset{pool/.style={black,draw=black,fill=pool,rectangle,minimum height=1cm}}
\tikzset{up/.style={black,draw=black,fill=up,rectangle,minimum height=1cm}}
\tikzset{view/.style={black,draw=black,fill=view,rectangle,minimum height=1cm}}
\tikzset{bn/.style={black,draw=black,fill=bn,rectangle,minimum height=1cm}}

\xpatchcmd{\linklayers}{\nn@lastnode}{\lastnode}{}{}
\xpatchcmd{\linklayers}{\nn@thisnode}{\thisnode}{}{}

\usepackage{etoolbox} 
\usepackage{listofitems} 

\tikzset{>=latex} 
\colorlet{myred}{red!80!black}
\colorlet{myblue}{blue!80!black}
\colorlet{mygreen}{green!60!black}
\colorlet{mydarkred}{myred!40!black}
\colorlet{mydarkblue}{myblue!40!black}
\colorlet{mydarkgreen}{mygreen!40!black}
\tikzstyle{node}=[very thick,circle,draw=myblue,minimum size=22,inner sep=0.5,outer sep=0.6]
\tikzstyle{connect}=[->,thick,mydarkblue,shorten >=1]
\tikzset{ 
  node 1/.style={node,mydarkgreen,draw=mygreen,fill=mygreen!25},
  node 2/.style={node,mydarkblue,draw=myblue,fill=myblue!20},
  node 3/.style={node,mydarkred,draw=myred,fill=myred!20},
}

\newcommand{\tensa}{\ensuremath{\mathbf{a}}}
\newcommand{\tensat}{\ensuremath{\tilde{\mathbf{a}}}}

\def\div{\ensuremath{\mathrm{div}}}
\newcommand{\grad}{\ensuremath{\mathbf{grad}}}

\newcommand{\myround}[2]{\num[group-separator=,round-mode=places,round-precision=#1]{#2}}

\newcommand{\SolNSGA}[2]{#1 ({\bf x}_{\rm #2})}

\begin{document}

\begin{frontmatter}

\title{Electromagnetic cloak design with mono-objective and bi-objective optimizers: seeking the best tradeoff between protection and invisibility}


\author[addressMarseille]{Ronald Aznavourian}
\author[addressMarseille]{Guillaume Demesy}
\author[addressLondres1,addressLondres2]{S\'ebastien Guenneau}
\author[addressMarseille]{Julien Marot\corref{mycorrespondingauthor}}
\cortext[mycorrespondingauthor]{Corresponding author}
\ead{julien.marot@fresnel.fr}



\address[addressMarseille]{Aix Marseille Univ, CNRS, Centrale Marseille, Institut Fresnel,
13397, Marseille, France}

\address[addressLondres1]
{UMI 2004 Abraham de Moivre-CNRS, Imperial College London, SW7 2AZ, London, UK}

\address[addressLondres2]
{The Blackett Laboratory, Department of Physics, Imperial College London, London SW7 2AZ, UK}


\begin{abstract}
We revisit the design of cloaks, without resorting to any geometric transform. Cancellation techniques and anomalous resonances have been applied for this purpose. Instead of a deductive reasoning, we propose a novel mono-objective optimization algorithm, namely a ternary grey wolf algorithm, and we adapt a bi-objective optimization algorithm. Firstly, the proposed chaotic ternary grey wolf algorithm searches three-valued spaces for all permittivity values in the cloak while minimizing the summation of a protection criterion and an invisibility criterion. Secondly, a bi-objective genetic algorithm is adapted to find pairs of optimal values of invisibility and protection.
\end{abstract}
\end{frontmatter}
\newpage

\newpage    

\section{Introduction}

Since the publication of works by the teams of Leonhardt \cite{leonhardt2010geometry} and Pendry \cite{Pendry2006} in the same issue of the Science magazine over 16 years ago, cloaking has become a mature research area optics. It is by now well known that one can design invisibility cloaks via geometric transforms that either lead to the anisotropic heterogeneous material parameters (e.g. rank-2 tensors of permittivity and permeability in optics), see \cite{Pendry2006}, or spatially varying, yet scalar valued, parameters \cite{leonhardt2010geometry}. The latter is achieved through conformal maps, hence constrained to the 2D case, and besides from that the cloak is of infinite extent. There is yet a third route to cloaking, that relaxes the severe material constraints in \cite{Pendry2006} (notably some infinite anisotropy on the inner boundary of cloak, rooted in the blow up of a point onto a ball of finite extent known as invisibility region, that can only be achieved over a narrow frequency bandwidth in practice \cite{cassier2017bounds}), and avoids the infinite extent of the cloak in \cite{leonhardt2010geometry}: so-called carpet cloaking is a combination of the previous two approaches that is based on quasi-conformal grids \cite{li2008hiding}. This third route requires only some moderate anisotropy, but only achieves invisibility for an object placed on a mirror.

Some works on cloaking focus on the mathematical aspects that are connected to famous inverse problems in particular on electric impedance tomography \cite{kohn1984determining,lee1989determining} wherein one wishes to uniquely determine the conductivity within a bounded region, by applying a known static voltage to the surface and recording the resulting current at the boundary (a Dirichlet-to-Neumann map).
The Dirichlet-to-Neumann map determines the conductivity \cite{kohn2008cloaking}, but this can only happen if the conductivity is scalar-valued, positive and finite.
However, if some of these conditions are not met (e.g. the conductivity is matrix valued) electric impedance tomography fails  \cite{sylvester1987global}.
This has been exploited to create non-unique conductivities sharing the same boundary measurements \cite{greenleaf2003anisotropic}.

\paragraph{Main contributions}
In the present work, we would like to revisit the design of cloaks, without resorting to any geometric transform. Not surprisingly, there is prior work that explored this alternative route, notably through scattering cancellation techniques \cite{alu2005achieving,chen2012invisibility} and anomalous resonances \cite{milton2005proof,milton2006cloaking}. However, our rational for the design of cloaks is not based on a deductive reasoning, but rather on some optimization algorithm, and more precisely on some nature-inspired optimizer known as the Grey Wolf Optimizer (GWO). Here again, one may point out former work on design of invisibility cloaks  \cite{bendsoe1999material,vial2015topology} based on topology optimization \cite{andkjaer2011topology}. In \cite{POMOT2020102413}, some mono-objective genetic optimization algorithm has been applied to estimate the best value, in terms of bias with respect to free-space propagation conditions, of 6 parameters which define the desired cloak. However, we stress that the nature of the optimization algorithms we shall use here is radically different: We aim at estimating a very elevated number of parameters, compared to some previous works such as \cite{POMOT2020102413}, and we aim at minimizing two contradictory criteria instead of one. 

\paragraph{Layout of the paper}
In section \ref{sec:cloakdesignproblem} we introduce the cloak design problem, pointing out the need for the minimization of two \textcolor{black}{antagonist} criteria: an invisibility criterion and a protection criterion. In section \ref{sec:backgroundstateoftheart}, we provide a state-of-the-art of mono-objective and bi-objective optimization algorithms. We focus on the mono-objective grey wolf optimizer, and on the bi-objective non dominated sorting genetic algorithm. In section \ref{sec:CTGWO} we propose a novel ternary version of the grey wolf algorithm, which is dedicated to search spaces with three values. We name it chaotic ternary grey wolf optimizer (CTGWO). In section \ref{sec:ResultsCloaks} we present the results obtained on cloak design with the mono-objective approach, including CTGWO, and with the bi-objective approach involving NSGA-II. In section \ref{sec:DiscussionCloaks} we discuss the results obtained. We point out the superiority of CTGWO over comparative algorithms in the mono-objective approach; and we emphasize the interest of the bi-objective approach for an end-user. Conclusions are drawn in section \ref{sec:ConclusionCTGWOCloakDesign}.

\paragraph{Notations}
The following notations are used throughout the paper:
Manifolds are denoted by blackboard bold, $\mainfold{A}{}$, matrices by boldface uppercase roman, $\mat{A}$. Vectors are denoted by boldface lowercase roman, $\myvector{a}$, and scalars by lowercase or uppercase roman, a, $b$ or $A$.
The $P$ scalar components of a vector $\myvector{a}$ are accessed ${\it via}$ ${a^{1},a^{2},\ldots,a^{P}}$, such that $\myvector{a}=\left[{a^{1},a^{2},\ldots,a^{P}}\right]^{T}$ with superscript $T$ denoting the transpose. The interval of real values between scalars $a$ and $b$ is denoted by $[a:b]$ with square brackets. A set of values is denoted by $\left\{a,\ldots,b\right\}$ with curly brackets.\\
The symbol $\circ$ denotes the Hadamard (also called component-wise) product of two vectors: $\myvector{a} \circ \myvector{b}$ is a vector whose $K$ components are equal to $a_1 b_1,a_2 b_2,\ldots,a_K b_K$. 
The symbol $|\myvector{a}|$ means element-wise absolute value and is a vector whose $K$ components are equal to $|a_1|, |a_2|,\ldots, |a_K|$.

\section{Cloak design problem}\label{sec:cloakdesignproblem}

\begin{figure}[h]
	\centering
	\begin{tabular}{c}
		\includegraphics[width=14cm]{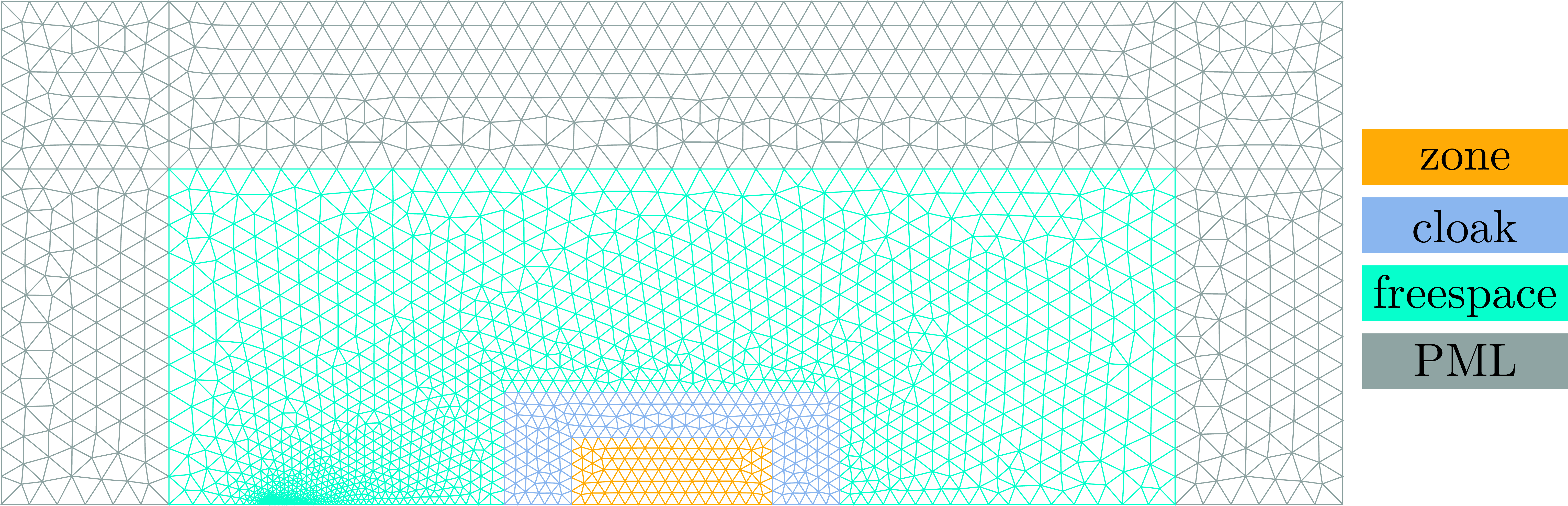}
	\end{tabular}
	\caption{Sketch of the scattering problem: Perfectly Matched Layers (PMLs, light grey) surround the free-space region $S_2$ (cyan) that contains the source, the cloak (blue) and the protected region $S_1$ (yellow)}
	\label{fig:cape}
\end{figure}

We consider the 2D scattering problem sketched in Fig.~\ref{fig:cape}(a). A point source is located at $(x_s,y_s)$ and radiates from freespace (in blue color) in the vicinity of the yellow zone $S_1$ which is the area to be protected. This yellow region is cloaked by the green rectangular zone. For computational purposes, the blue freespace zone is surrounded by Perfectly Matched Layers (PMLs) that model an unbounded medium \cite{teixeira1998general}. The scalar scattering problem amounts to finding the total scalar field $u$ such that:
\begin{equation}\label{eq:wave}
    \div\left[ \sigma\,\grad\,u\right]+k_0^2\,\chi\,u=\delta_S
\end{equation}
where $k_0$ is the freespace wavenumber (associated with the unbounded region of space outside of the object and its surrounding cloak and corresponding to a freespace wavelength of $\lambda_0=2\pi/k_0$), and $\sigma$ and $\chi$ represent the scalar material properties.
In the context of acoustic pressure waves in isotropic non-viscous fluids, $\sigma$ is the inverse of density and $\chi$ the inverse of compressibility, and $p$ is the amplitude of the pressure wave. For anti-plane shear waves in isotropic solids, $\sigma$ is the shear modulus and $\chi$ the density, whereas $u$ stands for the component of the displacement field perpendicular to the $(xy)$-plane. Finally, for electromagnetic waves in transverse magnetic (TM) polarization $\sigma$ is the inverse of the relative permittivity and $\chi$ the relative permeability. The relative permittivity will be denoted by $\epsilon$ in the rest of the paper. This is when $u$ represents the component of the magnetic field perpendicular to the $(xy)$-plane. The roles of permittivity and permeability are interchanged in transverse electric (TE) polarization, whereby the electric field is perpendicular to $(xy)$-plane. All these physical setups are equivalent from a mathematical standpoint. In the rest of the paper, the considered wavelength is $\lambda=500~nm$, the side of a triangle mesh has length $\lambda/6$ in the freespace; and $\lambda/12$ in the cloak and the protected zone.

In what follows, we focus on the TM case. In this scalar Helmholtz equation (\ref{eq:wave}), one can still consider a certain form of anisotropy \cite{renversez2012foundations}, provided that $\sigma$ can be written as a 2 by 2 matrix. 
Indeed, even if isotropic cloak and protected area only are considered here, the implementation of the PMLs relies on anisotropy (and absorption), see \cite{teixeira1998general}. As in standard topology optimization, our design space is constructed on the mesh of the cloak shown in Fig.~\ref{fig:cape}. Each triangular element constitutes a voxel that can be filled with a particular material whose isotropic physical properties are represented by the scalar quantity $\sigma$.
The practical Finite Elements discretization and implementation of the problem follow quite closely those described in Ref.~\cite{kuci2017design} and its associated ONELAB open source tutorials found in Ref.~\cite{onelab}. In short, a first dummy run allows to retrieve the constant-by-element table for a particular mesh, and this table is used to control the discrete values of $\gamma$ throughout the whole optimization procedure thanks to \texttt{GmshRead[]} and \texttt{ScalarField[]} functions \cite{docgetdp}. Then, Eq.~\ref{eq:wave} is solved using second order Lagrange elements in a standard manner. 

Finally, we are in a position to define the optimization criteria. The first one, $C_1$ is a protection criterion, the integral over the region to be protected of the square norm of the field :
\begin{equation}\label{def:cprotec}
	C_1 = \frac{1}{|S_1|}\int_{S_1}|u|^2\mathrm{d}S,
\end{equation}
where $|S_1|$ is the area of $S_1$ in Fig.~\ref{fig:cape}.

The second one, $C_2$ is an invisibility criterion, the integral over the freespace region of interest of the square norm of the difference between the field and a reference field denoted $u_0$ :
\begin{equation}\label{def:cinvis}
	C_2 = \frac{1}{|S_2|}\int_{S_2}|u-u_0|^2\mathrm{d}S,
\end{equation}
where $|S_2|$ is the area of $S_2$ \textcolor{black}{in Fig.~\ref{fig:cape}} and $u_0$ is the freespace solution to the problem (\emph{i.e.} the scalar Green function of the problem without any cloak or region to protect).

Our goal is to minimize $C_1$ and $C_2$ to achieve both protection and invisibility.  Intuitively, we expect there should be some trade-off between protection (i.e. a vanishing field $u$ inside the yellow region) and invisibility (i.e. $u$ as close as possible to $u_0$ in the blue region). For instance, surrounding the yellow region by an infinite conducting boundary would ensure a vanishing field $u$ inside the yellow region, but $u$ and $u_0$ would be very different in freespace blue region $S_2$ due to a large scattering. On the other hand, if we consider freespace in the green region, then $S_1$, the cloak and $S_2$ are impedanced matched at the interfaces between all three regions and thus $u=u_0$ in $S_2$ (the cloak and $S_1$ being transparent). But in that case there is no protection at all in $S_1$. Depending upon their need, cloak designers might just wish to give more weight to criterion $C_1$ or $C_2$. $C_1$ and $C_2$ depend both on $P$ parameters, where $P$ is equal to the number of voxels (equivalently triangles) in the cloak. These parameters, denoted by $K^{1},K^{2}, \ldots,K^{i}, \ldots,  K^{P}$, take their values in a so-called 'search space', which can be either discrete or continuous. In this paper we will firstly consider the realistic case where three possible permittivity values can be associated with each triangle. These values correspond to three different materials and yield the search space $\left\{7,10,12\right\}$. Equivalently, once these values are set, we may access them through the index values $\left\{0,1,2\right\}$. This will compound our 'ternary search space'.
Secondly, we will perform a study which is less realistic: in our simulations the permittivity for each voxel may be any real value in $[7:12]$ \textcolor{black}{(up to seven decimal digits)}.\\
In a nutshell, we notice that we face a bi-objective problem with either ternary, or continuous search spaces. In the rest of the paper, the criteria $C_1$ and $C_2$ will also be denoted by $f_{1}(\myvector{x})$: $\mainfold{R}{P} \mapsto \mainfold{R_{+}}{}$ and $f_{2}(\myvector{x})$: $\mainfold{R}{P} \mapsto \mainfold{R_{+}}{}$. Vector $\myvector{x}$ contains the parameter values $K^{1},K^{2}, \ldots,K^{i}, \ldots,  K^{P}$.

\section{Background and state-of-the-art of optimization methods}\label{sec:backgroundstateoftheart}
Optimization algorithms are meant to retrieve the location of the minimum value reached with a set of parameters.
In the case of single-objective optimization, only one function is considered for optimization; in the case of bi-objective optimization, two functions should be minimized simultaneously.\\
We assume that $P$ parameters should be estimated: $K^{1},K^{2}, \ldots,K^{i}, \ldots,  K^{P}$, where $P \geq 1$. We remind that, as explained in section \ref{sec:cloakdesignproblem}, in the considered problem, $P$ is equal to the number of voxels in the cloak.
The following notations will be used:\\
\noindent $\bullet$ $P$ is the number of expected parameters, which are indexed with $i$.\\
$\bullet$ ${\rm iter}$ denotes one iteration and ${\rm T_{max}}$ the maximum allowed number of iterations.\\
$\bullet$ $f(\cdot)$ is a function to be optimized, also called the criterion. It depends on the $P$ parameters mentioned above. In this paper, unless specified, minimization problems are considered.\\
In the case of a single-objective optimization, there is one function $f(\cdot)$ to be minimized. In the case of bi-objective optimization, there are two functions $f_{1}(\cdot)$ and $f_{2}(\cdot)$ to be minimized.\\
Both GWO and NSGA-II are agent-based algorithms.\\
$\bullet$ $\xqK{\rm iter}$ is a vector corresponding to an agent ${\rm q}=1,\ldots,Q$, at iteration ${\rm iter}$. It takes the form of a vector with a $P$-tuple of tested values $\xqK{\rm iter}=\left[K^{1},K^{2}, \ldots, K^{P}\right]^{T}$.\\
In subsection \ref{subsec:GWO}, we give a background on a single-objective optimization: the grey wolf algorithm (GWO). In subsection \ref{subsec:NSGAII}, we give a background on a bi-objective optimization algorithm, namely non-dominated sorting genetic algorithm (NSGA) \cite{srinivas1994muiltiobjective} and a fast version (NSGA-II) \cite{deb2002fast}.

\subsection{Background on single-objective optimization and Grey Wolf Optimizer}\label{subsec:GWO}

The GWO is a nature-inspired optimizer based on the observation of the social life of grey wolves in nature~\cite{Mirjalili201446}. In this method an agent is called a wolf. It simulates the common behaviour and hunting strategies of grey wolves in their environment. 
The seminal GWO searches a continuous space \cite{Mirjalili201446}. Among the search agents, there are three leaders $\alpha$, $\beta$, and $\delta$. All other agents are the $\omega$ wolves.\\ 
$\bullet$ $\xK{\alpha}{iter}$, $\xK{\beta}{iter}$, and $\xK{\delta}{iter}$ denote the position of the leaders $\alpha$, $\beta$, and $\delta$ respectively, at iteration ${\rm iter}$.\\  
The position of any wolf at iteration ${\rm iter+1}$ is calculated as:
\begin{equation}\label{eq:GWO_Xmethod}
\xqK{iter+1} = \frac{1}{3}(\textbf{y}_{\alpha}({\rm iter}) + \textbf{y}_{\beta}({\rm iter}) + \textbf{y}_{\delta}({\rm iter}))
\end{equation}
It results from the equal contribution of the $\alpha$, $\beta$, and $\delta$ wolves. These contributions are computed at each iteration ${\rm iter}$ as follows, for any leader ${\rm l}$, either ${\rm \alpha}$, ${\rm \beta}$, or ${\rm \delta}$: 
\begin{equation}
\ylK{iter} = \xlK{iter} - \arA \circ \arD{l}
\end{equation}
with: $\arD{l} = |\arC \circ \xlK{iter} -  \xqK{iter}|$, $|\cdot|$ denoting absolute value.\\
The vectors $\arA$ and $\arC$ are calculated as $\arA=2\ara \circ \arr_{1} -  \ara$ and $\arC=2 \arr_{2}$. In these expressions, vectors $\arr_{1}$, $\arr_{2}$ have random components between 0 and 1.\\ 
For the $i^{\rm th}$ parameter ($i=1,\ldots,P$):\\

The component $b^{i}$ of $\arA$ is provided by: 
\begin{equation}\label{Eq:bi}
b^{i}=2 a r_{1} - a, 
\end{equation}
the same whatever $i$.\\
The component $\ddimi{l}$ of $\arD{l}$ is provided by: 
\begin{equation}\label{Eq:dli}
\ddimi{l}=|2 r_{2} \xdimi{l} - \xqi{i}{iter}|
\end{equation}

where $r_{1}$ and $r_{2}$ are two random values between 0 and 1;
$\xqi{i}{iter}$ is the $i^{\rm th}$ component of the $q^{\rm th}$ agent at iteration ${\rm iter}$; $\xdimi{l}$ is the $i^{\rm th}$ component of leader ${\rm l}$.\\

The component $\ydimi{l}$ of $\ylK{iter}$ is provided by: 

\begin{equation}\label{Eq:yli}
\ydimi{l} = \xdimi{l} - b^{i} \ddimi{l}
\end{equation}
During the hunt, the wolves firstly diverge from each other to search for the prey, or equivalently to encircle it. Secondly, they converge to kill the prey. This is mathematically modeled through the deterministic vector $\ara$. The components of vector $\ara$ are all equal to $a$, a scalar value which is a key parameter in the algorithm. When $a>1$, the search agents are obliged to diverge from the prey: this is the exploration phase. Conversely, when $a \leq 1$, the search agents are obliged to attack towards the prey: this is the exploitation phase. In the vanilla version of GWO \cite{Mirjalili201446}, the key parameter $a$ decreased regularly from 2 to 0:
\begin{equation}\label{eq:originalexpression_a}
a=2(1 - \frac{{\rm iter} }{{\rm T_{max}}})
\end{equation}
In more recent works, various expressions have been proposed for $a$ such as a quadratic \cite{Mittal2016} or adaptive \cite{Martin_MarotAppliedSoftComputing_18} function. Whatever the version \cite{Martin_MarotAppliedSoftComputing_18,Mittal2016} the exploration phase lasts until $a=1$, then the exploitation phase lasts from $a=1$ to $a=0$.
Storing all values of $f(\xK{\alpha}{iter})$ across all iterations from 1 to ${\rm T_{max}}$ yields a so-called convergence curve. The outcomes of a single-objective optimization method are mainly the solution $\xK{\alpha}{{\rm T_{max}}}$, but also the convergence curve.  

\subsection{Background on bi-objective optimization and non-dominated sorting genetic algorithm}\label{subsec:NSGAII}
Non-dominated sorting genetic algorithms (NSGA and NSGA-II) are inspired by Darwin's rules of evolution. In this method an agent is called a chromosome. The interest of multi-objective optimization has already been emphasized in physical phenomenon modeling for instance in \cite{WASCHKOWSKI2022110922}, where NSGA-II is used to model turbulence; or in the design of finite 3D periodic structures \cite{CHEN2010806}, etc. The main steps of the algorithm are as follows:
generate a random population of chromosomes, calculate function values $f_{1}$ and $f_{2}$ for each chromosome, sort chromosomes in the population, choose parents in the next generation by tournament algorithm, generate children by crossover and mutation, extract a new generation through ranking, and repeat the process from parent choice. The expected outcome of NSGA-II is different from the outcom of GWO: For a single-objective method, we will represent the results as convergence curves, and the solution is a single set of values extracted from the search spaces. We will try to minimize our cost function so that our objective tends as much as possible towards zero. It is up to the user to determine a threshold value for our cost function, from which we will retrieve one optimal solution. For a bi-objective method, we will no longer have convergence curves, but Pareto fronts. The solution is composed of several sets of values extracted from the search space and located on the Pareto front. The principle of the Pareto front is that we will represent the value of the first cost function on the horizontal axis, and the value of the second cost function on the vertical axis. Thus, visually, we will see very quickly if a solution favors either $f_1$ or $f_2$. The main method to estimate the quality of a solution is the 'domination' principle. In fact, for each solution, we will calculate the distance between the point of origin of coordinates (0,0), and the considered solution, with its coordinates. Of course, the solution which has the smallest distance will 'dominate' the solution which has a larger distance. Considering two solutions ${\bf x}_{\rm q1}$ and ${\bf x}_{\rm q2}$, we can say that ${\bf x}_{\rm q1}$ 'dominates' ${\bf x}_{\rm q2}$, if the following condition is respected:
\begin{gather*} 
{(({\SolNSGA{f_{1}}{q1}} \leq {\SolNSGA{f_{1}}{q2}}) }~ \mathrm{and}~ { ({\SolNSGA{f_{2}}{q1}} \leq {\SolNSGA{f_{2}}{q2}})) }\\
~\mathrm{and}~\\ 
{(({\SolNSGA{f_{1}}{q1}} < {\SolNSGA{f_{1}}{q2}}) }~ \mathrm{or}~{ ({\SolNSGA{f_{2}}{q1}} < {\SolNSGA{f_{2}}{q2}})) }
\end{gather*}
It is up to the user to choose the best solution(s) to keep. Indeed, the user may very well want to favour one of the two objectives, or seek the best compromise between the two objectives.
Actually, in the last decades, different methods to determine the 'domination' principle have been proposed, and then, the 'non-domination' principle emerged, notably thanks to Srinivas and Deb \cite{srinivas1994muiltiobjective} who proposed the NSGA algorithm \cite{srinivas1994muiltiobjective} and then an improved version, called NSGA-II \cite{deb2002fast}. This fast sorting method by 'non-domination' has been widely spread by other algorithms as an efficient technique. The particularity of NSGA-II is to hierarchize the levels of 'domination', with a first Pareto front containing only the non-dominated solutions, a second Pareto front with the solutions dominated by one or two solutions, and finally, a third Pareto front with all the other solutions, those dominated by more than two solutions. For this last category, we compute 'crowding distances' for the solutions of this category, then we sort the set of results thus obtained. The 'crowding distance' is calculated criterion by criterion. For example, for the criterion represented on the horizontal axis, we will first determine the extreme solutions, which we will call '$\mathrm{min}$' and '$\mathrm{max}$', it being understood that '$\mathrm{min}$' will be the solution which will have the smallest value on the horizontal axis, and '$\mathrm{max}$', the solution which will have the largest value on the horizontal axis. Considering that we have $Q'<Q$ solutions on the Pareto front, we will then assign an index to each solution, with the index $1$, for '$\mathrm{min}$', and the index $Q'$, for '$\mathrm{max}$'. For each solution of index $q$ with $1 < q < Q'$, we will calculate the distance on the horizontal axis between the $(q-1)^{\rm th}$ solution and the solution $(q+1)^{\rm th}$ solution, and we will divide this distance by the distance on the horizontal axis between '$\mathrm{max}$' and '$\mathrm{min}$'. Thus, it is this result that will be considered as the crowding distance of each solution.\\
The next step is to create a 'descendant'. To do this, we first organize a selection tournament, \emph{i.e.} we will randomly draw pairs of solutions in the set of solutions, and for each pair of solutions, we will determine which one can become 'parent', by comparing the hierarchical levels of the Pareto front. For example, if the first solution belongs to the second Pareto front and the second solution belongs to the third Pareto front, then the selection tournament will be won by the first solution because the second Pareto front contains solutions that are better than the third Pareto front. If two solutions belonging to the same Pareto front were drawn, then the solution with the smallest crowding distance would be selected as the 'parent'. Then, for each pair of 'parents', we will generate a child which will have the values of the first 'parent' for some unknowns, and the values of the other "parent" for the other unknowns. This step is called 'cross-over'. Finally, according to a percentage defined beforehand, the values of some unknowns of the child will be slightly modified. This last step is called the 'mutation'. The whole process is repeated, as many times as there are iterations, and in the end, we obtain the set of 'optimized' solutions.

\section{Chaotic ternary grey wolf algorithm}\label{sec:CTGWO}

We propose a ternary version of GWO, which searches specifically ternary spaces with enhanced exploration abilities.\\
Our first motivation is that, in the considered application, the search space is associated to three values of epsilon. But this method could be applied to other situations and applications, involving for instance sensors with three possible states.\\ 
Our second motivation is to propose a method with enhanced exploration properties. Indeed metaheuristics with enhanced exploration properties are of great interest to cope with applications where the objective function exhibits an elevated number of local minima. We aim at achieving such enhanced exploration properties while proposing a ternary map which evolves across iterations, and inserting chaotic sequences in the update rules of the agents.\\ 
Our third motivation is to improve the diversity of the agents behavior.
Indeed, GWO exhibits premature convergence due to poor diversity of the population of wolves. So we propose to divide the wolf pack into two groups: the first with enhanced 'exploration' abilities, and the second with 'exploitation' abilities.\\
The proposed chaotic ternary GWO is denoted by CTGWO. In subsection \ref{subsec:proposedternaryalgo}, we derive the update rules which relies on specific transform functions depending on a parameter $a$.
We wish to preserve the original philosophy of GWO: the number of leaders ruling the update of the agents is superior to 1, and the parameter $a$ permits to distinguish between an exploration phase at the beginning of the algorithm and an exploitation phase at the end. In subsection \ref{subsec:proposedternaryalgo}, we just assume about parameter $a$ that it decreases from 2 to 0 across the iterations. 
Then in subsection \ref{subsec:chaoticversionTGWO}, we investigate a chaotic expression for $a$.

\subsection{Ternary update rules based on dedicated transform maps}
\label{subsec:proposedternaryalgo}
We propose here innovative update rules, dedicated to a ternary search space, performed with the help of an ad hoc transform function. Firstly propose a novel manner to compute the contribution of a leader, and the mean contribution of several leaders. Secondly, we propose an update rule depending on this mean contribution.

\paragraph{Contribution of a leader}
We remind that in the continuous case, the contribution $\ydimi{l}$ of a leader $l$ is computed as in Eq. (\ref{Eq:yli}). In this case $\ydimi{l}=\xdimi{l} - b^{i} \ddimi{l}$ depends on the product $b^{i} \ddimi{l}$, which decreases to $0$ simultaneously with $a$, reaching $0$ when ${\rm iter=T_{max}}$. So $\ydimi{l}$ is a real value which gets closer to $\xdimi{l}$ across the iterations. In the following we propose to compute a contribution $\ydimi{l}$ which is a real value between $0$ and $2$. We propose, as expression of the contribution of leader $l$:
\begin{equation}\label{eq:Contributionternary}
\ydimi{l}=(\xdimi{l} - b^{i} \ddimi{l})~\rm{mod}~2
\end{equation}
where $b^{i}$ is defined as in Eq. (\ref{Eq:bi}) and $\ddimi{l}$ is defined as in Eq. (\ref{Eq:dli}). Based on the hypothesis that the maximum value of $a$ is 2, we deduce from Eqs. (\ref{Eq:bi}) and (\ref{Eq:dli}) that the values of $b^{i} \ddimi{l}$ are between -8 and 8. So the values of $\xdimi{l} - b^{i} \ddimi{l}$ can be out of the interval $[0:2]$.

The 'modulo' operator, denoted by $mod$ is meant to enforce the contributions $\ydimi{l}$ to remain into the interval $[0:2]$. It is defined as follows: whatever the values $z1 \in \mainfold{R}{}$ and $z2 \in \mainfold{R}{*}$:
\begin{equation}\label{eq:defmodulo}
z1~\rm{mod}~z2=\left \{
        \begin{array}{ccc}
      z1 - z2 \lfloor z1/z2 \rfloor  & {\rm if} & z1\neq z2 \\
      z2  & {\rm if} & z1=z2,~{\rm or}~z1=0
   \end{array}
\right.
\end{equation}
where $\lfloor z1 \rfloor$ denotes the largest value in $\mainfold{Z}{}$ which is smaller than the scalar $z1~\in~\mainfold{R}{}$. 

\paragraph{Weighted contribution of the leaders}
We denote by $\overline{\ydimi{}}$ the weighted contribution of four leaders $\alpha$, $\beta$, $\delta$, and $\rho$: 

\begin{equation}\label{eq:weightedleadercontribution}
\overline{\ydimi{}}=\left \{
        \begin{array}{ccc}
      \frac{1}{3}(\ydimi{\alpha}+\ydimi{\beta}+((1-a/2)\ydimi{\delta}+a/2\ydimi{\rho}))  & {\rm if} & a>1 \\
      \frac{1}{3}(\ydimi{\alpha}+\ydimi{\beta}+\ydimi{\delta})  & {\rm if} & a \leq 1
   \end{array}
\right.
\end{equation}
In Eq. (\ref{eq:weightedleadercontribution}), leader $\rho$ is a wolf which is selected at random among the wolf pack.

We notice that $\overline{\ydimi{}}$ gets closer to $\frac{1}{3}(\ydimi{\alpha}+\ydimi{\beta}+\ydimi{\delta})$ when the iteration index increases.\\

We can now derive a ternary rule which updates the position of any wolf in a ternary search space.
\paragraph{Ternary update rule}
We propose an evolving map which enhances exploration at the beginning of the algorithm, and exploitation at the end of the algorithm. Compared to recent works about the binary GWO \cite{EMARY2016371,HU2020105746}, the new feature in the proposed transform map is of course the division of the map into three regions instead of 2, but also, the fact that the map evolves across iterations: we will introduce a term which is proportional to $a$ in the transform functions and which emphasizes exploration at the beginning of the algorithm and exploitation at the end of the algorithm.\\
The proposed novel process dedicated to ternary search spaces permits to select either value $0$, $1$, or $2$. In dimension $i$ ($i=1,\ldots,P$), wolf ${\rm q}$ is updated from iteration ${\rm iter}$ to iteration ${\rm iter+1}$ as follows:\\ 
\vspace{0.2cm}
\begin{equation}\label{eq:choice0or1or2randomselection}
\xqi{i}{iter+1}=
\left \{
        \begin{array}{ccc}
      0  & {\rm if} & r \geq \varphi^{u}(\overline{\ydimi{}},a) \\
      1  & {\rm if} & r < \varphi^{u}(\overline{\ydimi{}},a)~{\rm and}~ 
                      r \geq \varphi^{d}(\overline{\ydimi{}},a)\\
      2  & {\rm if} & r < \varphi^{d}(\overline{\ydimi{}},a)
   \end{array}
\right.
\end{equation}
where the scalar $r$ is a random value between 0 and 1 and taken from a normal distribution.
In Eq. (\ref{eq:choice0or1or2randomselection}) we introduce two functions, which are necessary to define the ternary map. These functions are denoted by $\varphi^{u}$ and $\varphi^{d}$:
\begin{equation*}
\varphi^{u}:~[0:2] \times \mathbb{R}_{+} \to [0:1];~y \mapsto \varphi^{u}(y,a)   
\end{equation*}
\begin{equation*}
\varphi^{d}:~[0:2] \times \mathbb{R}_{+} \to [0:1];~y \mapsto \varphi^{d}(y,a)   
\end{equation*}
Function $\varphi^{u}$ separates the uppermost part of the map from the rest of the map; and function $\varphi^{d}$ separates the lowermost part of the map from the rest of the map. The basic idea is that if the random number $r$ leads to the region in-between, the value 1 will be chosen as an updated value. If $r$ leads to the region which is above $\varphi^{u}$ (resp. below $\varphi^{d}$), the value $0$ (resp. 2) will be selected. We will now detail the shape of the frontiers between regions 0, 1, and 2. We set a ternary map based on a 'power function'.
The basic function we use is a power function applied to any scalar $y$ and depending on $5$ parameters $c1$, $c2$, $c3$, $c4$, and $a$:\\
\vspace{0.1cm}
$\varphiTernaryone{y}{c1}{c2}{c3}{c4}{a}=$
\begin{equation}\label{eq:transformfunctionPowerscaleshift}
\varphitSternaryAllParams{y}{c1}{c2}{c3}{c4} + \varphitGaussian{y}{a}
\end{equation}

In Eq. (\ref{eq:transformfunctionPowerscaleshift}), the first term depending on $c1$, $c2$, $c3$, and $c4$  gives the overall shape of the function. The second term depending on $a$ permits to get $\varphiTernaryone{0}{c1}{c2}{c3}{c4}{a}\geq0$ and $\varphiTernaryone{2}{c1}{c2}{c3}{c4}{a}\leq1$. We use two versions of this function to define the ternary map. The first one, with $c3=2$ and $c4=1$; the second one, with $c3=0$ and $c4=0$: 

\begin{equation}\label{eq:varphipowerup}
\varphi^{u}(y,a)=\varphiTernaryone{y}{2}{3}{2}{1}{a}
\end{equation}

\begin{equation}\label{eq:varphipowerdown}
\varphi^{d}(y,a)=\varphiTernaryone{y}{2}{3}{0}{0}{a}
\end{equation}
The second term depending on $a$ (see Eq. (\ref{eq:transformfunctionPowerscaleshift})) permits to get values on $y=0$ which are slightly larger than 0, and values on $y=2$ which are slightly smaller than 1:\\
$\varphiTernaryone{0}{c1}{c2}{c3}{c4}{a}=\frac{a}{5}(1-exp(-2))$, 
and $\varphiTernaryone{2}{c1}{c2}{c3}{c4}{a}=1-\frac{a}{5}(1-exp(-2))$ in both Eqs. (\ref{eq:varphipowerup}) and (\ref{eq:varphipowerdown}).\\
So: 
\begin{equation}\label{eq:powerequalon0}
\varphi^{u}(0,a)=\varphi^{d}(0,a)=\frac{a}{5}(1-exp(-2))
\end{equation}
and
\begin{equation}\label{eq:powerequalon2}
\varphi^{u}(2,a)=\varphi^{d}(2,a)=1-\frac{a}{5}(1-exp(-2))
\end{equation}

The functions $\varphi^{u}(y,a)$ and $\varphi^{d}(y,a)$ defined in Eqs. (\ref{eq:varphipowerup}) and (\ref{eq:varphipowerdown}) can then be used in Eq. (\ref{eq:choice0or1or2randomselection}) to get the 'Power' transform map.

\begin{figure}[H]
	\centering{
		\begin{tabular}{c@{\,}c@{\,}c@{\,}}
			\includegraphics[width=4cm]{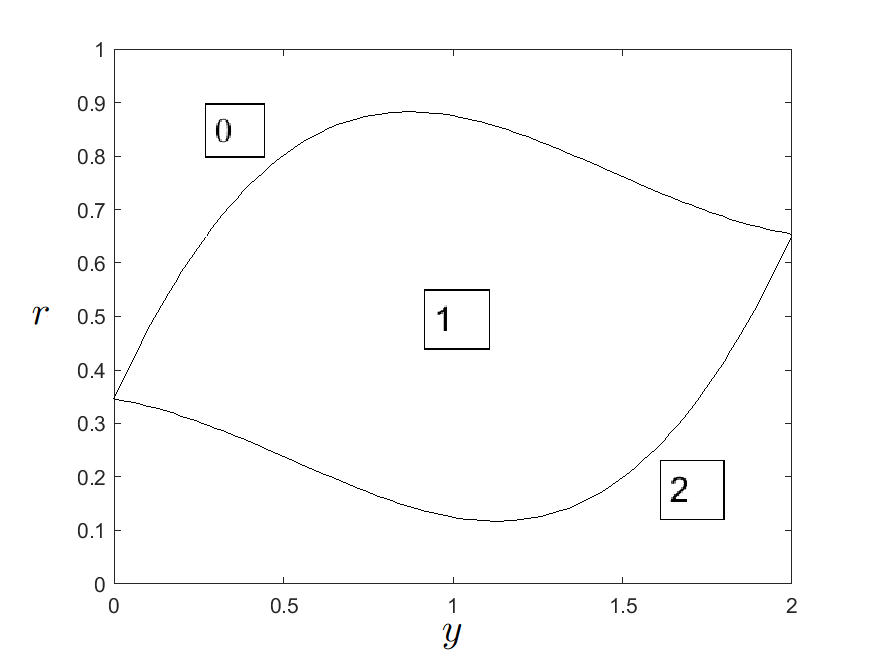}
			&\includegraphics[width=4cm]{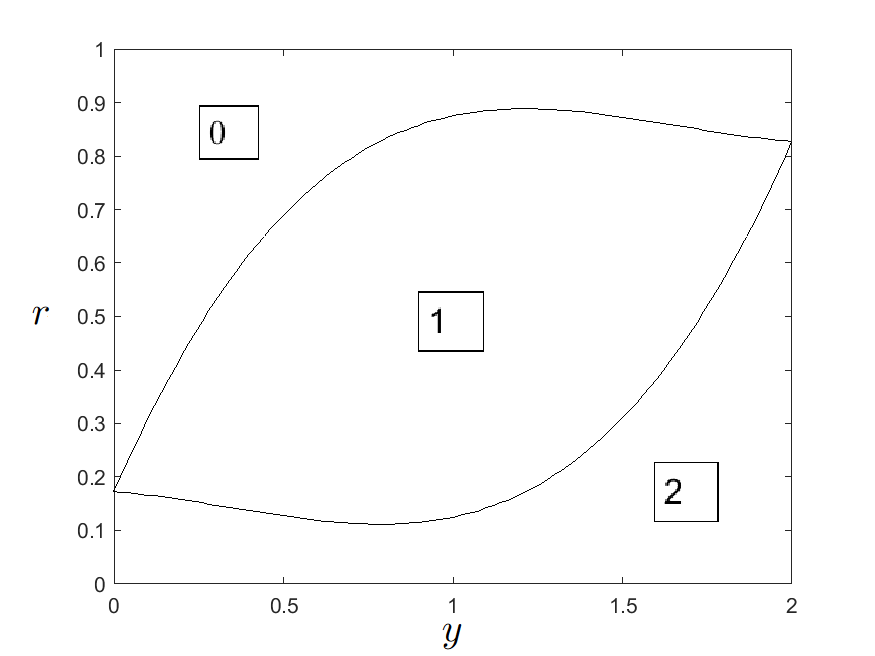}&
			\includegraphics[width=4cm]{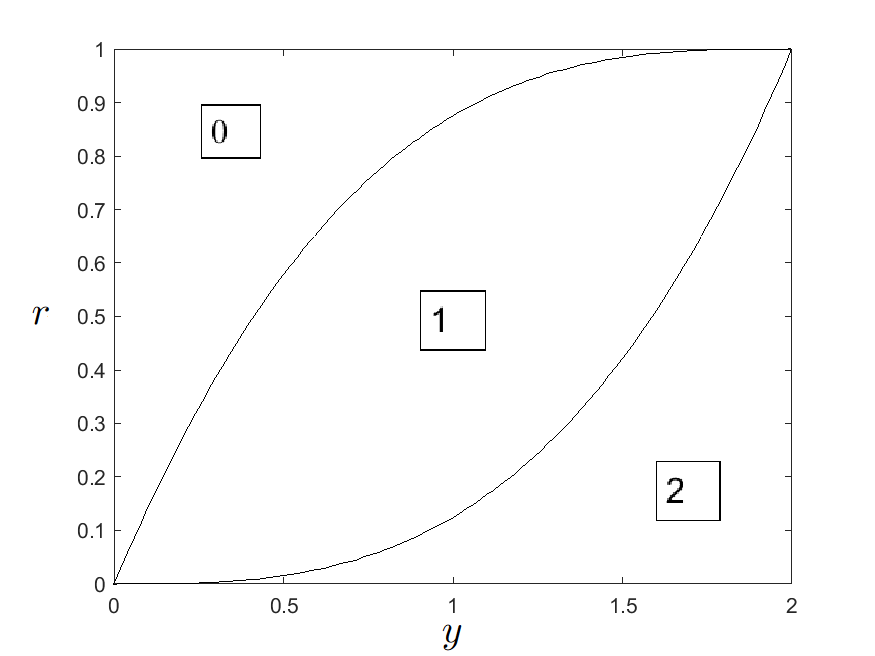}\\
			(a) $\varphi^{u}(y,2)$ and $\varphi^{d}(y,2)$ & (b) $\varphi^{u}(y,1)$ and $\varphi^{d}(y,1)$&(c) $\varphi^{u}(y,0)$ and $\varphi^{d}(y,0)$
		\end{tabular}
		\caption{Ternary 'Power' selection map with different values of parameter $a$. (a) $a=2$; (b) $a=1$; (c) $a=0$. \label{fig:ternarymapv1}}}
\end{figure}

\paragraph{Representation of the ternary transform map and interpretation}
Fig. \ref{fig:ternarymapv1} shows the 'Power' ternary map. In each case the uppermost region maps for 0, the central region maps for 1, and the lowermost region maps for 2. It can be noticed that the shape of the functions $\varphi^{u}$ and $\varphi^{d}$ is consistent with the derivations in Eqs. (\ref{eq:powerequalon0}), and (\ref{eq:powerequalon2}).

%
%

We can check that: 

\begin{itemize}\label{items:values2and0}
\item the term which is proportional to $a$ in Eq. (\ref{eq:transformfunctionPowerscaleshift}) yields a map which evolves across the iterations; this is an important difference compared to the binary map presented in \cite{EMARY2016371};
\item for a value of parameter $a$ which is 2, either the value $0$ or $2$ may be selected with an elevated probability when all leader contributions are equal to $0$, or $2$;
\item for a value of parameter $a$ which is 0, the value $0$ (resp. $2$) is selected with probability 1 when $\overline{\ydimi{}}=0$ (resp. $\overline{\ydimi{}}=2$);
\item whatever the value of $a$, either the values 0, 1, or 2 may be selected  when $\overline{\ydimi{}}=1$.  
\end{itemize}
Indeed, as defined in Eq. (\ref{eq:Contributionternary}) the values of $\ydimi{l}$ are real and between 0 and 2, whatever $i$ and ${\rm l}$. In subsection \ref{subsec:chaoticversionTGWO} we embed chaotic sequences in the expression of $a$ to improve again the exploration abilities of our algorithm. 

\subsection{Embedding chaotic sequences in the ternary grey wolf optimizer}
\label{subsec:chaoticversionTGWO}

\paragraph{Chaotic expression of parameter '$a$' for improved exploration abilities}

We modify the expression of parameter $a$ with respect to other versions of GWO, and propose:


%

\begin{equation}\label{eq:expressionachaotic}
a=2(1- (\frac{{\rm iter}}{{\rm T_{max}}})^{(\eta_{\rm q} (1+ \Gamma(c_{\rm q}^1, {\rm iter})))} )
\end{equation}
In Eq. (\ref{eq:expressionachaotic}), we notice that, for the first time in this paper, the expression of $a$ depends on the agent index ${\rm q}$.
Firstly, we reposition the worst agents closer to the three leaders, with a value of $\eta$ which depends on the score of the agent: for the worst half of the agents (associated with the largest scores), $\eta_{\rm q}=\frac{2}{3}$;
for the best half of the agents (associated with the smallest scores), $\eta_{\rm q}=\frac{3}{2}$.
Secondly, inserting a chaotic sequence $\Gamma(c_{\rm q}^1, {\rm iter})$ enhances the variability of the behavior of the agents. $c_{\rm q}^1$ is the initial value of the chaotic sequence, which is different for every agent ${\rm q}$. 
The principles of the proposed method is that the $a$ parameter of the GWO which rules the displacement of the agents is perturbed through the value of a chaotic sequence.

Choosing a different value of $c_{\rm q}^1$ for each agent permits to emphasize diversity in the displacement of the agents. Meanwhile, we choose a sequence with one attractor to ensure that $\Gamma(c_{\rm q}^1 , {\rm T_{max}})$ is the same whatever $q$. 

\paragraph{Construction of the chaotic sequences}

\begin{figure}[H]
\centering{
 \begin{tabular}{c@{\,}c@{\,}}
  \includegraphics[width=8cm]{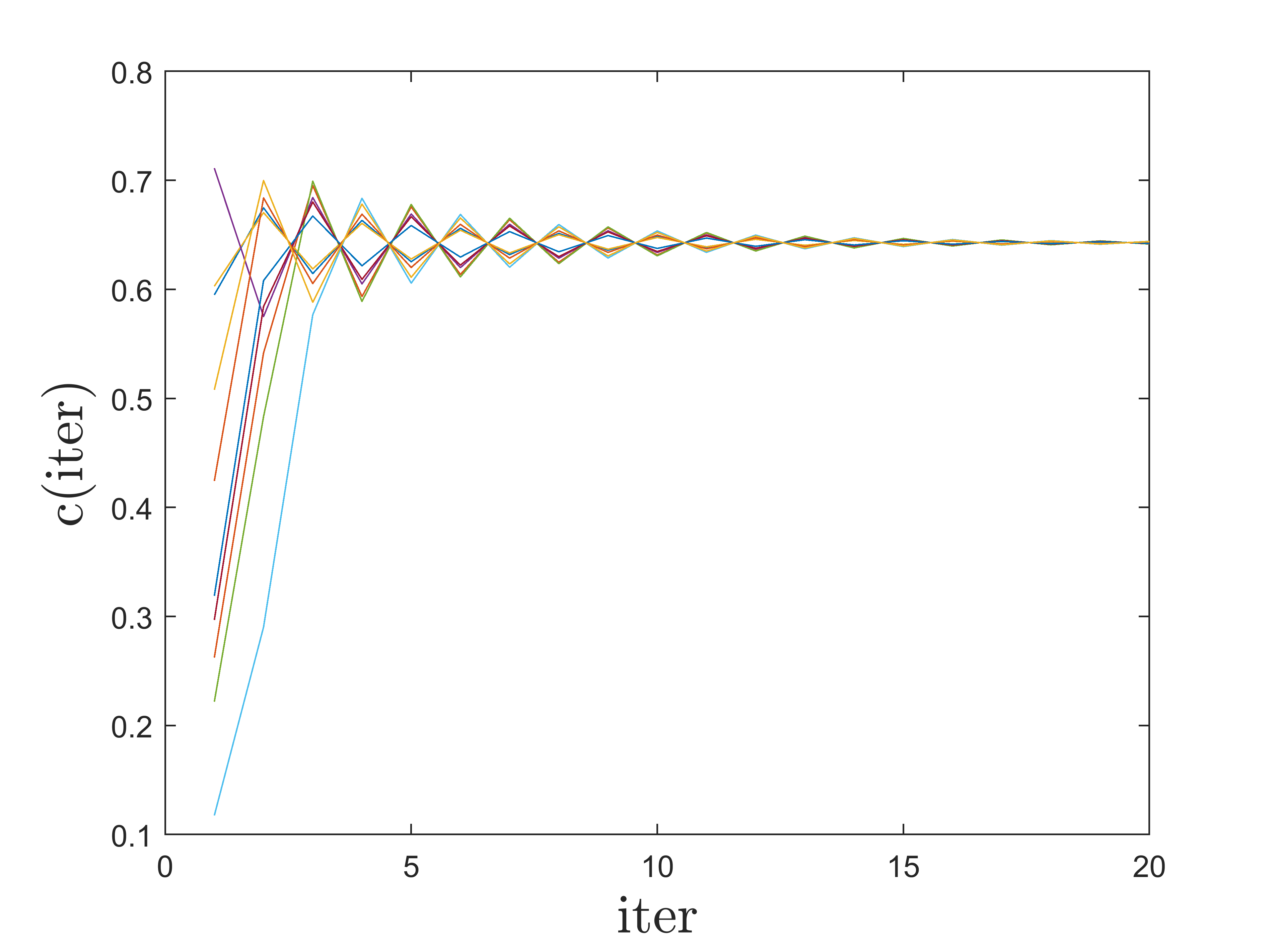}
  &\includegraphics[width=8cm]{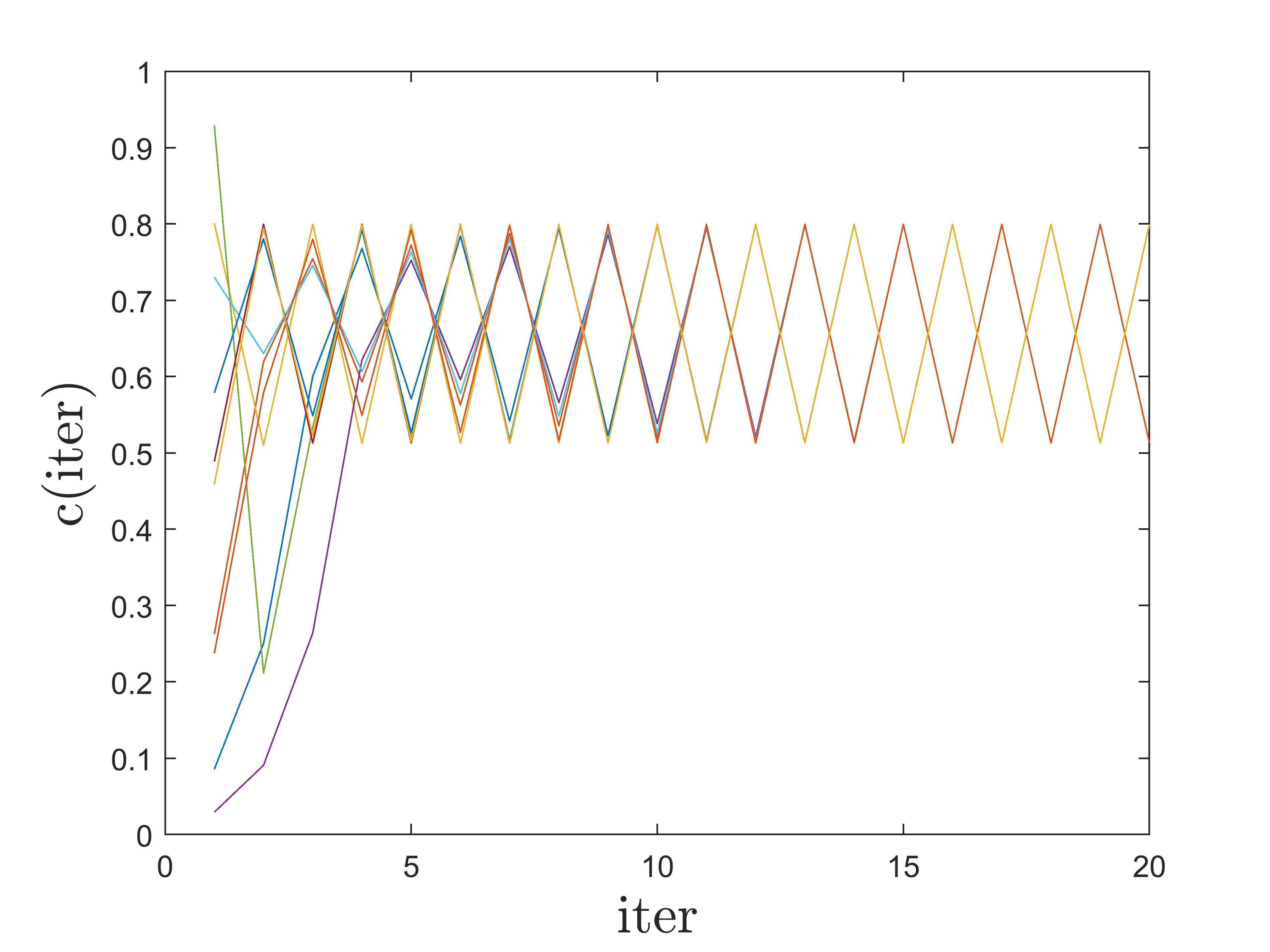}\\
  (a) $\kappa=2.80$ & (b) $\kappa=3.20$ \\
  \includegraphics[width=8cm]{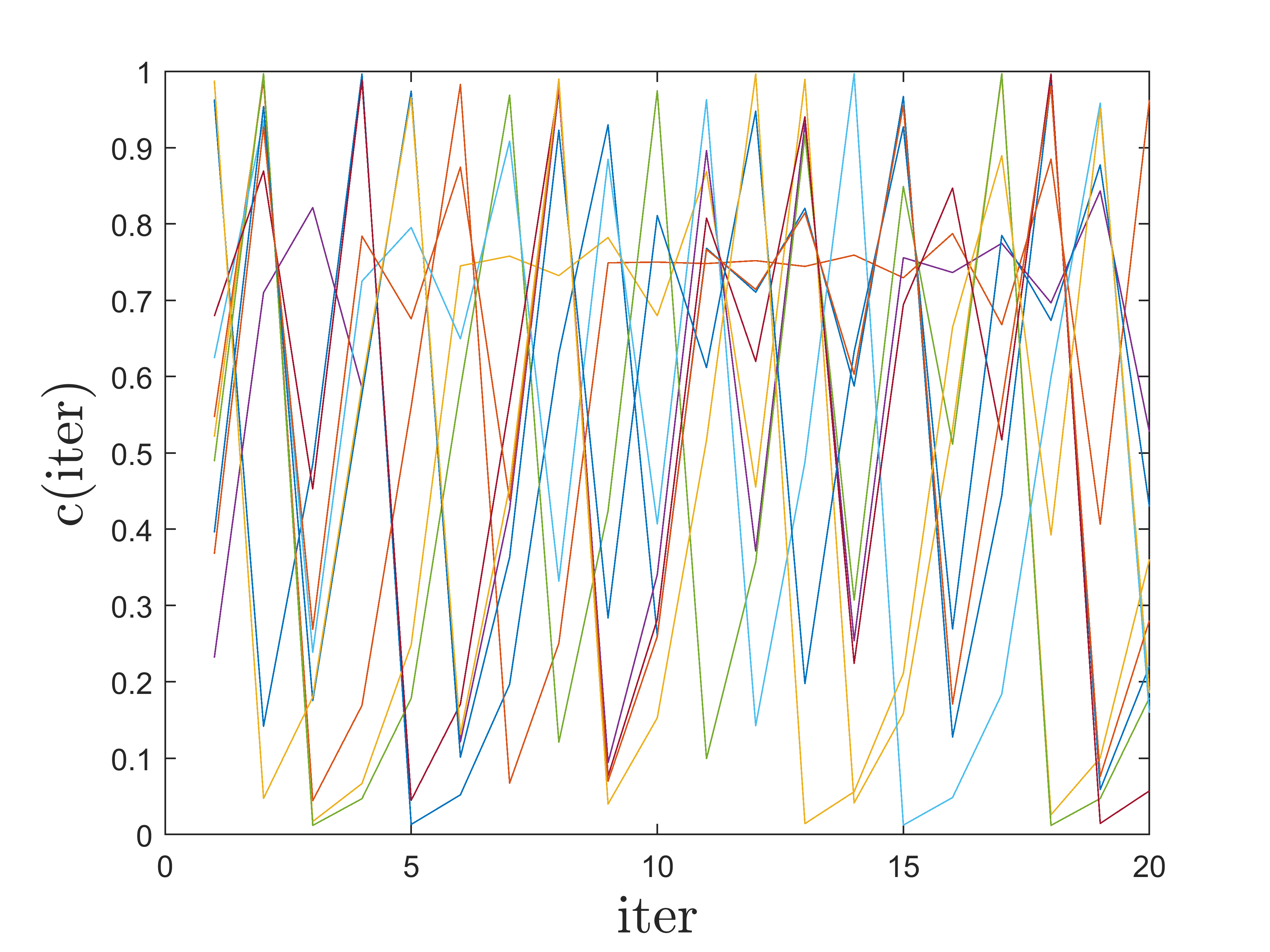}
  &\includegraphics[width=7cm]{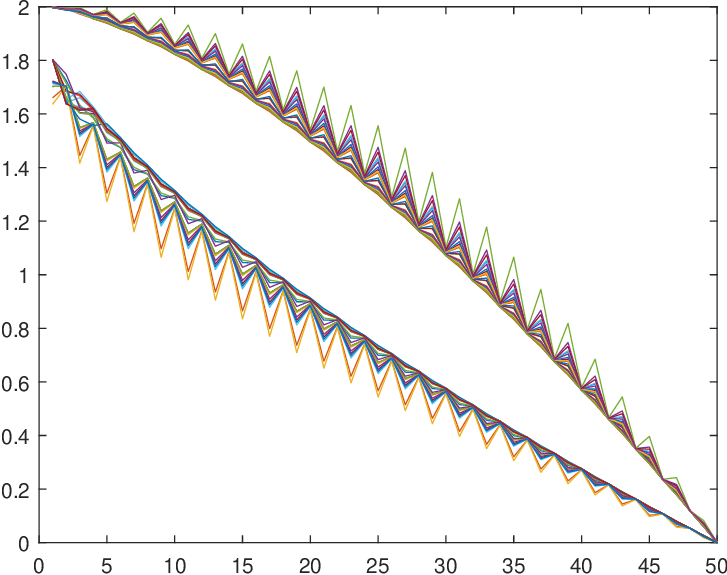}\\
  (c) $\kappa=3.99$ & (d) $a$ values, all agents 
 \end{tabular}
\caption{Chaotic sequences with various values of chaos parameter $\kappa$. (a) $\kappa=2.80$, one attractor; (b) $\kappa=3.20$, two attractors; (c) $\kappa=3.99$, an infinite number of attractors. \label{fig:chaoticmapsattractors}}}
\end{figure}

To privilege exploration abilities, the behaviors of the agents should differ one from the other. To privilege exploitation abilities at the end of the algorithm, the agents behavior should get closer to each other while the iteration index increases. So we set the following constraints on the chaotic sequences:

\begin{itemize}
\item the values of $\Gamma(c_{\rm q}^1, {\rm iter})$ are positive, in the interval $[0:1]$;
\item the last value should be $\Gamma(c_{\rm q}^1, {\rm T_{max}}) = 0$ whatever the initial value $\Gamma(c_{\rm q}^1,1)$. In this way, at the last iteration ${\rm iter}={\rm T_{max}}$, the behavior of all agents is the same.
\end{itemize}
To fulfill easily those constraints, we have to choose a sequence with one known attractor. 
We base our sequence $\Gamma$ on a logistic sequence, denoted by $c({\rm iter})$. Given an initial term $c({\rm 1})$, each subsequent term (for ${\rm iter}=2,\ldots,{\rm T_{max}}$) is defined as: 
\begin{equation}\label{eq:adhoclogisticmapbase}
c({\rm iter+1}) = \kappa c({\rm iter})(1-c({\rm iter}))
\end{equation}
where $\kappa$ $\in~\mathbb{R}_{+}^{*}$. The number of attractors for this sequence depends on the value of $\kappa$. Fig. \ref{fig:chaoticmapsattractors} shows chaotic sequences with one (Fig. \ref{fig:chaoticmapsattractors}(a)), two (Fig. \ref{fig:chaoticmapsattractors}(b)), or an infinite number of attractors (Fig. \ref{fig:chaoticmapsattractors}(c)). We chose $\kappa=2.8$ in Fig. \ref{fig:chaoticmapsattractors}(a), $\kappa=3.2$ in Fig. \ref{fig:chaoticmapsattractors}(b), $\kappa=3.99$ in Fig. \ref{fig:chaoticmapsattractors}(c). In Figs. \ref{fig:chaoticmapsattractors}(a),(b), and (c), each plot with a given color corresponds to a different value for $c({\rm 1})$. There are ten chaotic sequences in each case.
We choose a sequence such as in Fig. \ref{fig:chaoticmapsattractors}(a), where $c({\rm T_{max}})$ bears the same value, equal to $0.65$ approximately, whatever the sequence. For this we set with $\kappa=2.8$.
 
For any agent ${\rm q}$, and for ${\rm iter}=1,\ldots,{\rm T_{max}}$, we define the ad hoc chaotic sequence $\Gamma(c_{\rm q}^1 , {\rm iter})$ as follows, based on the logistic sequence of Eq. (\ref{eq:adhoclogisticmapbase}):
\begin{equation}\label{eq:adhoclogisticmap}
\Gamma(c_{\rm q}^1 , {\rm iter}) = 0.1(c({\rm iter})-c({\rm T_{max}}))
\end{equation}
Hence, for any agent ${\rm q}$ $\Gamma(c_{\rm q}^1 , {\rm T_{max}})=0$. We notice that the initial term $c_{\rm q}^1$ is defined as follows:
\begin{equation}\label{eq:cq1}
c_{\rm q}^1 = 0.1(c({\rm 1})-c({\rm T_{max}}))
\end{equation}
We set the initial term $c({\rm 1})$ of the logistic sequence as a random value between 0 and 1, taken from a normal distribution. 
As $c({\rm 1})$ is a random value, $c_{\rm q}^1$ is also random and a different sequence is generated for each agent. Fig. \ref{fig:chaoticmapsattractors}(d) shows the sequences of values for $a$ for all agents.
We clearly distinguish two families of agents: the first family with $\eta_{\rm q}=\frac{3}{2}$ and rather elevated values of $a$, and the second family with $\eta_{\rm q}=\frac{2}{3}$ and smaller values of $a$, which tend more rapidly towards 0.

\section{Results: mono-objective and bi-objective approaches}\label{sec:ResultsCloaks}
\subsection{Experimental conditions and metrics}\label{subsec:experconditionssyntheticfunctions}
In this section, the test environment is a server running Linux, equipped with 4 Intel(R) Xeon(R) CPU X7560 @ 2.27GHz (64 processors, Hyper-threading activated) and 1000 GB RAM.
The software is Python.\\ We consider cloaks with $P=329$ voxels (of triangular shape) and the expected outcomes of the algorithms are vectors $\myvector{x}$ containing permittivity values $K^{1},K^{2}, \ldots,K^{i}, \ldots,  K^{P}$.\\
In subsection \ref{subsec:monoobjectiveapproach} we display the results obtained with a monoobjective approach. The criterion which is minimized is $f(\myvector{x})$: $\mainfold{R}{P} \mapsto \mainfold{R_{+}}{}$: $f(\myvector{x})=\frac{1}{2}(f_{1}(\myvector{x})+f_{2}(\myvector{x}))$. We remind (see section \ref{sec:cloakdesignproblem}) that $f_{1}$ stands for protection, and $f_{2}$ stands for invisibility.\\
We remind that topological optimization limits the number of possible materials to two. Instead, and for the first time in this paper, we adapt our new variant CTGWO of the GWO which searches solutions among three materials. We compare the proposed CTGWO with the adaptive mixed GWO in discrete mode \cite{Martin_MarotAppliedSoftComputing_18} (denoted by amixedGWO), and the vanilla continuous GWO \cite{Mirjalili201446}. The computational time for one trial of either $f_{1}$ or $f_{2}$  is $0.99$ sec. by trial.
For GWO and amixedGWO, we use the expression of $a$ presented in Eq. (\ref{eq:originalexpression_a}). For CTGWO, we use the expression of $a$ presented in Eq. (\ref{eq:expressionachaotic}). The search space for the values of the permittivity $\epsilon$ is $\left\{7,10,12\right\}$. CTGWO and amixedGWO access these values \emph{via} indices retrieved from the discrete search space $\left\{0,1,2\right\}$. GWO access these values \emph{via} rounded indices retrieved from the continuous search space $[0:2]$. We run the three algorithms with $Q=100$ agents and ${\rm T_{max}}=250$ iterations, that is, $25~10^{3}$ trials of the objective function. The agents are initialized with random integers between 0 and 2.\\ 
In subsection \ref{subsec:biobjectiveapproach} we display the results obtained with a biobjective approach. The couple of criteria which are jointly minimized are $f_{1}(\myvector{x})$: $\mainfold{R}{P} \mapsto \mainfold{R_{+}}{}$, and $f_{2}(\myvector{x})$: $\mainfold{R}{P} \mapsto \mainfold{R_{+}}{}$. We display the results of GWO, amixed GWO, and the proposed CTGWO; as well as the results obtained by NSGA-II in four situations.
These five experimental conditions are summarized in Table \ref{Table:studyconditionsmonobiobjectivecloaks}.
The ternary search space mentioned in Table \ref{Table:studyconditionsmonobiobjectivecloaks} is $\left\{7,10,12\right\}$: NSGA-II accesses these values \emph{via} rounded indices retrieved from the continuous search space $[0:2]$. The continuous search space mentioned in Table \ref{Table:studyconditionsmonobiobjectivecloaks} is $[7:12]$. This last situation is prospective in the sense that we assume that we afford any material with any permittivity value between $7$ and $12$. 

\begin{table*}[h!]
	\centering
\begin{tabular}{ccccc}
\hline
Approach& search space & ${\rm T_{max}}$ & $Q$ & Link \\
\hline
Mono-objective&ternary& 250 & 100 & \ref{subsec:monoobjectiveapproach}\\
\hline
Bi-objective&ternary& 250 & 100 & \ref{subsubsec:biobjectiveternary250_100}\\
&ternary& 1000 & 200 & \ref{subsubsec:biobjectiveternary1000_100}\\
&continuous& 250 & 100 & \ref{subsubsec:biobjectivecontinuous250_100}\\
&continuous& 1000 & 200 & \ref{subsubsec:biobjectivecontinuous1000_200}\\
\hline
\end{tabular}
\caption{cloak design experimental conditions: mono-objective approach (GWO, amixedGWO, CTGWO); and bi-objective approach (NSGA-II): \label{Table:studyconditionsmonobiobjectivecloaks}}
\end{table*}

\newpage

\subsection{Mono-objective approach}\label{subsec:monoobjectiveapproach}

\begin{figure}[H]
	\centering{
		\includegraphics[scale=0.6]{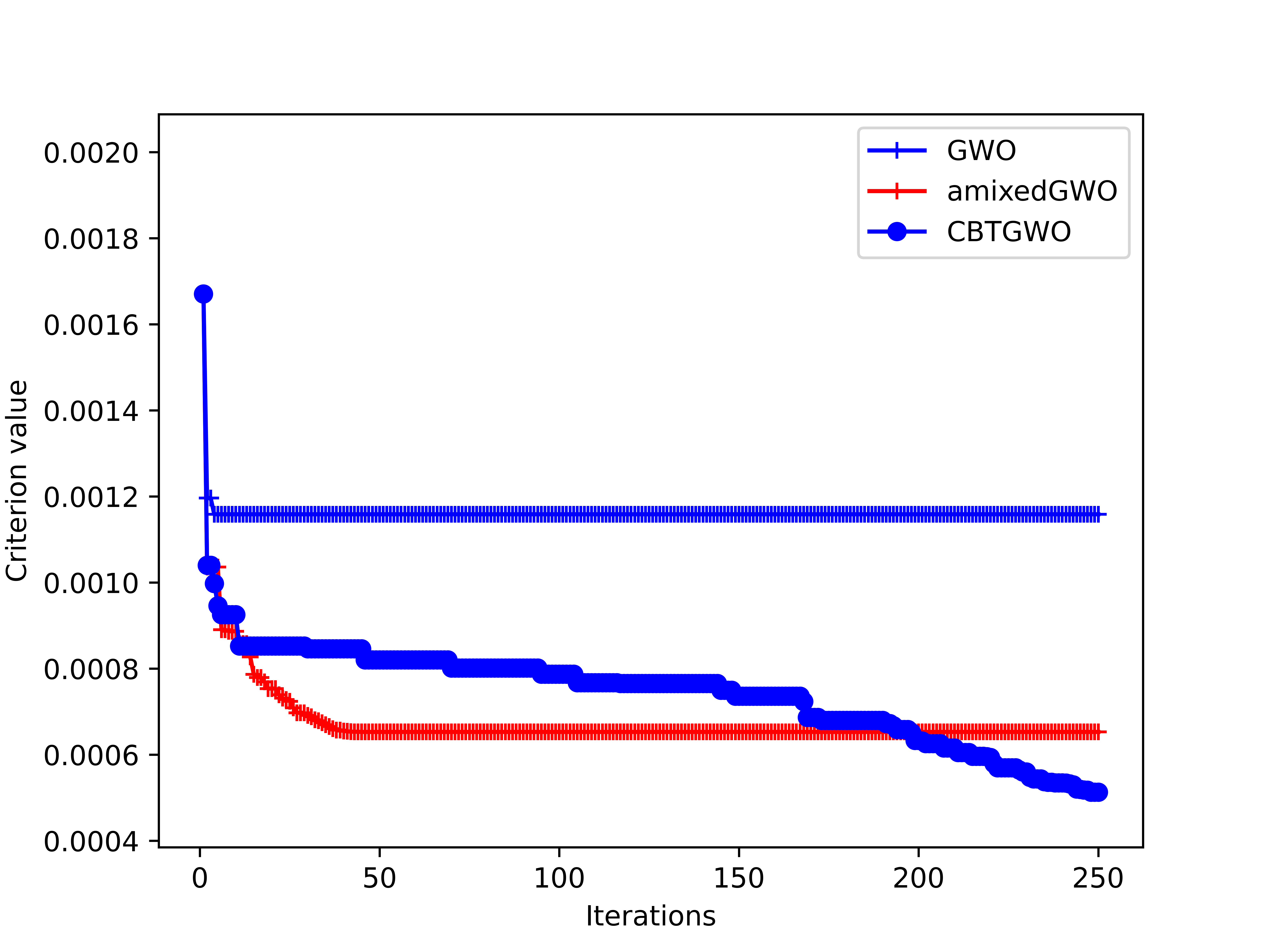} 
		\caption{Convergence curve of GWO, amixedGWO, and CTGWO algorithms \label{fig:convcurvesmonoobjcloaks}}}
\end{figure}

\begin{figure}[H]
    \centering{
        \begin{tabular}{c@{\,}c@{\,}}
             Cloak design & Wave propagation field\\
            \includegraphics[width=0.49\linewidth]{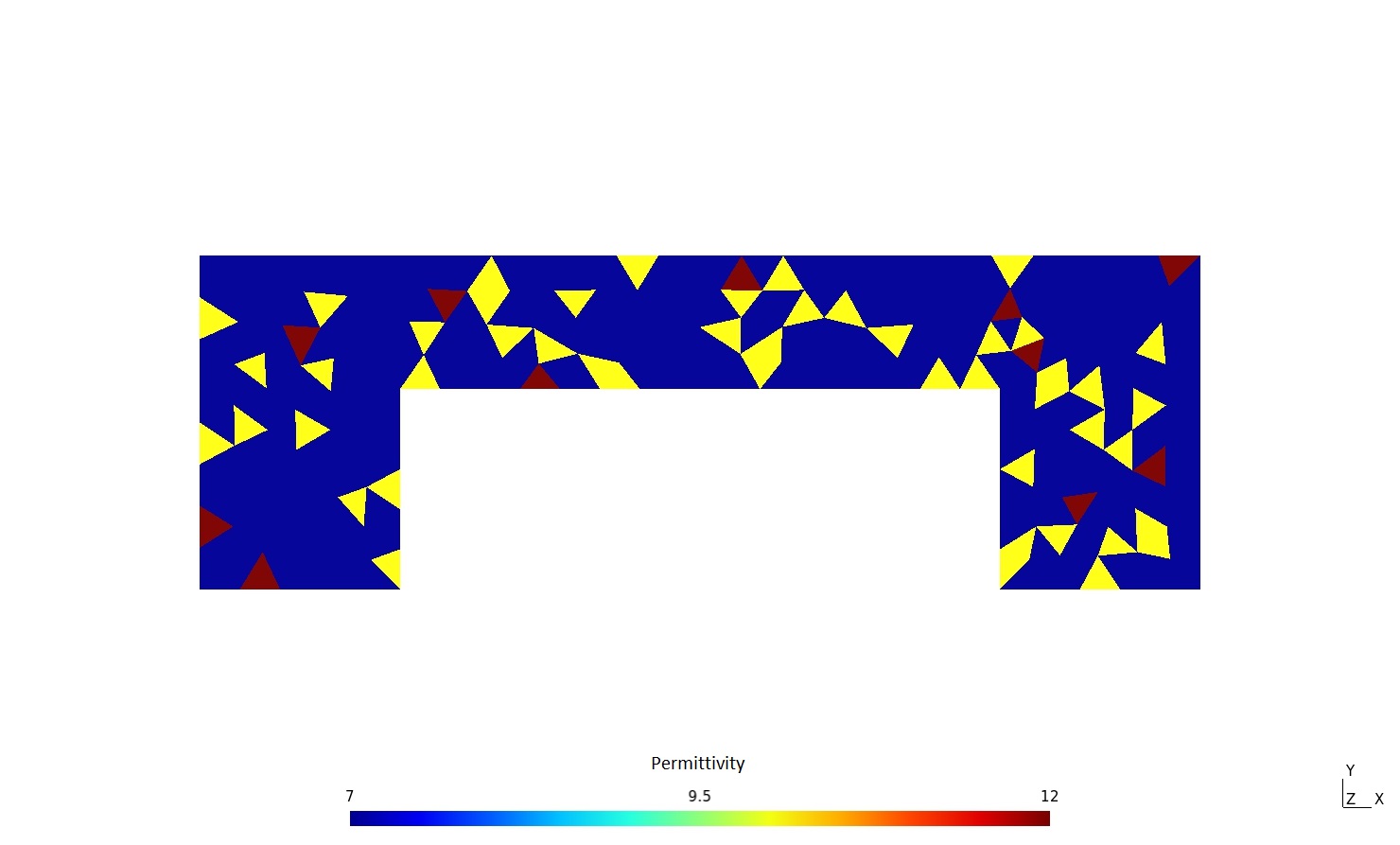}    &\includegraphics[width=0.49\linewidth]{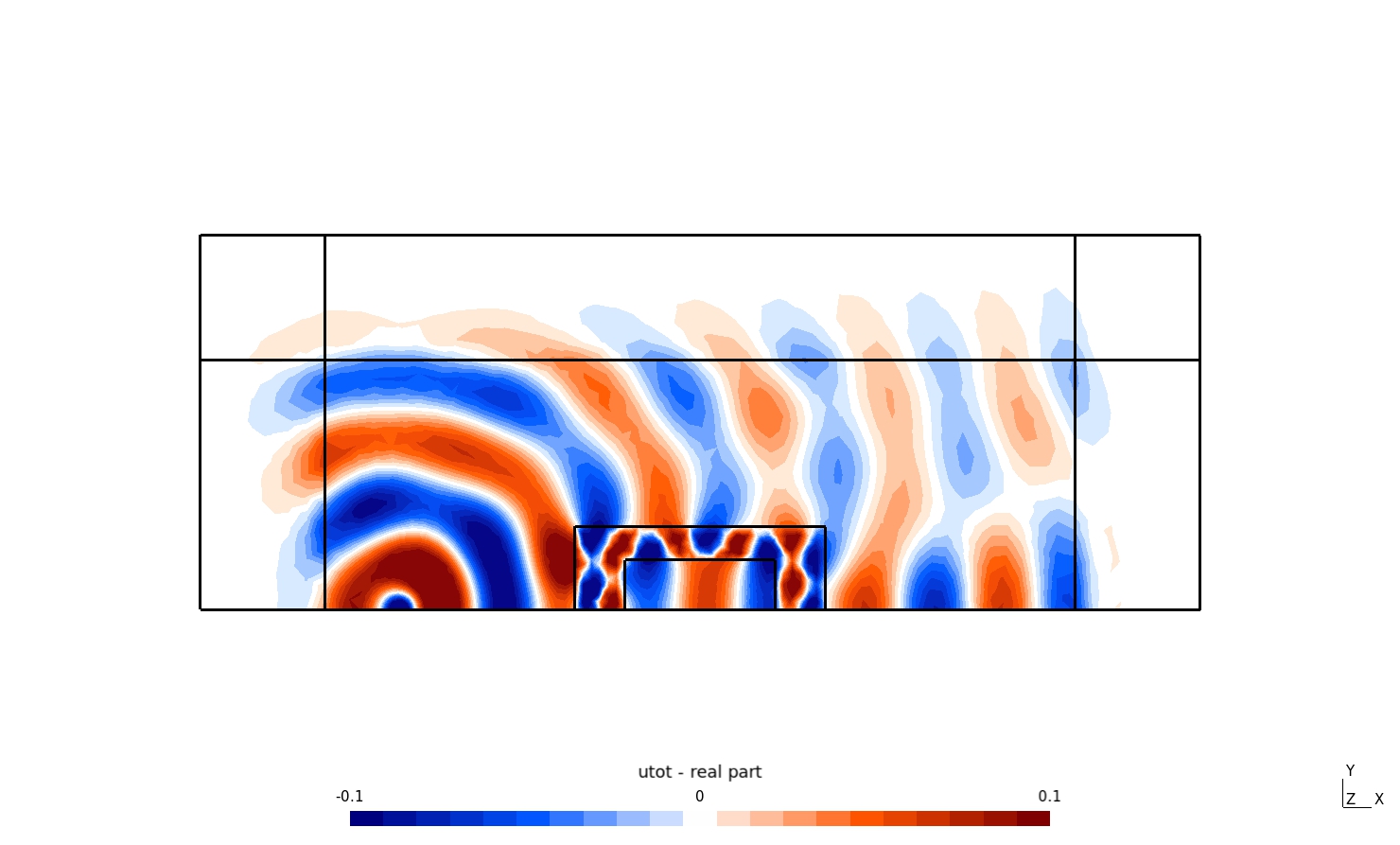}\\
         (a) & (d) \\
        \includegraphics[width=0.49\linewidth]{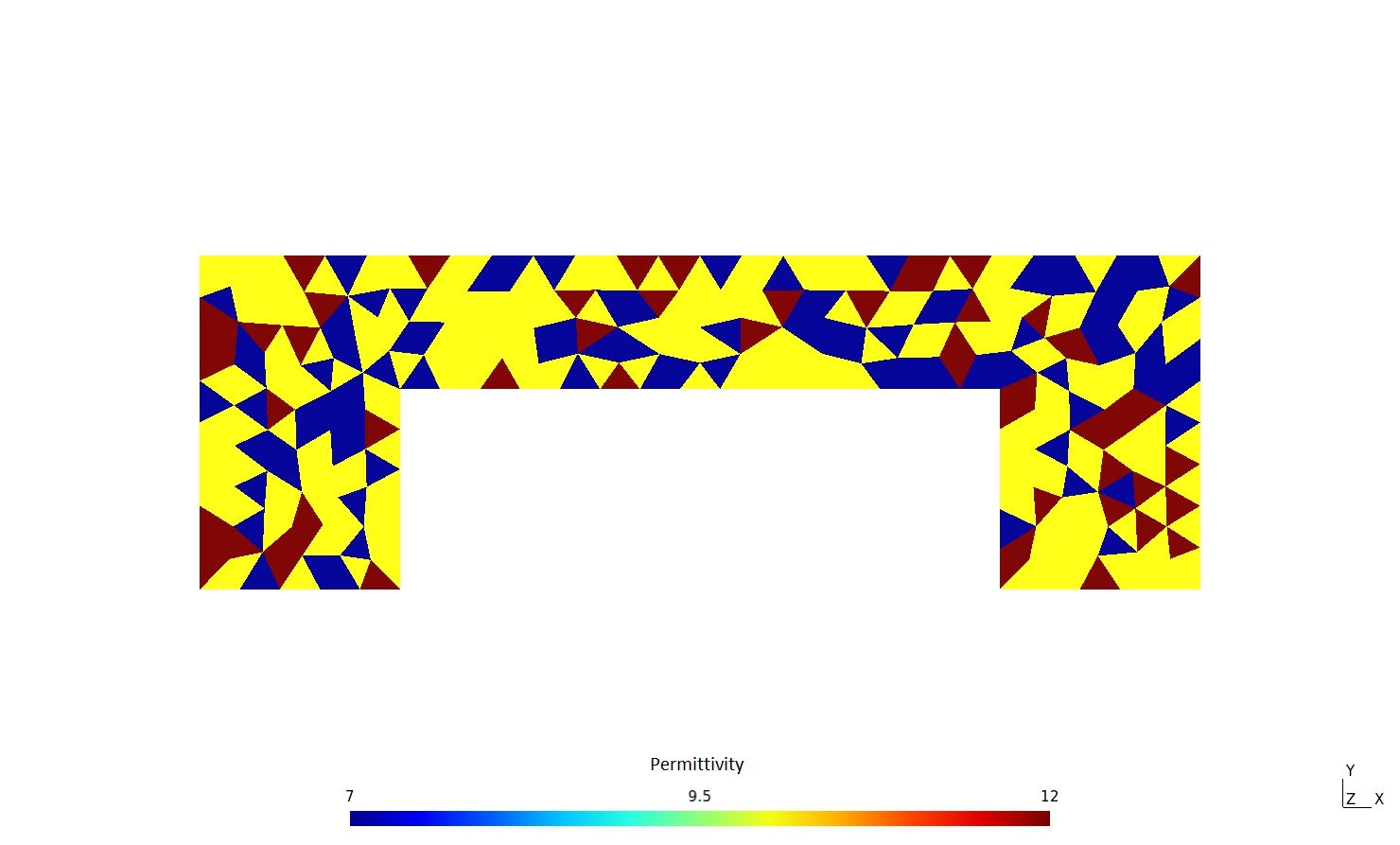}    &\includegraphics[width=0.49\linewidth]{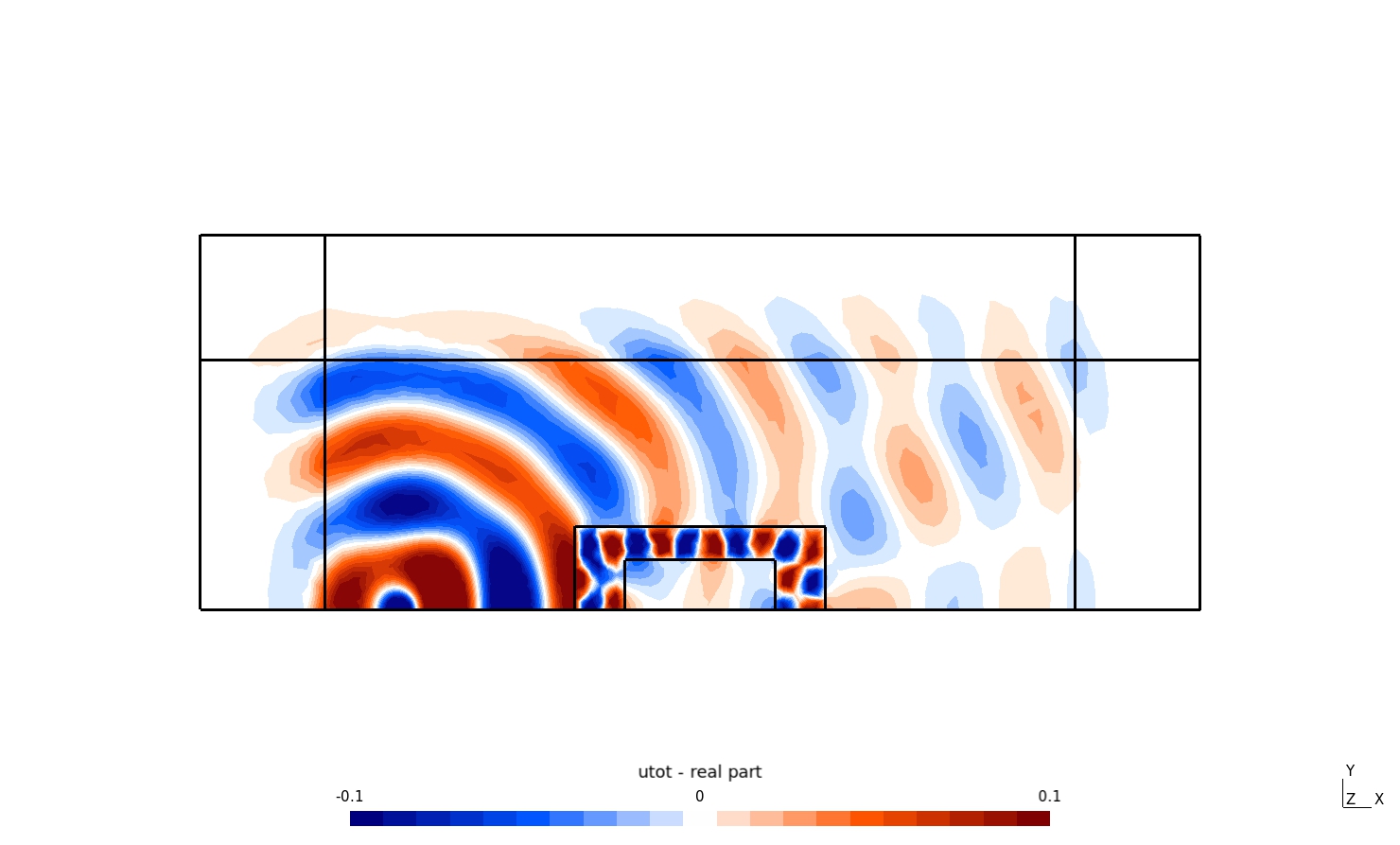}\\
         (b) & (e) \\
        \includegraphics[width=0.49\linewidth]{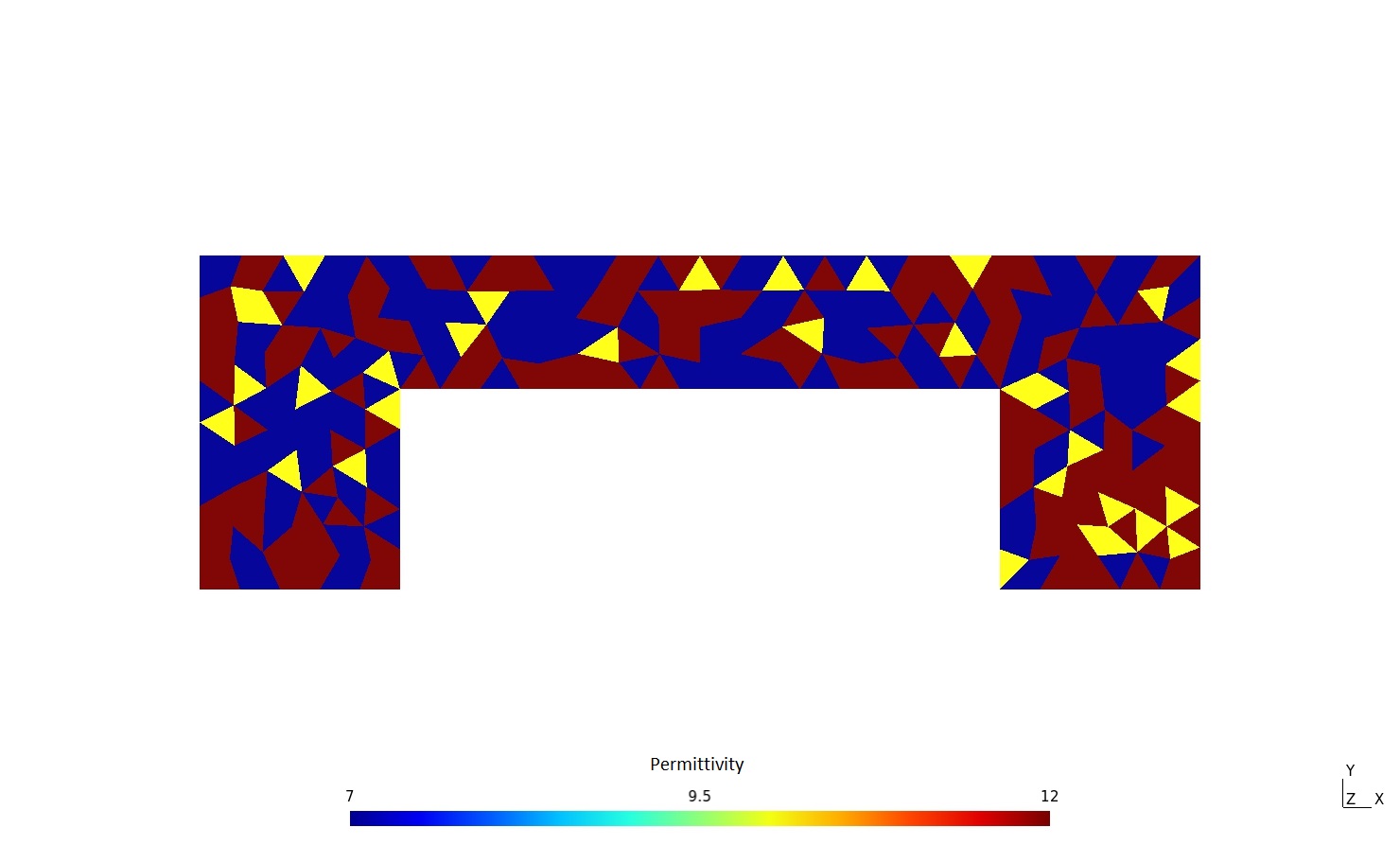}    &\includegraphics[width=0.49\linewidth]{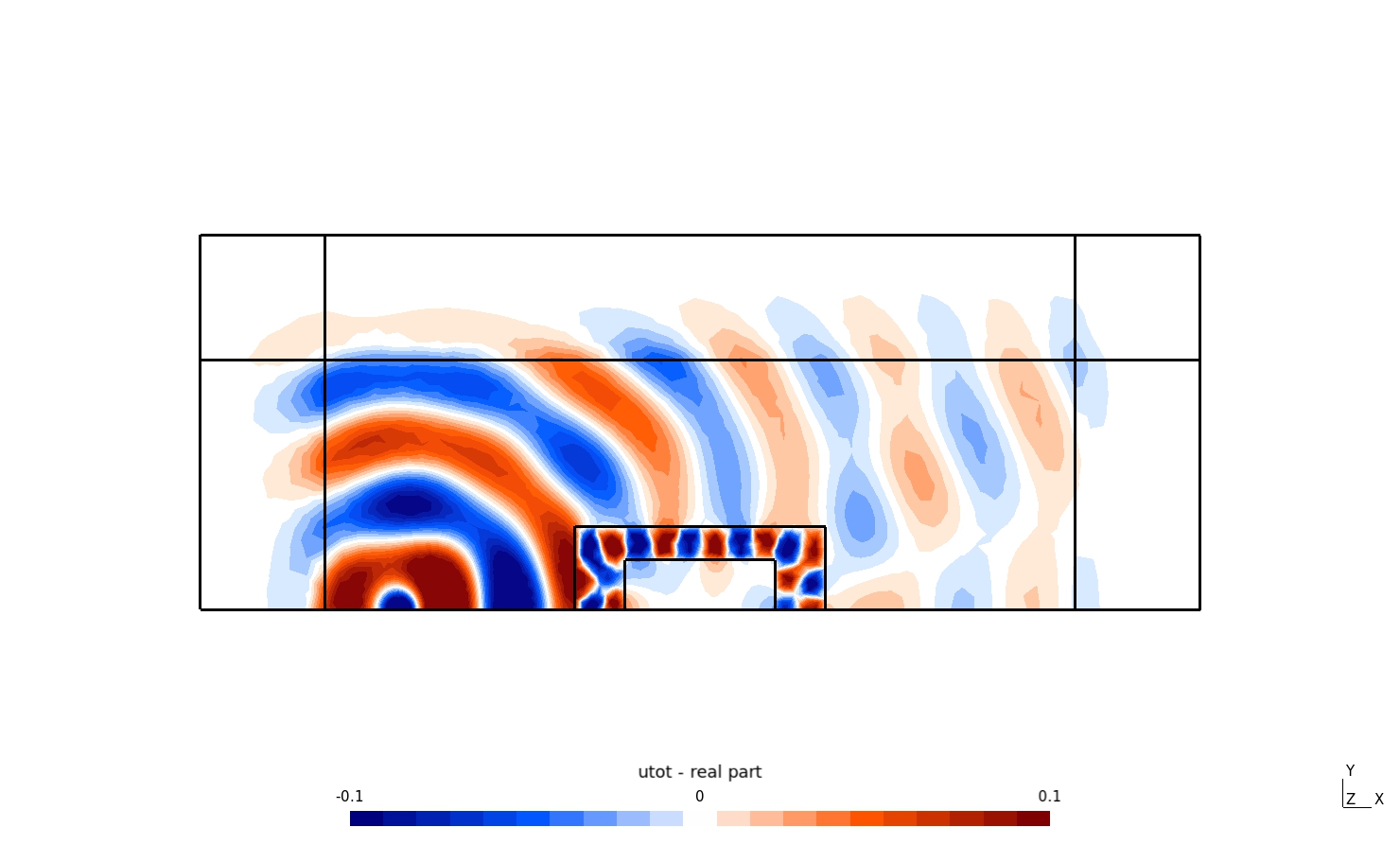}\\
         (c) & (f) \\
        \end{tabular}
        \caption{Column 1: in (a) GWO optimized cloak, in (b) Discrete GWO optimized cloak, and in (c), CTGWO optimized cloak; Column 2: in (d) GWO result, in (e) Discrete GWO result, and in (f), CTGWO result  \label{fig:f5}}}
\end{figure}

In this subsection, we display the results obtained by CTGWO, amixedGWO, and GWO: the convergence curves in Fig. \ref{fig:convcurvesmonoobjcloaks}; the scores reached by each method, and corresponding protection and invisibility performances in Table \ref{Table:numresmonoobjective}; the cloak design and wave propagation field in Fig. \ref{fig:f5}. 

\begin{table*}[h!]
	\centering
	\resizebox{1.0\textwidth}{!}{
\begin{tabular}{c|ccc|c}
\hline
 & Protection $f_{1}(\xK{\alpha}{\rm T_{max}})$ & Invisibility $f_{2}(\xK{\alpha}{\rm T_{max}})$ & Score $f(\xK{\alpha}{\rm T_{max}})$&Links to figures\\
\hline
 GWO&\myround{7}{0.0042343101366780255}&\myround{7}{0.0017399327049968023}&\myround{7}{0.002987121420837414}& fig. \ref{fig:f5}(d)\\
 amixedGWO&\myround{7}{0.0003481234829584469}&\myround{7}{0.0009572061288221007}&\myround{7}{0.0006526648058902739}& fig. \ref{fig:f5}(e)\\
 CTGWO&\bf{\myround{7}{0.00015608238359645862}}&\bf{\myround{7}{0.0008695749972742106}}&\bf{\myround{7}{0.0005128286904353347}}& fig. \ref{fig:f5}(f)\\
			\hline
		\end{tabular}
	}
	\caption{Comparison of GWO, amixedGWO and CTGWO mono-objective methods in protection ($f_{1}$) and invisibility ($f_{2}$) for a rectangular and half-rectangular cloak. ${\rm T_{max}}=250$, $Q=100$}\label{Table:numresmonoobjective}
\end{table*}

\newpage

In Table \ref{Table:numresmonoobjective}, we display, for GWO, amixedGWO, and CTGWO, the score of wolf $\alpha$ at the last iteration $f(\xK{\alpha}{\rm T_{max}})$. For instance, the solution provided by CTGWO is $\xK{\alpha}{\rm T_{max}}=\left[7,12,12,\ldots,10\right]^{T}$. We notice that, in CTGWO, the power map is such that there is still a small probability that the number of the area in which we are going to place ourselves does not correspond to the value obtained, used as a position on the horizontal axis. The interest of such a mismatch is to avoid our algorithm to be locked in a local minimum. The chaotic sequences also play a role in the good behavior of CTGWO shown by the convergence curve in Fig. 
\ref{fig:convcurvesmonoobjcloaks}. Fig. \ref{fig:f5}(f) indicates that the protection abilities of the cloak designed by CTGWO are clearly better than for GWO and amixedGWO. Values in \ref{Table:numresmonoobjective} confirm this impression: the protection criterion is two times smaller, while the mean criterion is also significantly smaller.  

\newpage

\subsection{Bi-objective approach:}\label{subsec:biobjectiveapproach}
In a second step, we used a bi-objective genetic algorithm as an optimization method.
Thus, we no longer work on convergence curves, but on Pareto fronts, to balance the two minimized criteria.\\
The parameter values for NSGA-II are as follows: mutation and crossover probabilities are $p_{m} = 1/P \simeq 0.003$ and $p_{c} = 0.9$;
mutation and crossover distribution indices are $\eta_{m}=20$ and $\eta_{c}=15$.

\subsubsection{Bi-objective ternary optimization with 250 iterations, 100 research agents.}\label{subsubsec:biobjectiveternary250_100}

\begin{figure}[H]
	\centering{
			\includegraphics[scale=0.4]{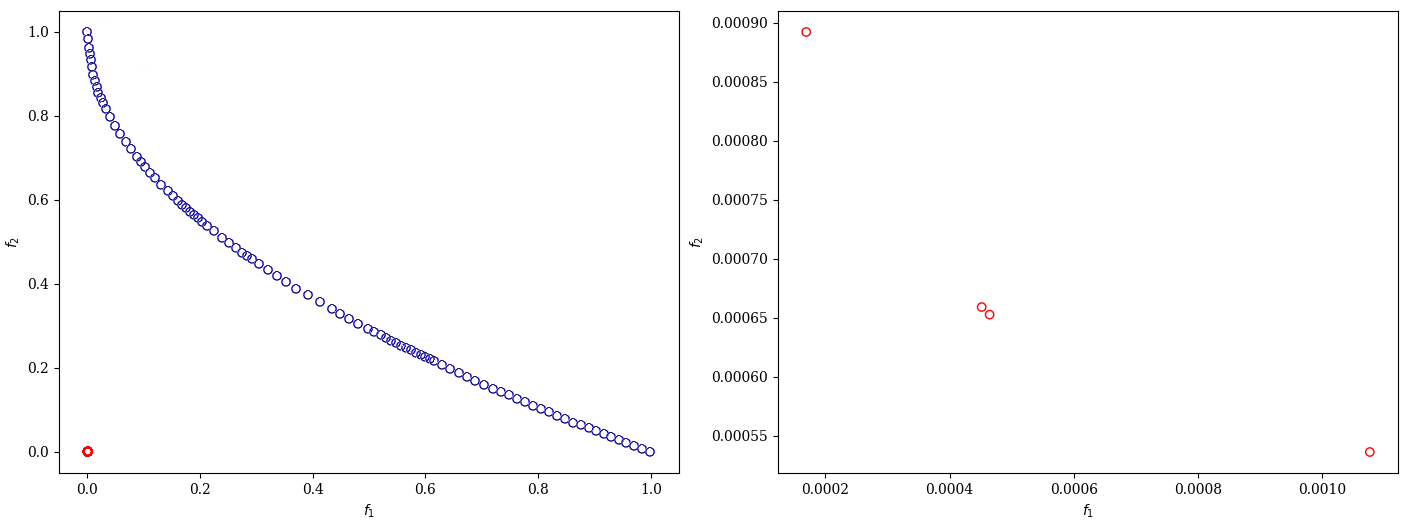}
		\caption{Left: Pareto front for  NSGA-II Table with 250 iterations, 100 research agents and four selected solutions marked in red. \textcolor{black}{Right: Magnified view of the four selected solutions, three of which appear in \ref{Table:NSGAII250_100}.}\label{fig:paretofront250_100} }}
\end{figure}

\begin{figure}[H]
    \centering{
        \begin{tabular}{c@{\,}c@{\,}}
             Cloak design & Wave propagation field\\
            \includegraphics[width=0.49\linewidth]{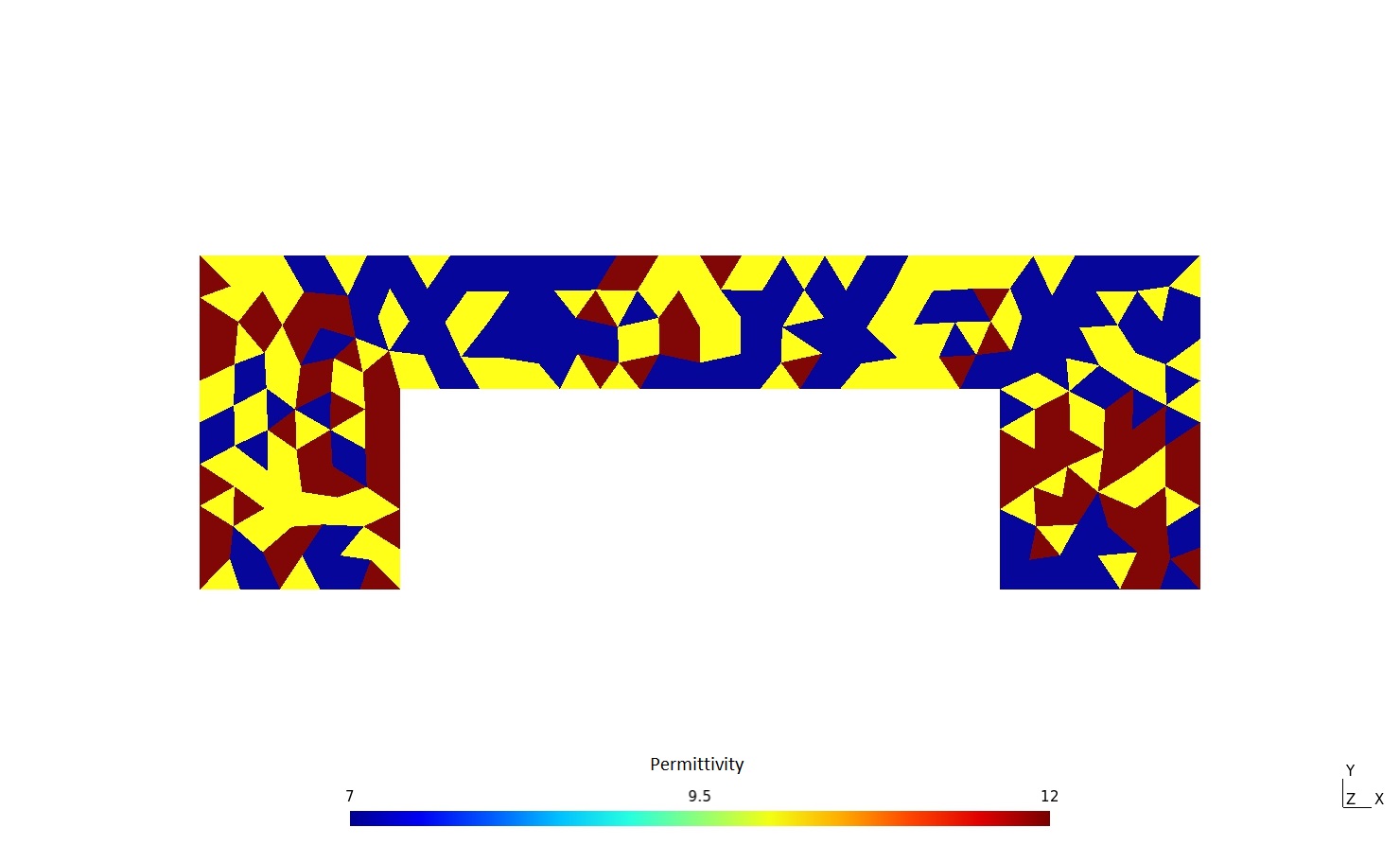}    &\includegraphics[width=0.49\linewidth]{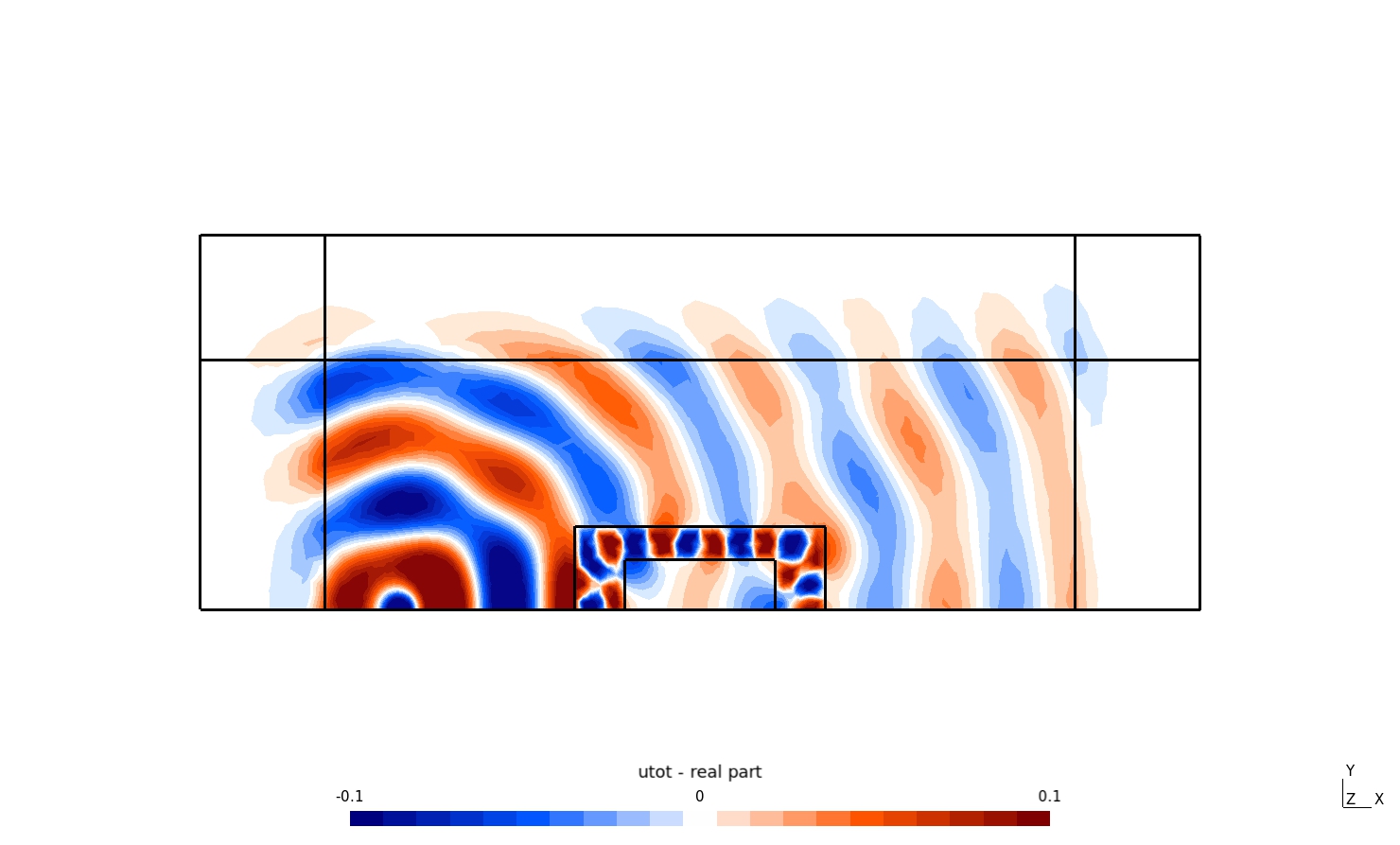}\\
         (a) & (d) \\
        \includegraphics[width=0.49\linewidth]{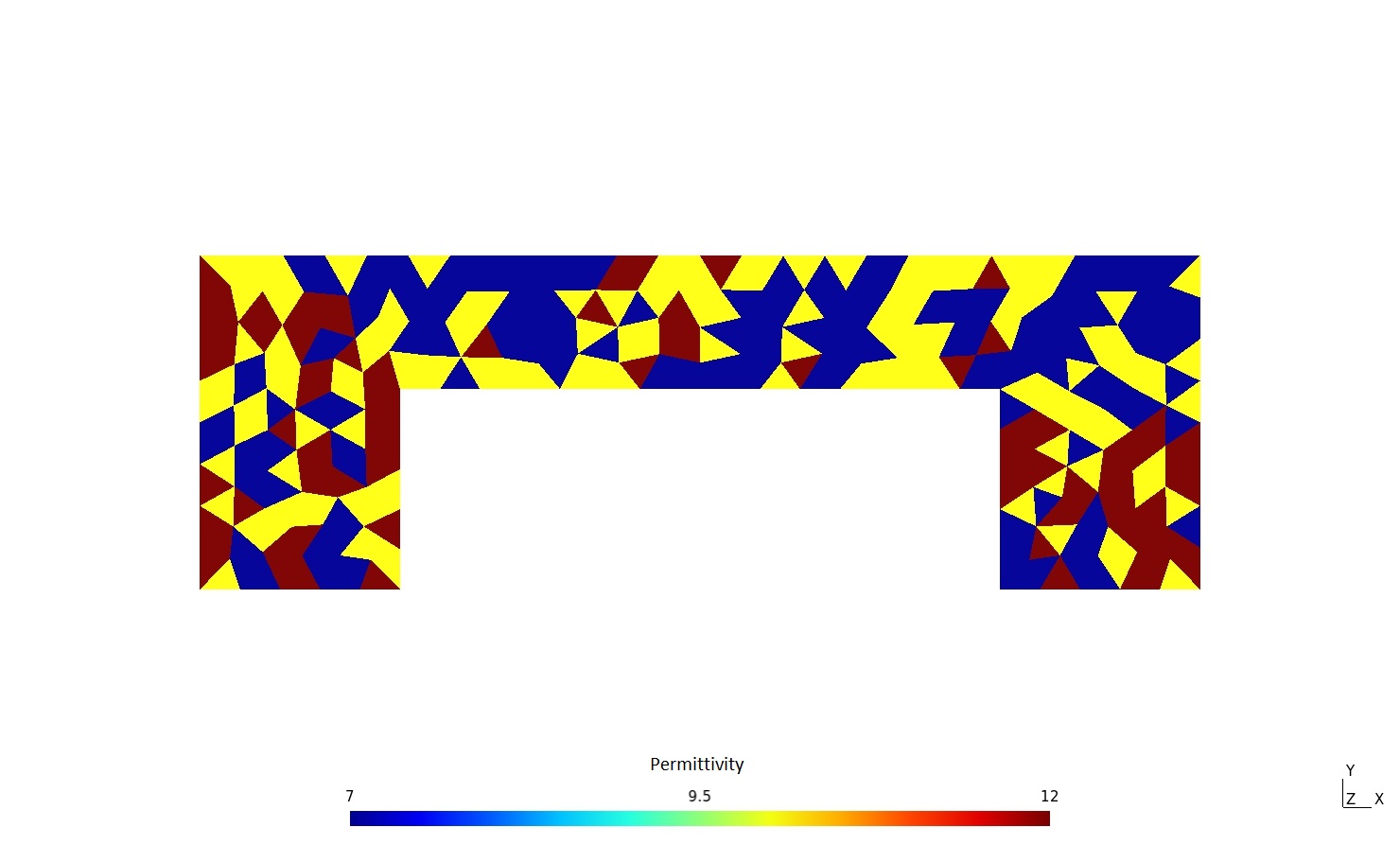}    &\includegraphics[width=0.49\linewidth]{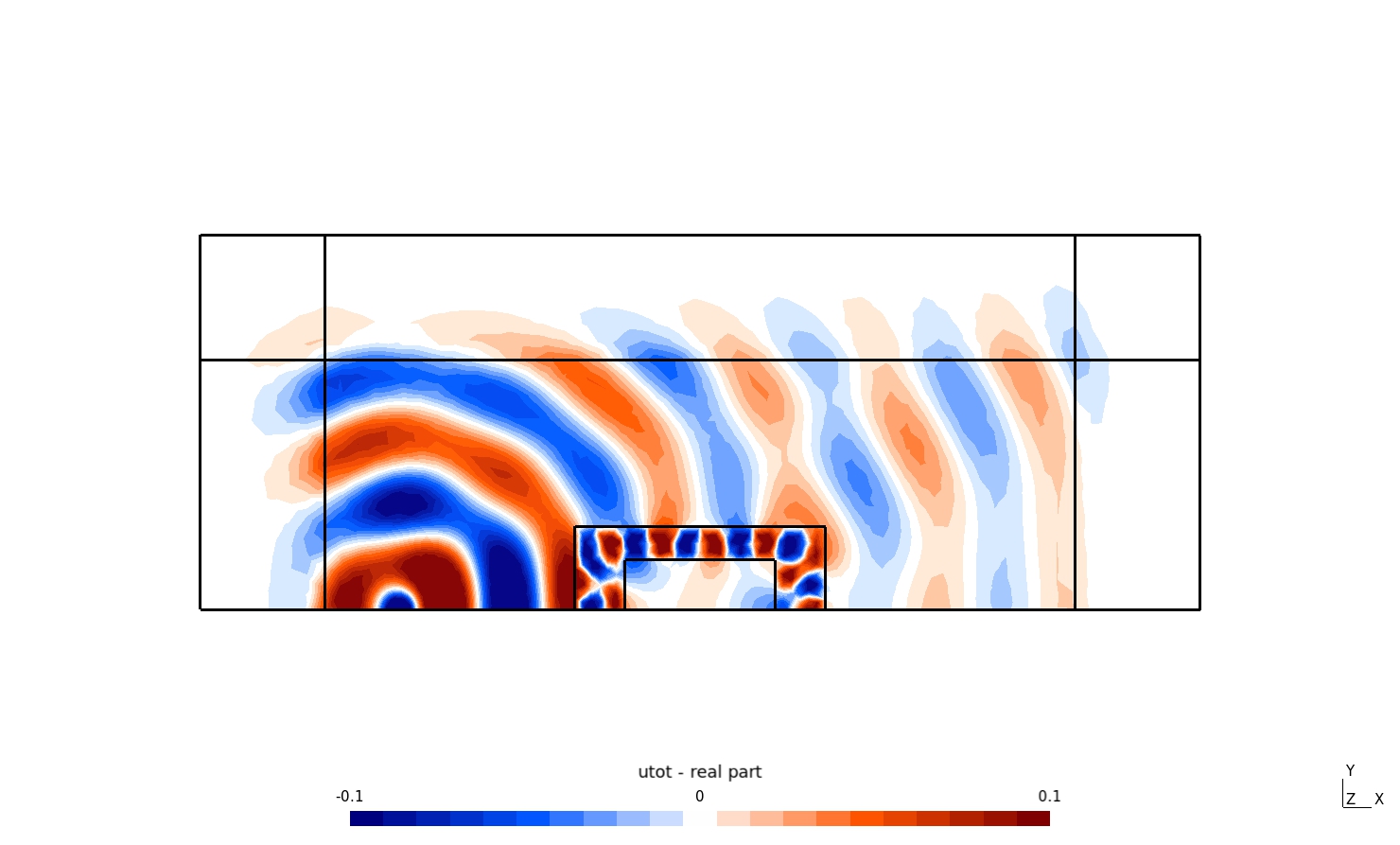}\\
         (b) & (e) \\
        \includegraphics[width=0.49\linewidth]{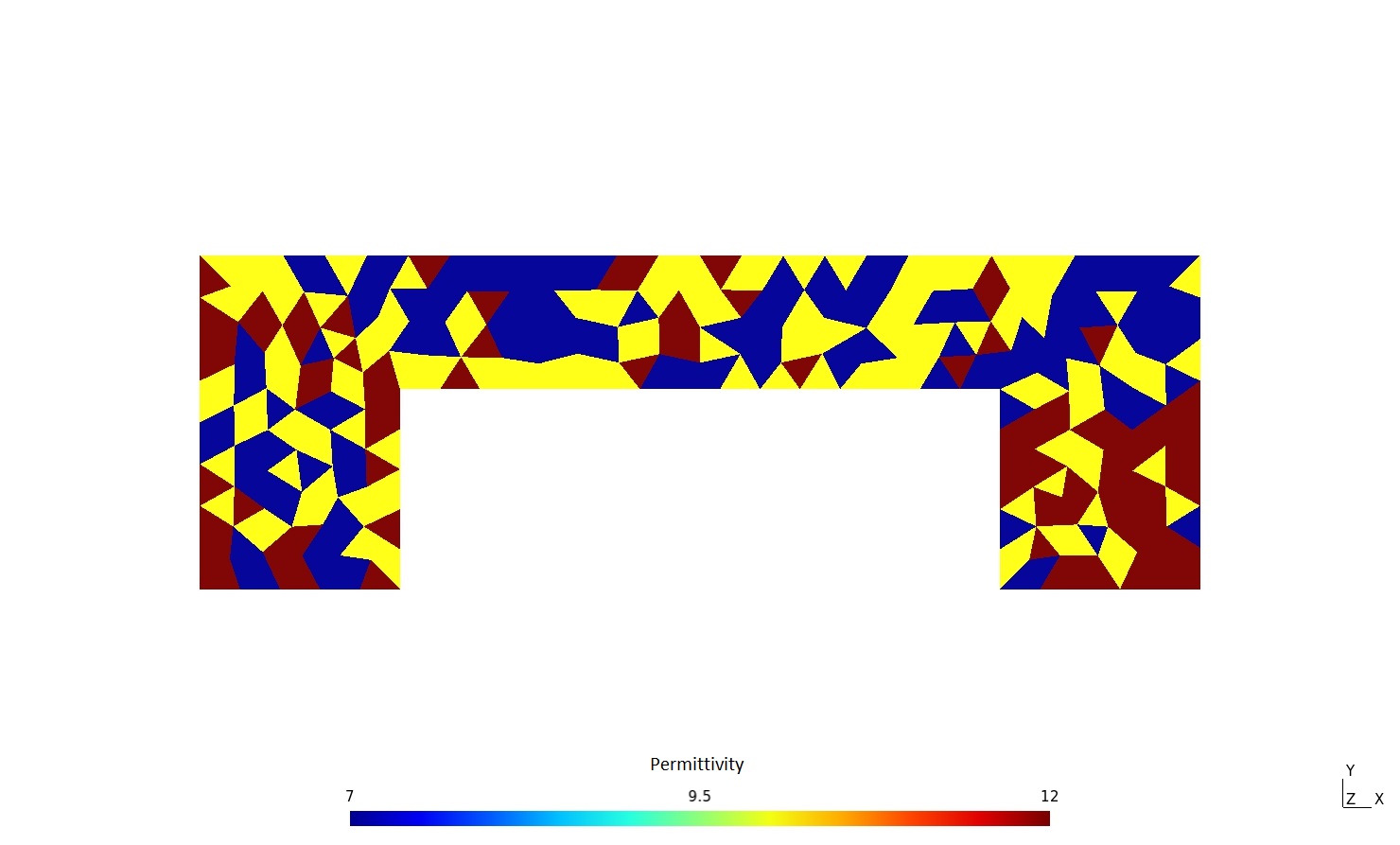}    &\includegraphics[width=0.49\linewidth]{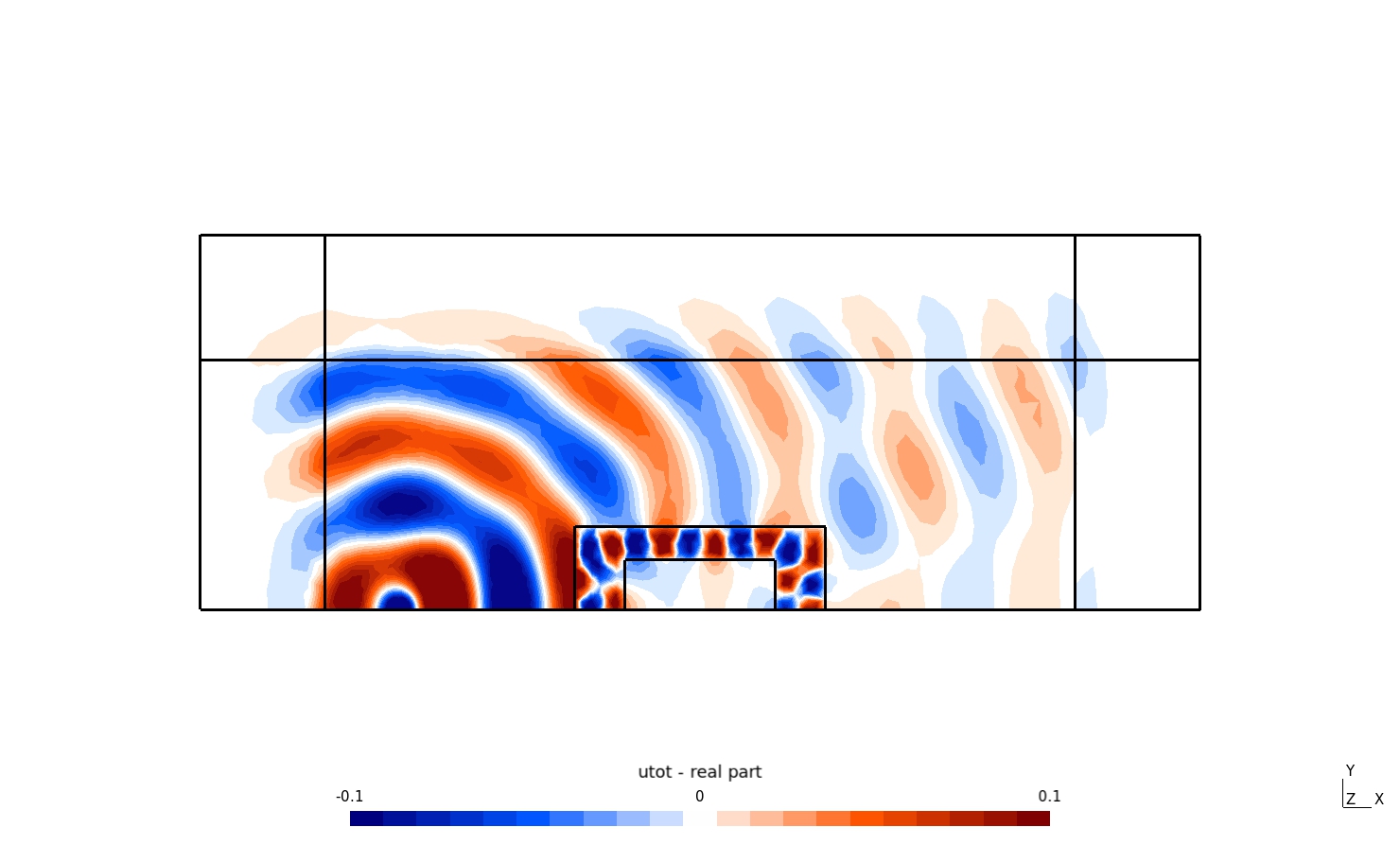}\\
         (c) & (f) \\
        \end{tabular}
        \caption{ Column 1: in (a) Best invisibility cloak, in (b) Best compromise between invisibility and protection cloak, and in (c), Best protection cloak; Column 2: in (d) Best invisibility result, in (e) Best compromise between invisibility and protection result, and in (f), Best protection result. See Table \ref{Table:NSGAII250_100} for numerical values.}
        \label{fig:f7}}
\end{figure}

In this subsubsection, we display the results obtained by NSGA-II: the Pareto front in Fig. \ref{fig:paretofront250_100}; the protection and invisibility performances in Table \ref{Table:NSGAII250_100}; the cloak design and wave propagation field in Fig. \ref{fig:f7}. 

\begin{table*}[h!]
	\centering
	\resizebox{1.0\textwidth}{!}{
\begin{tabular}{c|ccc|c}
\hline
 & Protection $f_{1}$ & Invisibility $f_{2}$ & $\frac{f_{1}+f_{2}}{2}$&Links to figures\\
\hline
 Best Protection&\bf{\myround{7}{0.00016903308539718457}}&\myround{7}{0.0008922496715084679}& \bf{\myround{7}{0.0005306413784528263}}&fig. \ref{fig:f7}(f)\\
 Best Invisibility&\myround{7}{0.0010765244967287531}&\bf{\myround{7}{0.0005362717180578936}}&\myround{7}{0.0008063981073933233}&fig. \ref{fig:f7}(d)\\
 Best Compromise&\myround{7}{0.00046434484599299766}&\myround{7}{0.000652728258292644}&\myround{7}{0.0005585365521428208}&fig. \ref{fig:f7}(e)\\
\hline
        	\end{tabular}
         }
			\caption{Comparison of a NSGA-II bi-objective method in protection and invisibility for a half-rectangular cloak. ${\rm T_{max}}=250$, $Q=100$, \textcolor{black}{see fig. \ref{fig:paretofront250_100} for the Pareto front and fig. \ref{fig:f7} for 2D plots of cloak design and wave fields.}  \label{Table:NSGAII250_100} }
\end{table*}

\newpage

\newpage

\subsubsection{Bi-objective ternary optimization with 1000 iterations, 200 research agents.}\label{subsubsec:biobjectiveternary1000_100}

\begin{figure}[H]
	\centering{
			\includegraphics[scale=0.45]{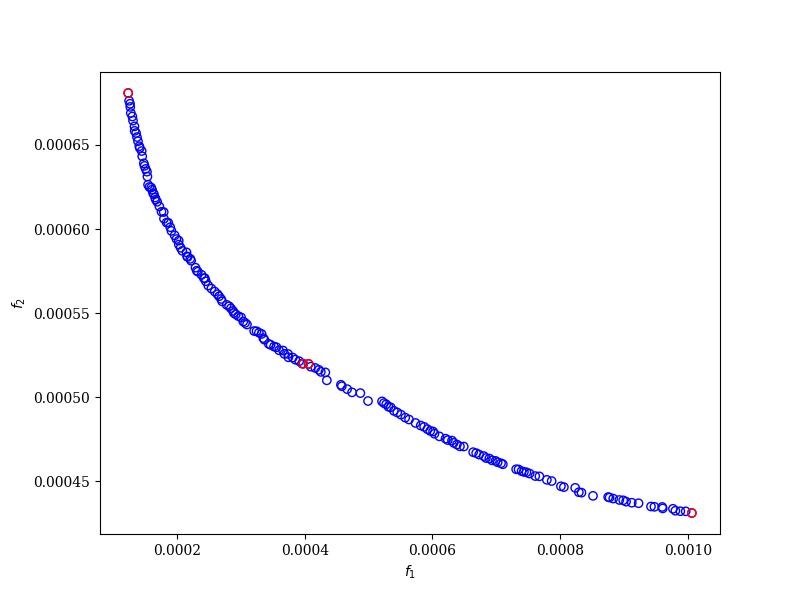}

		\caption{Pareto front for for  NSGA-II Table with 250 iterations, 100 research agents, with four selected solutions out of a total of 175. \label{fig:paretofront1000_200}}}
\end{figure}

\begin{figure}[H]
    \centering{
        \begin{tabular}{c@{\,}c@{\,}}
             Cloak design & Wave propagation field\\
            \includegraphics[width=0.49\linewidth]{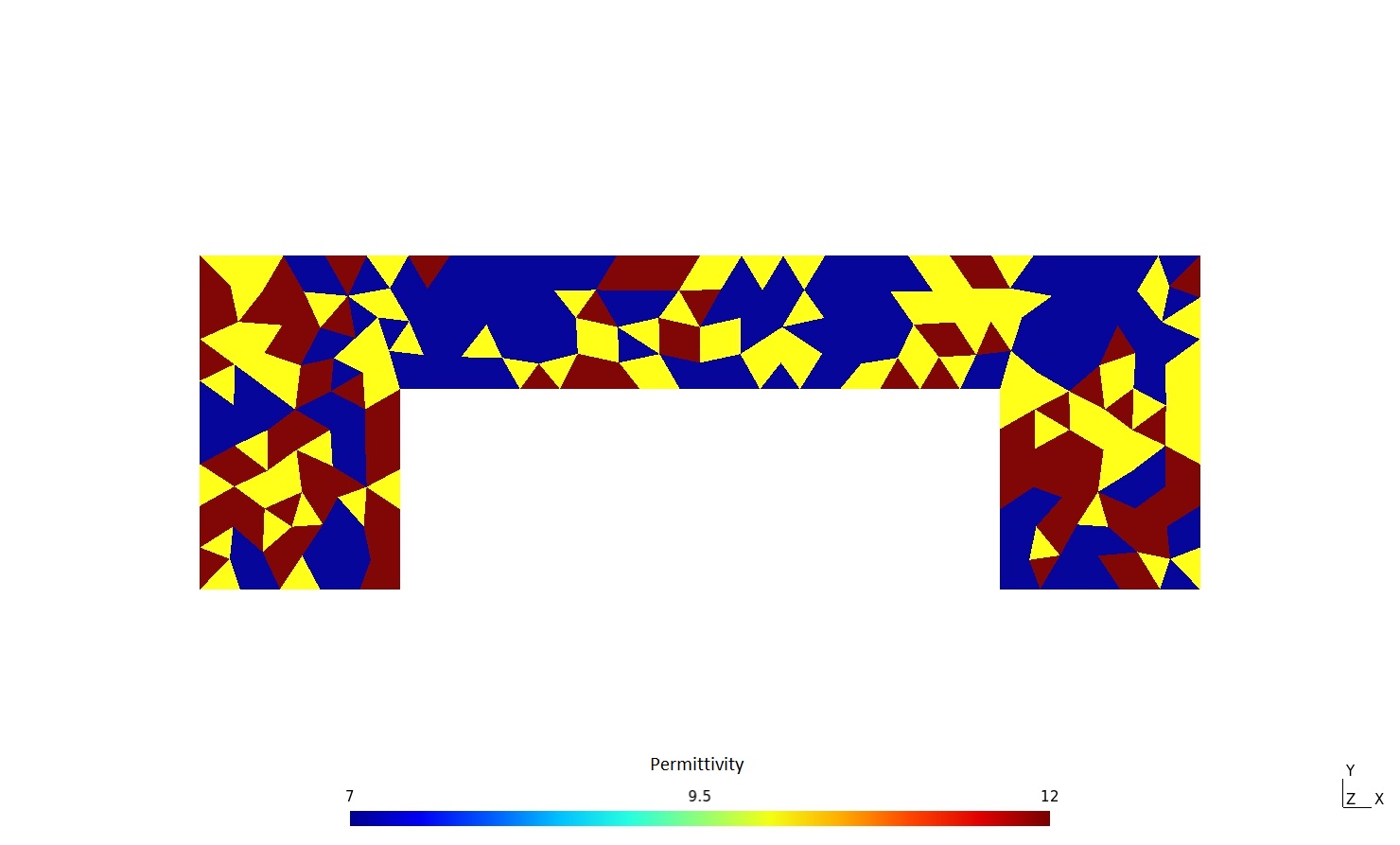} &\includegraphics[width=0.49\linewidth]{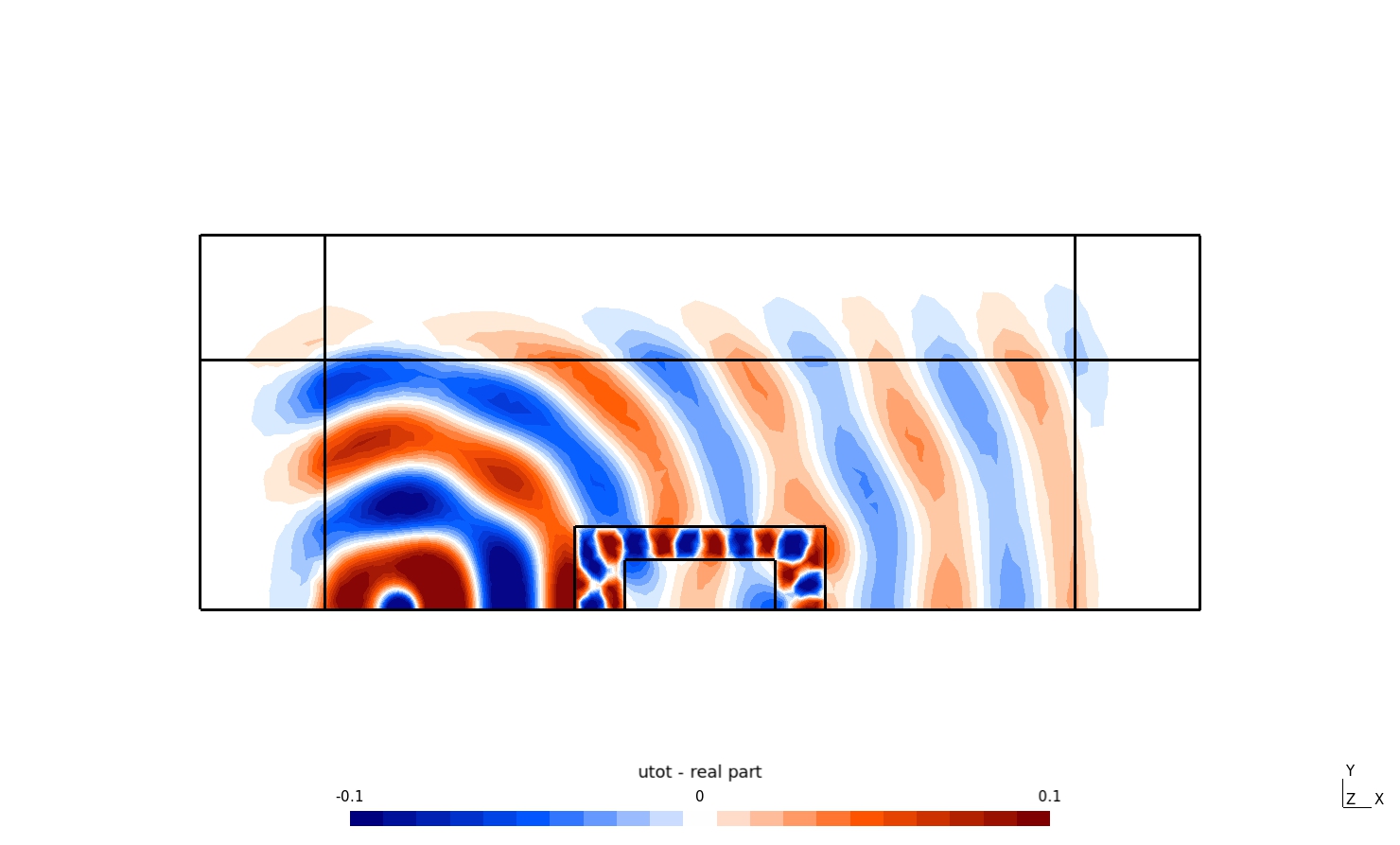}\\
         (a) & (d) \\
        \includegraphics[width=0.49\linewidth]{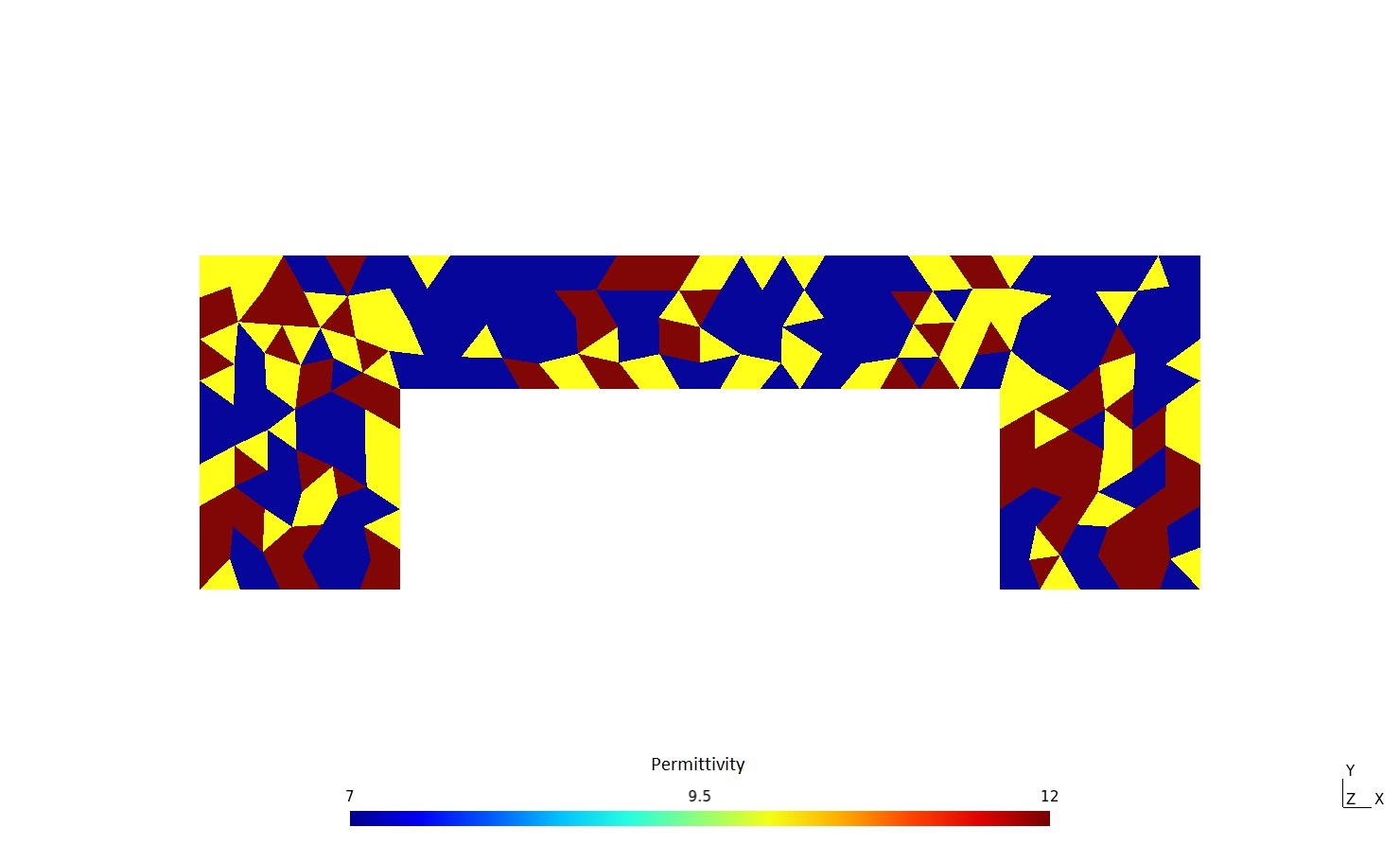}    &\includegraphics[width=0.49\linewidth]{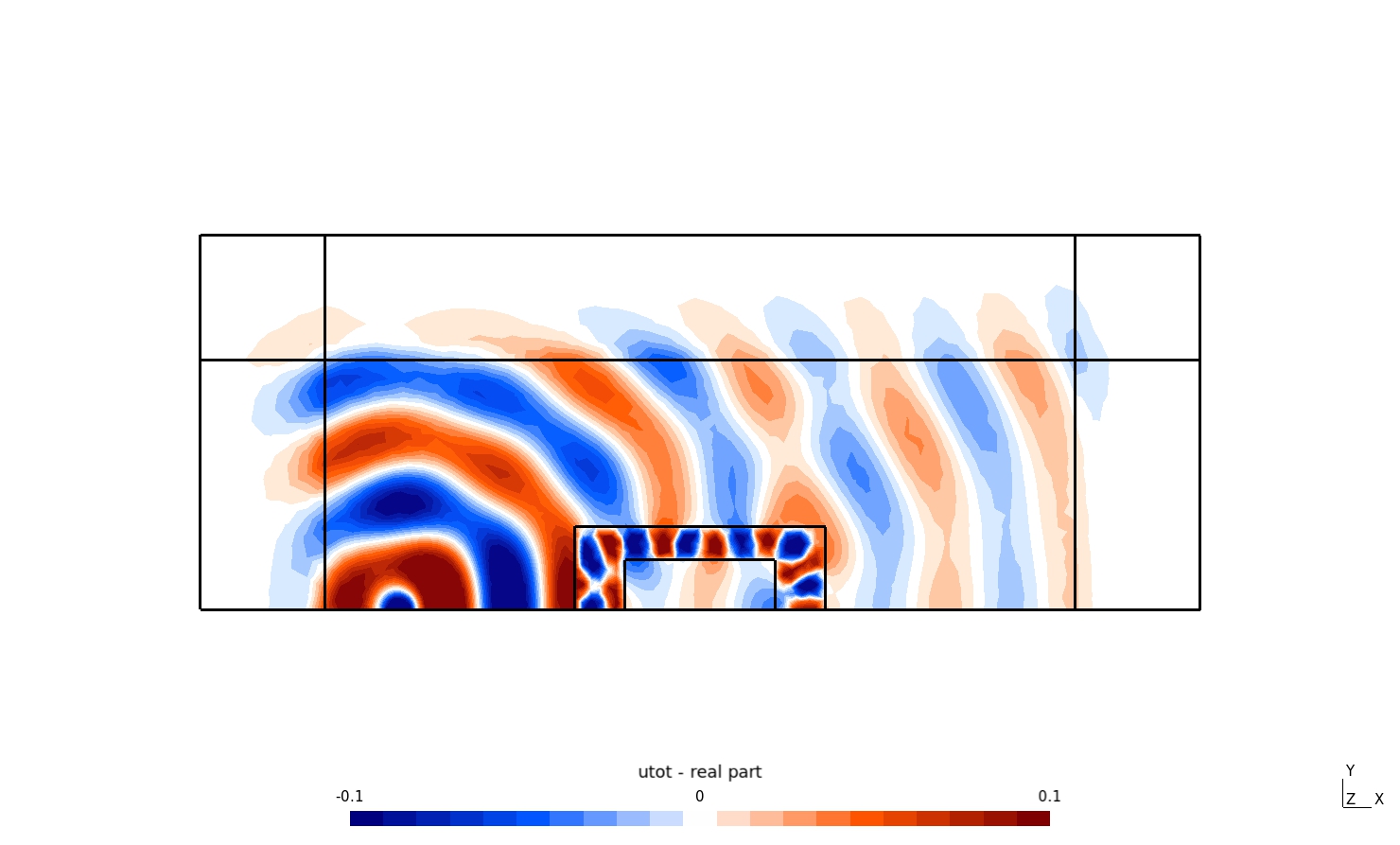}\\
         (b) & (e) \\
        \includegraphics[width=0.49\linewidth]{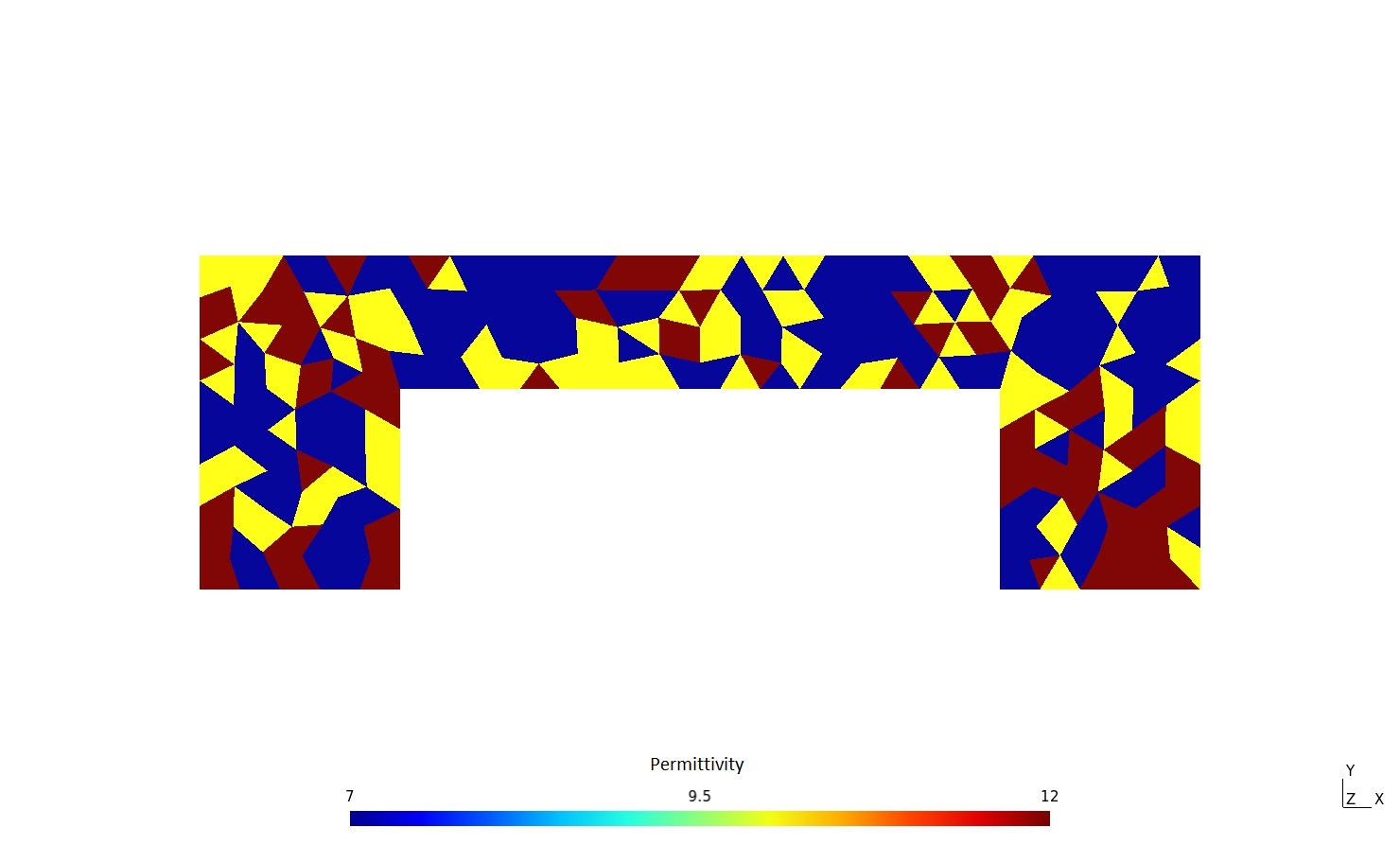}    &\includegraphics[width=0.49\linewidth]{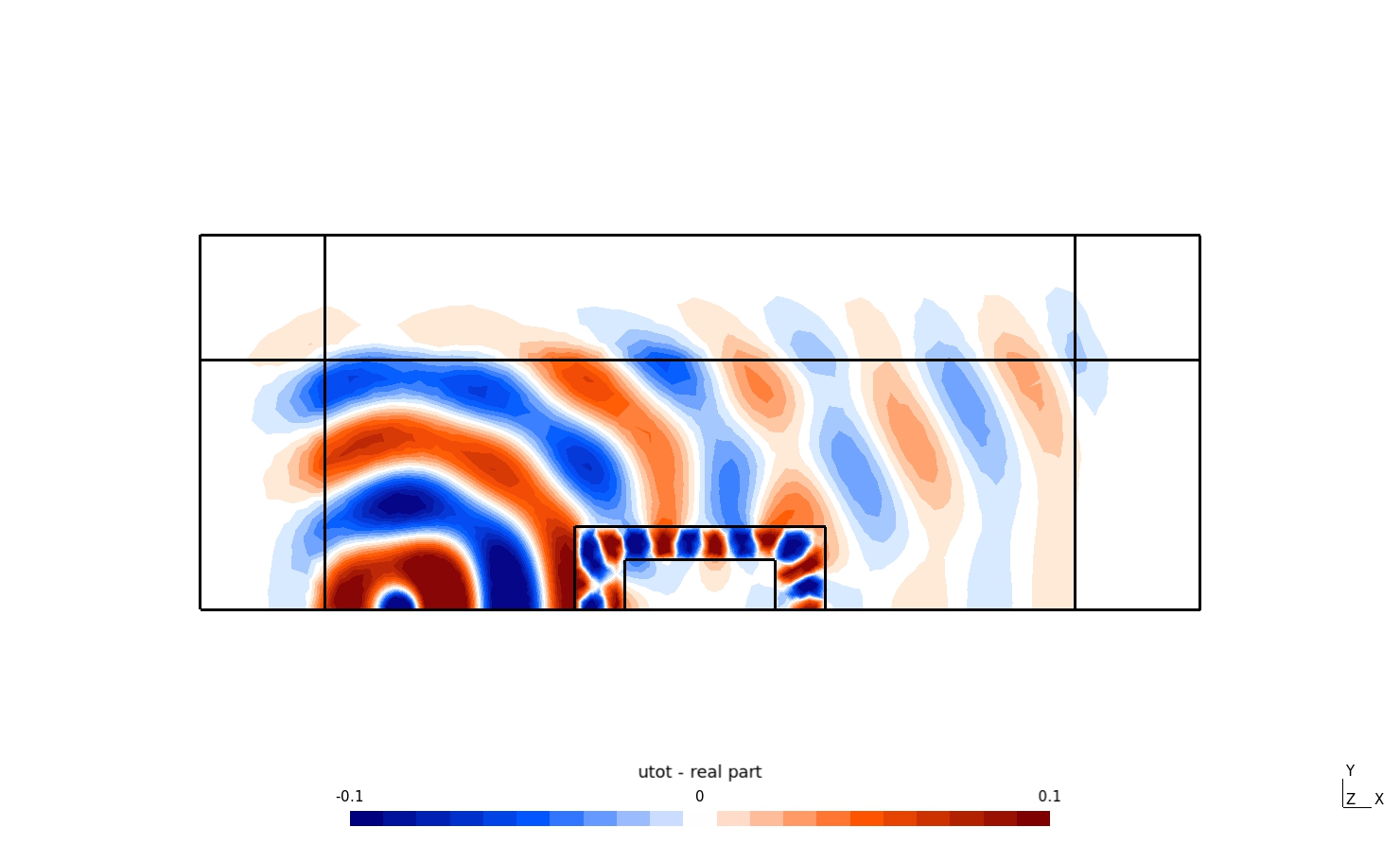}\\
         (c) & (f) \\
        \end{tabular}
        \caption{ Column 1: in (a) Best invisibility cloak, in (b) Best compromise between invisibility and protection cloak, and in (c), Best protection cloak; Column 2: in (d) Best invisibility result, in (e) Best compromise between invisibility and protection result, and in (f), Best protection result. \label{fig:f9}}}
\end{figure}

In this subsubsection, we display the results obtained by NSGA-II: the Pareto front in Fig. \ref{fig:paretofront1000_200}; the protection and invisibility performances in Table \ref{Table:NSGAII1000_200}; the cloak design and wave propagation field in Fig. \ref{fig:f9}. 

\begin{table*}[h!]
	\centering
	\resizebox{1.0\textwidth}{!}{
\begin{tabular}{c|ccc|c}
\hline
 & Protection $f_{1}$ & Invisibility $f_{2}$ & $\frac{f_{1}+f_{2}}{2}$&Links to figures\\
\hline
 Best Protection&\bf{\myround{7}{0.0001233556779295822}}&\myround{7}{0.0006808233349312879}& \bf{\myround{7}{0.0004020895064304350}} &fig. \ref{fig:f9}(f)\\
 Best Invisibility&\myround{7}{0.001006414416092534}&\bf{\myround{7}{0.00043123852989015136}}& \myround{7}{0.0007188264729913427} &fig. \ref{fig:f9}(d)\\
 Best Compromise&\myround{7}{0.00040597975114207725}&\myround{7}{0.0005198314758624204}& \myround{7}{0.0004629056135022488} &fig. \ref{fig:f9}(e)\\
				\hline
			\end{tabular}
   }
			\caption{Comparison of a NSGA-II bi-objective method in protection and invisibility for a half-rectangular cloak. ${\rm T_{max}}=1000$, $Q=200$. See fig. \ref{fig:paretofront250_100} for the Pareto front and fig. \ref{fig:f7} for 2D plots of cloak design and wave fields. \label{Table:NSGAII1000_200} }
\end{table*}

\newpage

\newpage

\subsubsection{Bi-objective continuous optimization with 250 iterations, 100 research agents.}\label{subsubsec:biobjectivecontinuous250_100}

\begin{figure}[H]
	\centering{
			\includegraphics[scale=0.4]{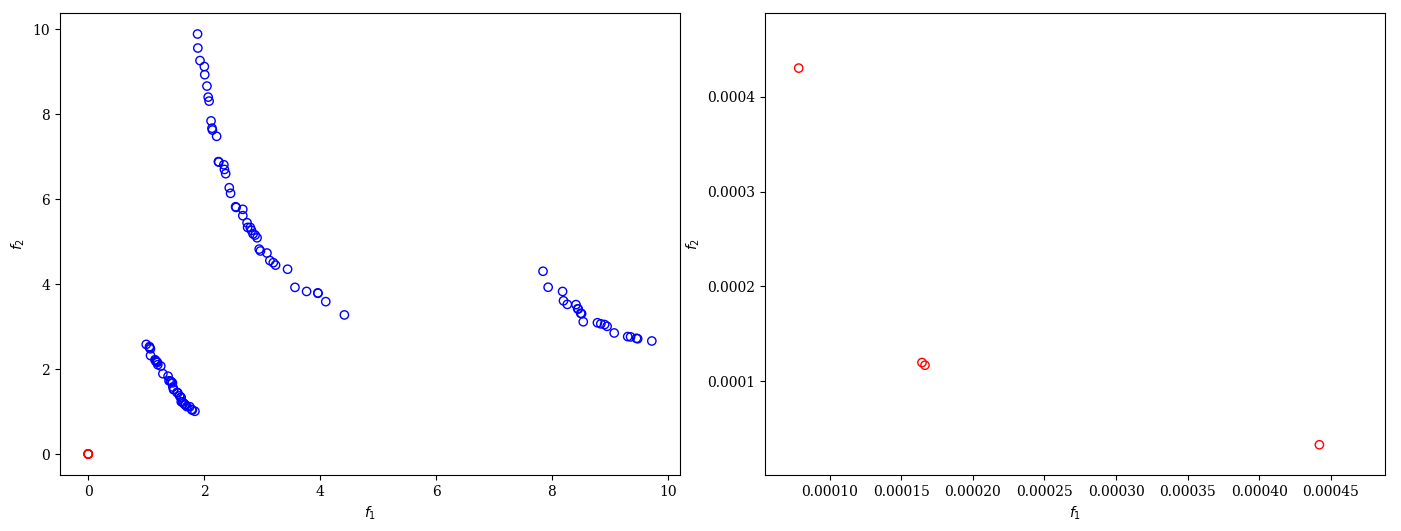}

		\caption{ In red, the four selected solutions. \label{fig:continuousparetofront250_100}}}
\end{figure}

\begin{figure}[H]
    \centering{
        \begin{tabular}{c@{\,}c@{\,}}
             Cloak design & Wave propagation field\\
            \includegraphics[width=0.49\linewidth]{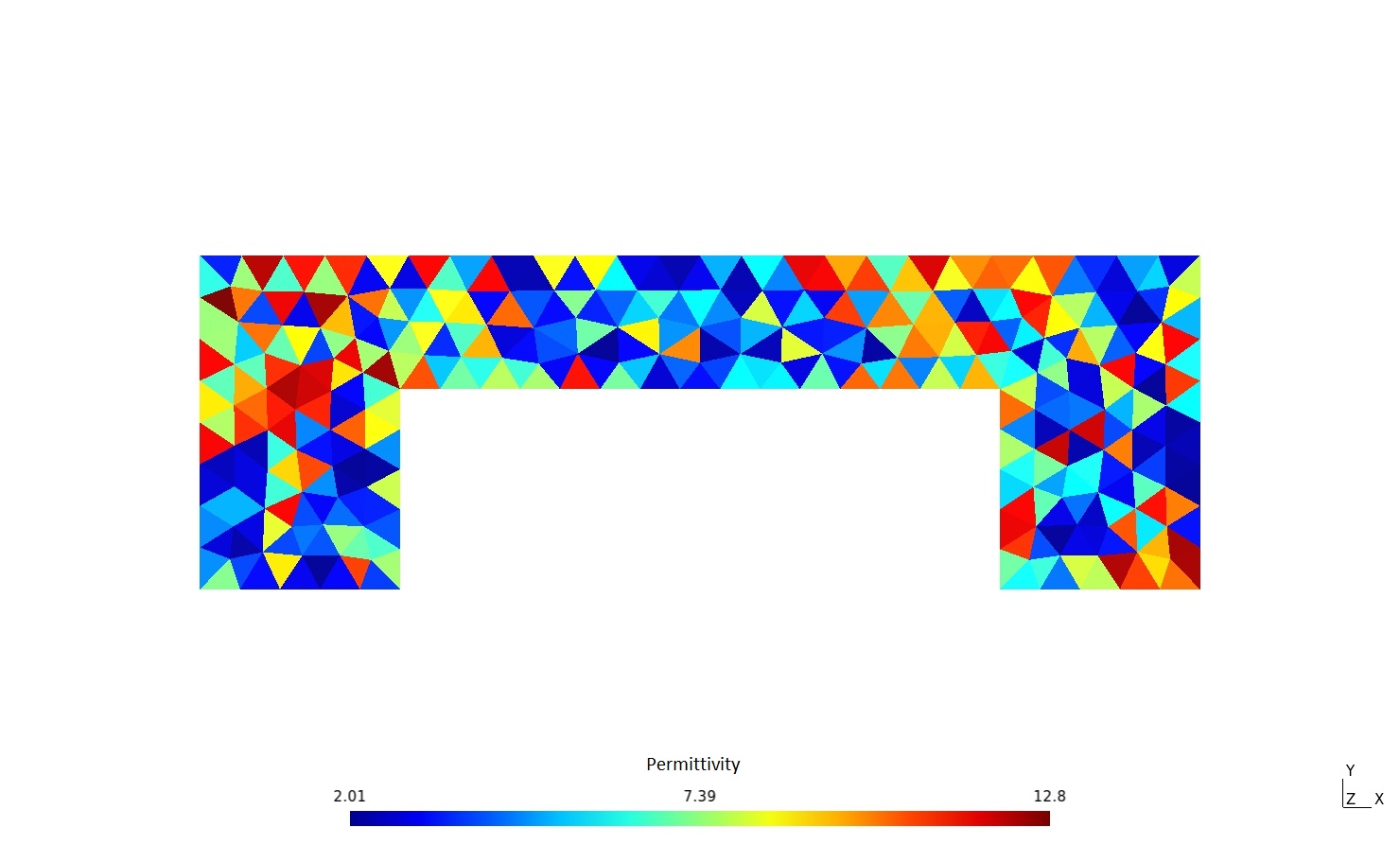}    &\includegraphics[width=0.49\linewidth]{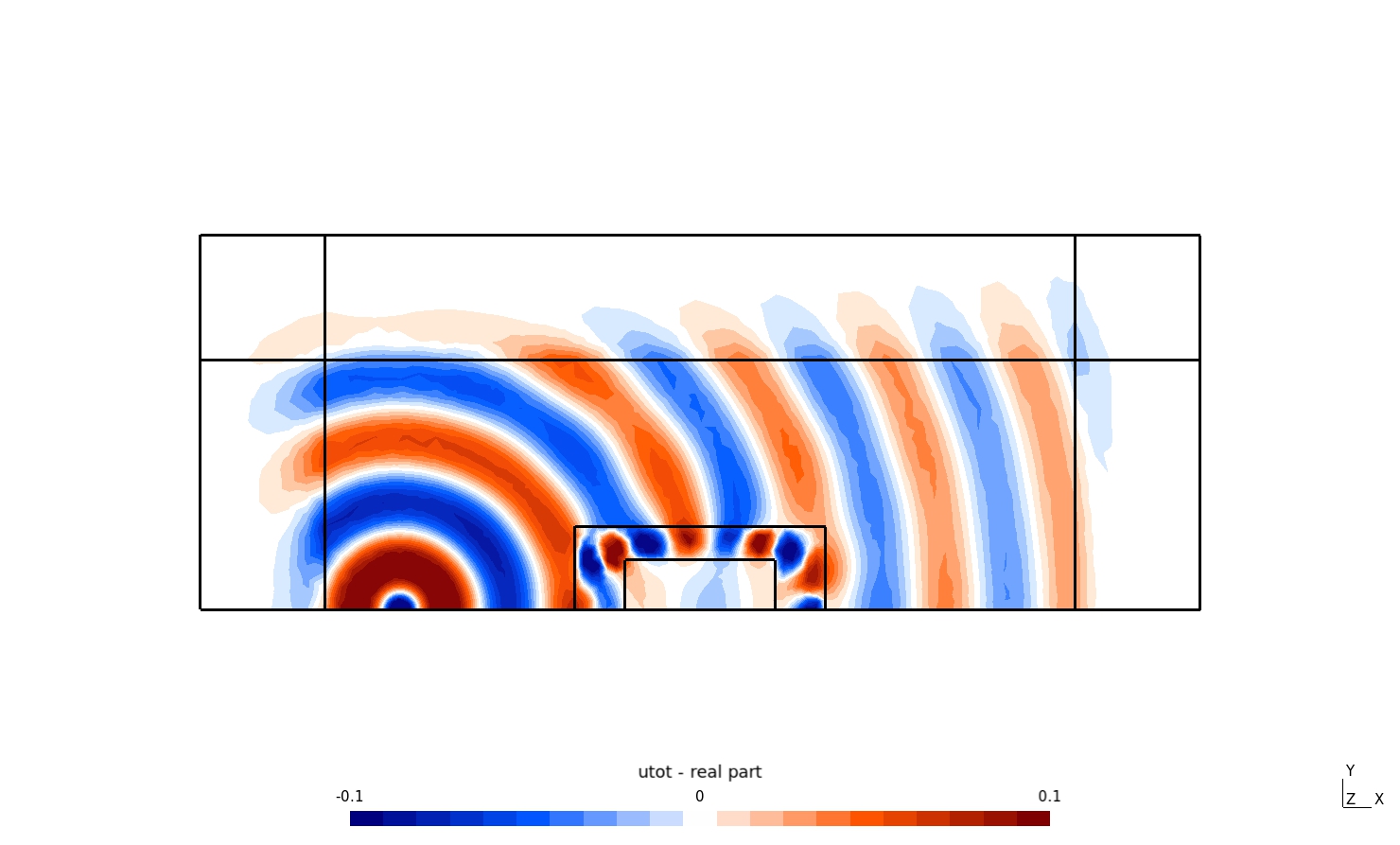}\\
         (a) & (d) \\
        \includegraphics[width=0.49\linewidth]{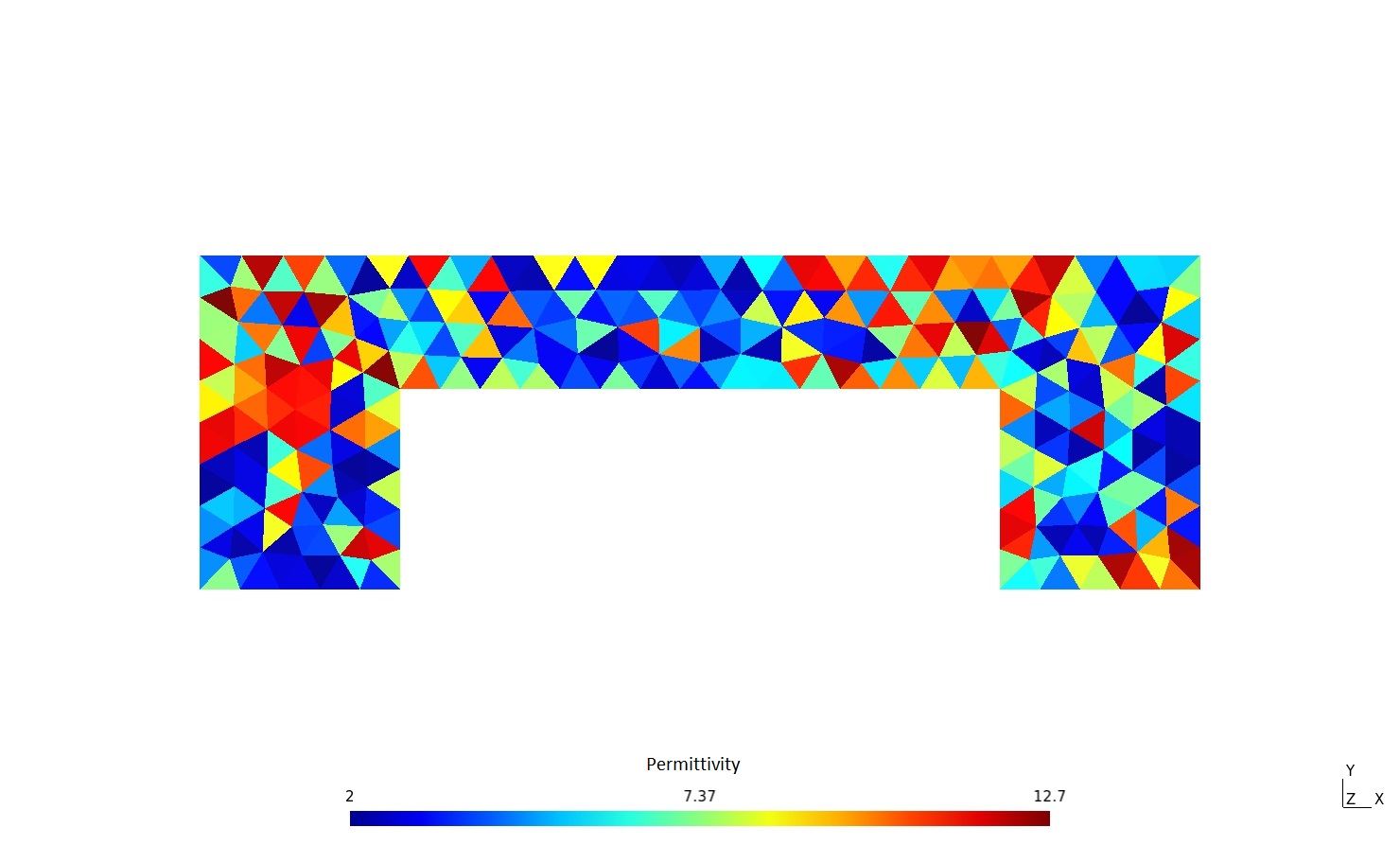}    &\includegraphics[width=0.49\linewidth]{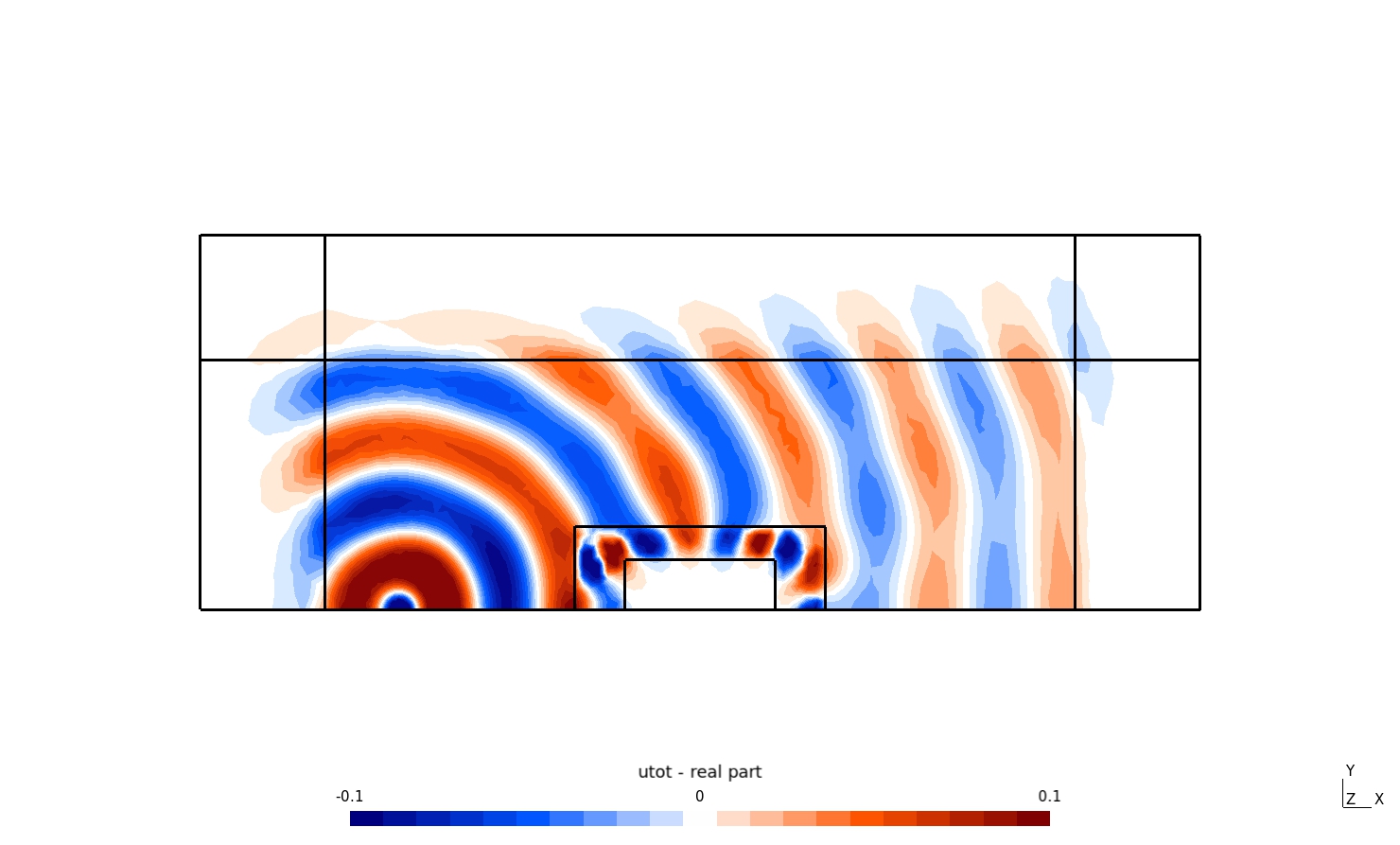}\\
         (b) & (e) \\
        \includegraphics[width=0.49\linewidth]{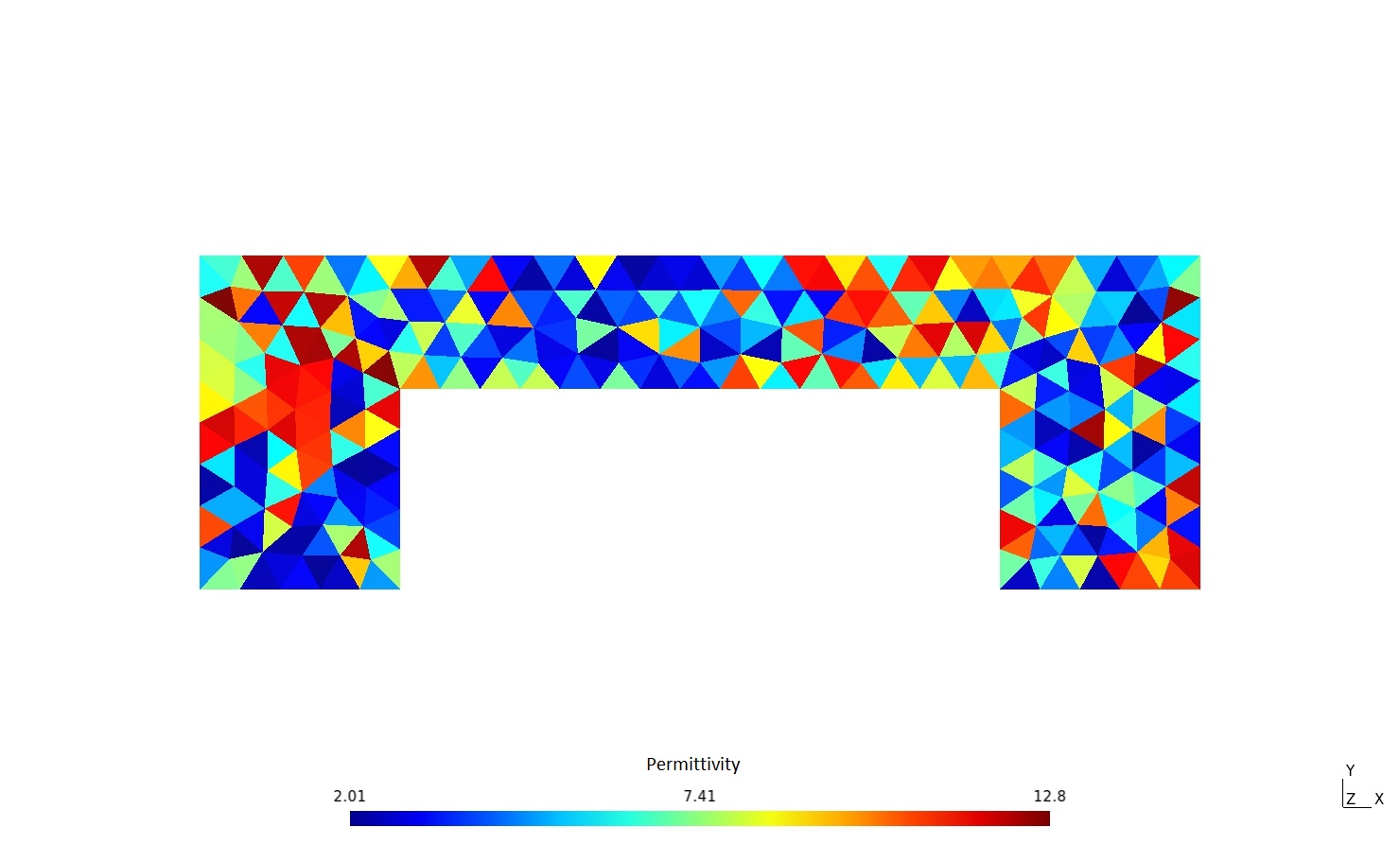}    &\includegraphics[width=0.49\linewidth]{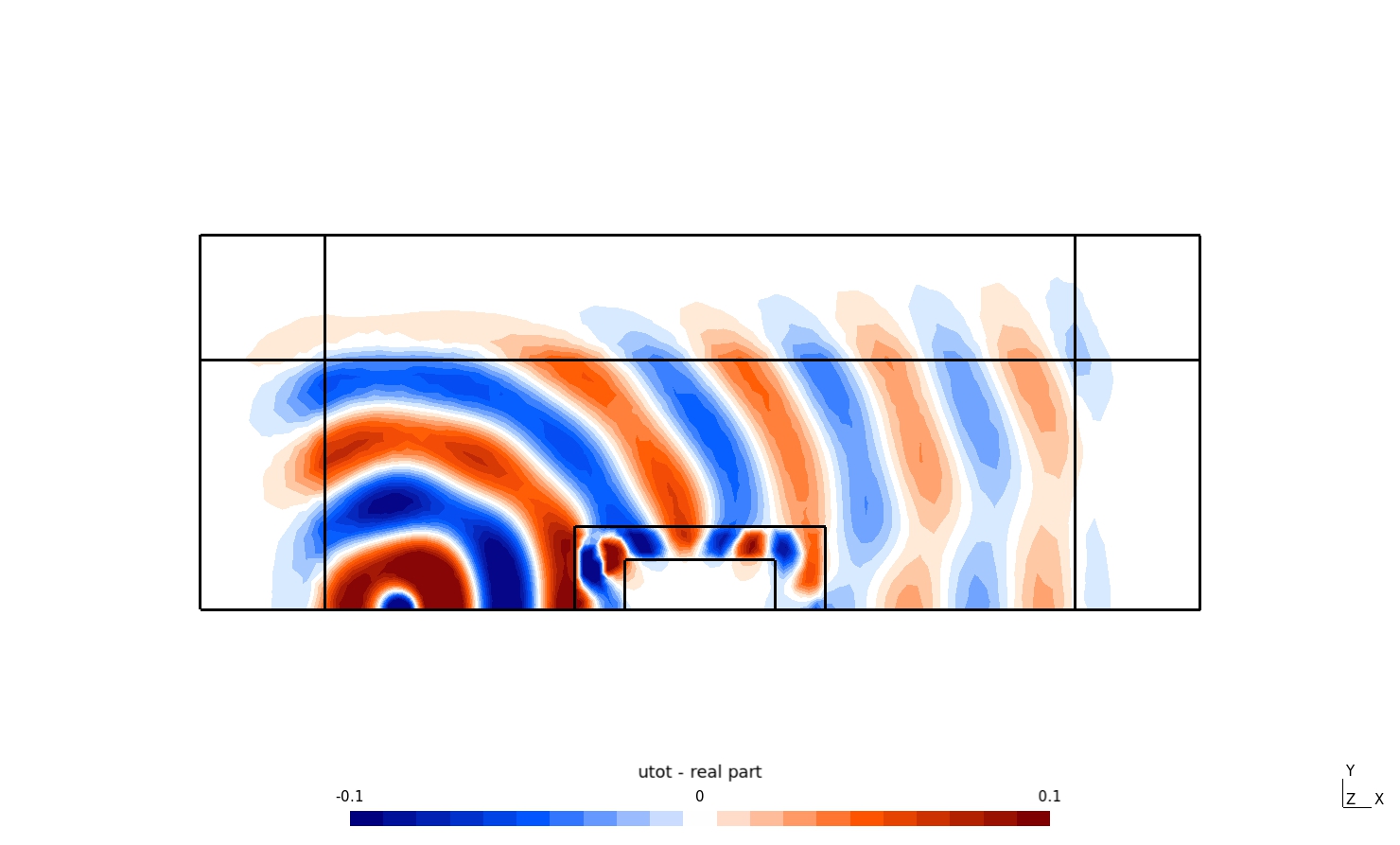}\\
         (c) & (f) \\
        \end{tabular}
        \caption{ Column 1: in (a) Best invisibility cloak, in (b) Best compromise between invisibility and protection cloak, and in (c), Best protection cloak; Column 2: in (d) Best invisibility result, in (e) Best compromise between invisibility and protection result, and in (f), Best protection result. \label{fig:f11}}}
\end{figure}

In this subsubsection, we display the results obtained by NSGA-II: the Pareto front in Fig. \ref{fig:continuousparetofront250_100}; the protection and invisibility performances in Table \ref{Table:continuousNSGAII250_100}; the cloak design and wave propagation field in Fig. \ref{fig:f11}. 

\begin{table*}[h!]
	\centering
	\resizebox{1.0\textwidth}{!}{
\begin{tabular}{c|ccc|c}
\hline
 & Protection $f_{1}$ & Invisibility $f_{2}$ & $\frac{f_{1}+f_{2}}{2}$&Links to figures\\
\hline
 Best Protection&\bf{\myround{7}{0.00007845140595910772}}&\myround{7}{0.0004302410353088411}& \myround{7}{0.0002543462206339744}&fig. \ref{fig:f11}(f)\\
 Best Invisibility&\myround{7}{0.00044190355957334074}&\bf{\myround{7}{0.000032741951781281986}}&\myround{7}{0.0002373227556773114}&fig. \ref{fig:f11}(d)\\
 Best Compromise&\myround{7}{0.00016660076580053048}&\myround{7}{0.00011668570132422142}&\bf{\myround{7}{0.0001416432335623760}}&fig. \ref{fig:f11}(e)\\
				\hline
			\end{tabular}}
			\caption{Comparison of a NSGA-II bi-objective method in protection and invisibility for a half-rectangular cloak. ${\rm T_{max}}=250$, $Q=100$\label{Table:continuousNSGAII250_100} }
	\end{table*}

\newpage

\subsubsection{Bi-objective continuous optimization with 1000 iterations, 200 research agents.}\label{subsubsec:biobjectivecontinuous1000_200}

\begin{figure}[H]
	\centering{
			\includegraphics[scale=0.4]{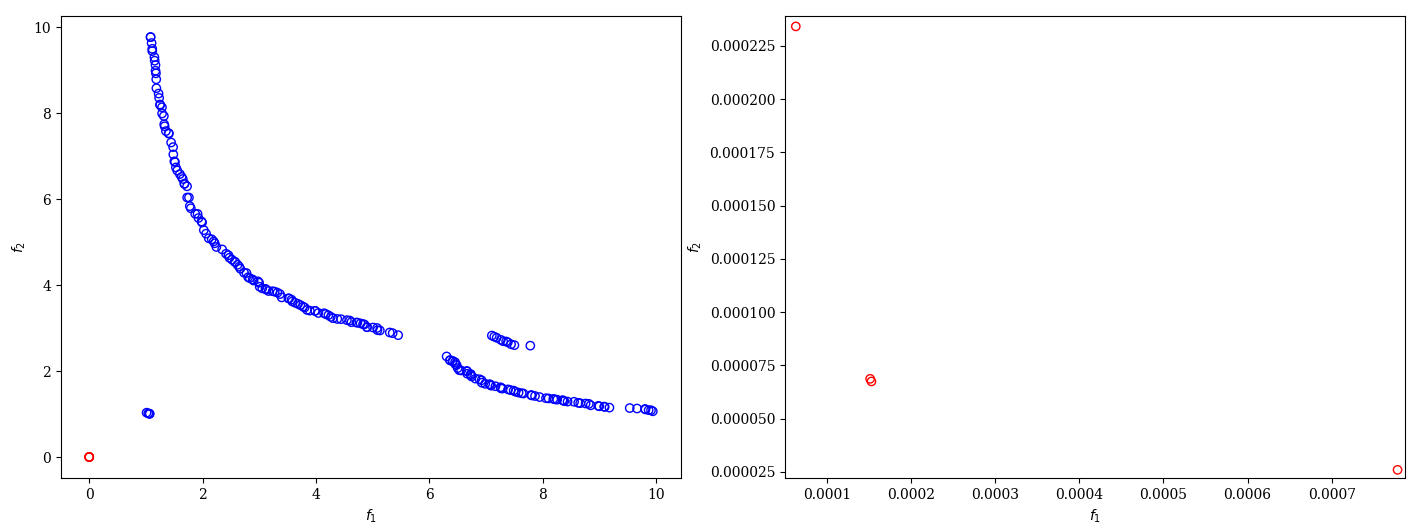}
		\caption{ In red, the four selected solutions. \label{fig:continuousparetofront1000_200}}}
\end{figure}

\begin{figure}[H]
    \centering{
        \begin{tabular}{c@{\,}c@{\,}}
             Cloak design & Wave propagation field\\
            \includegraphics[width=0.49\linewidth]{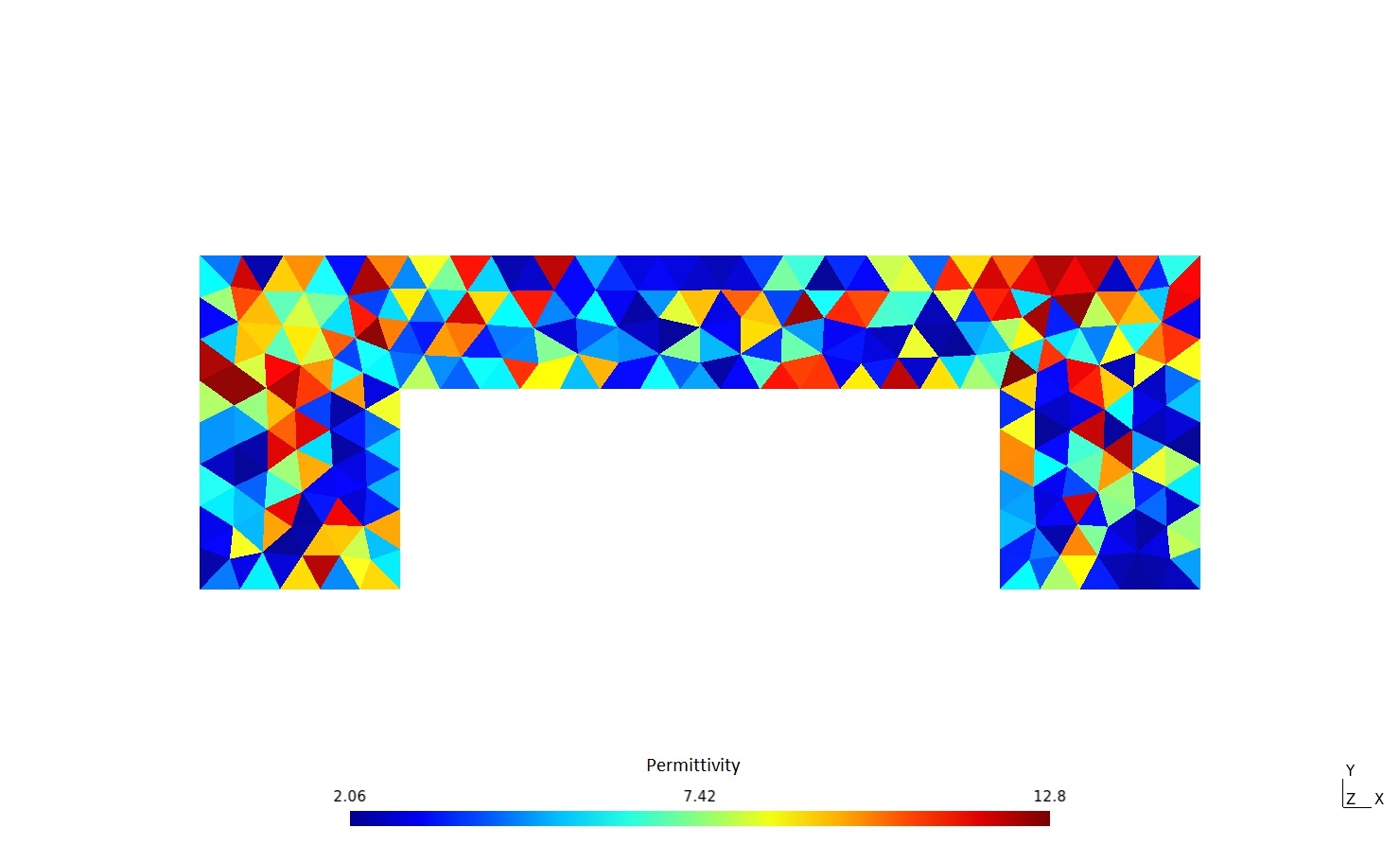} &\includegraphics[width=0.49\linewidth]{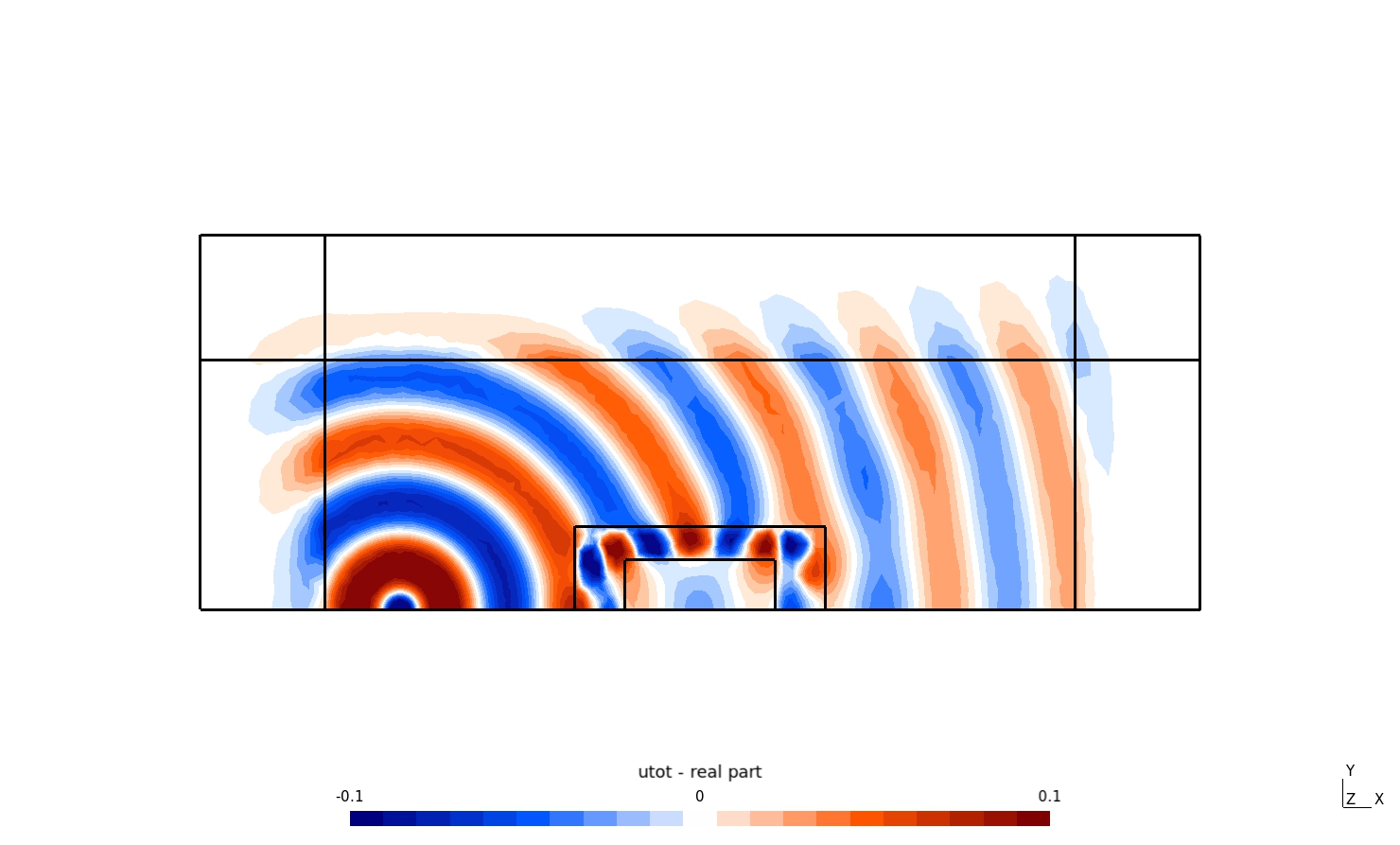}\\
         (a) & (d) \\
        \includegraphics[width=0.49\linewidth]{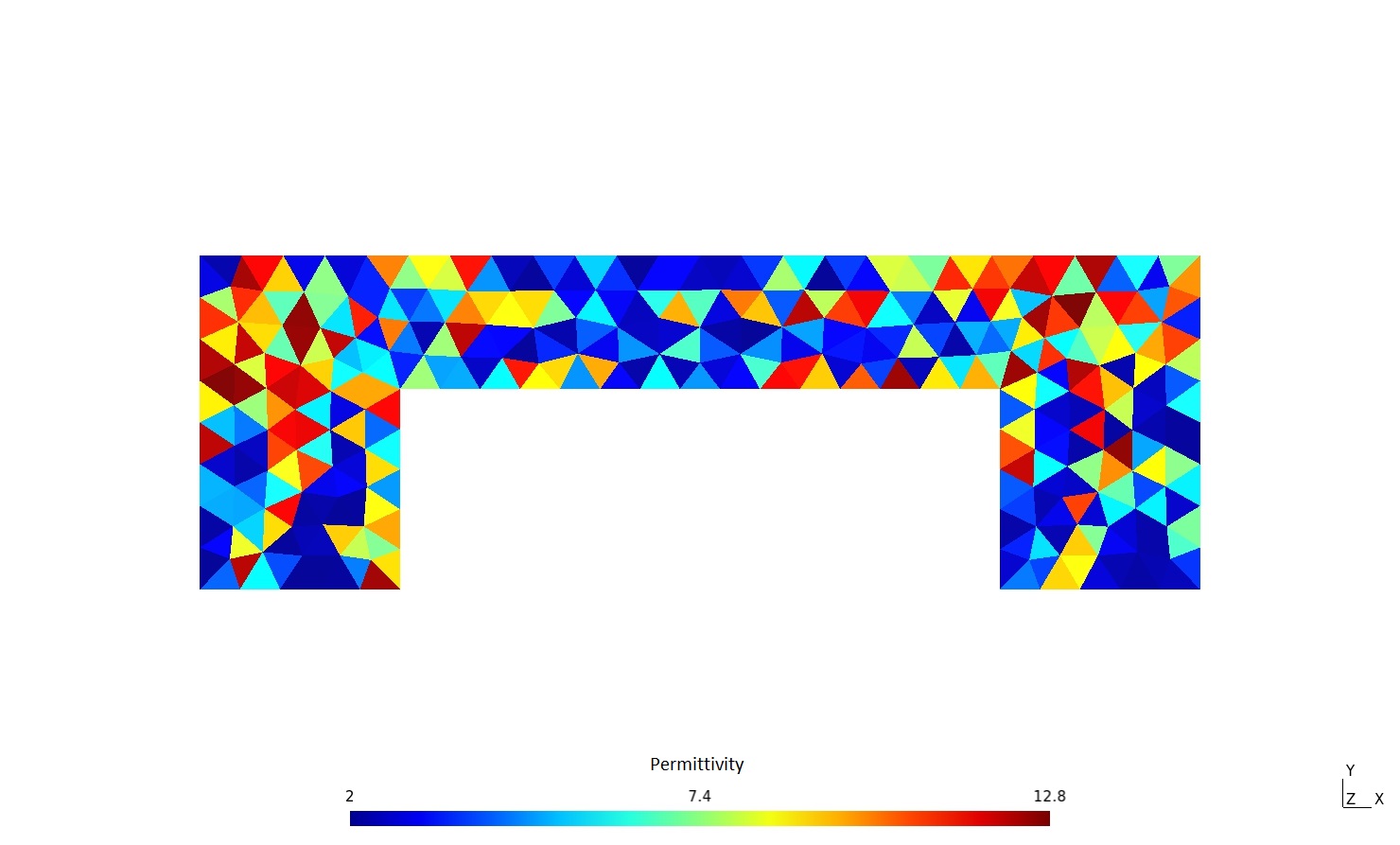}    &\includegraphics[width=0.49\linewidth]{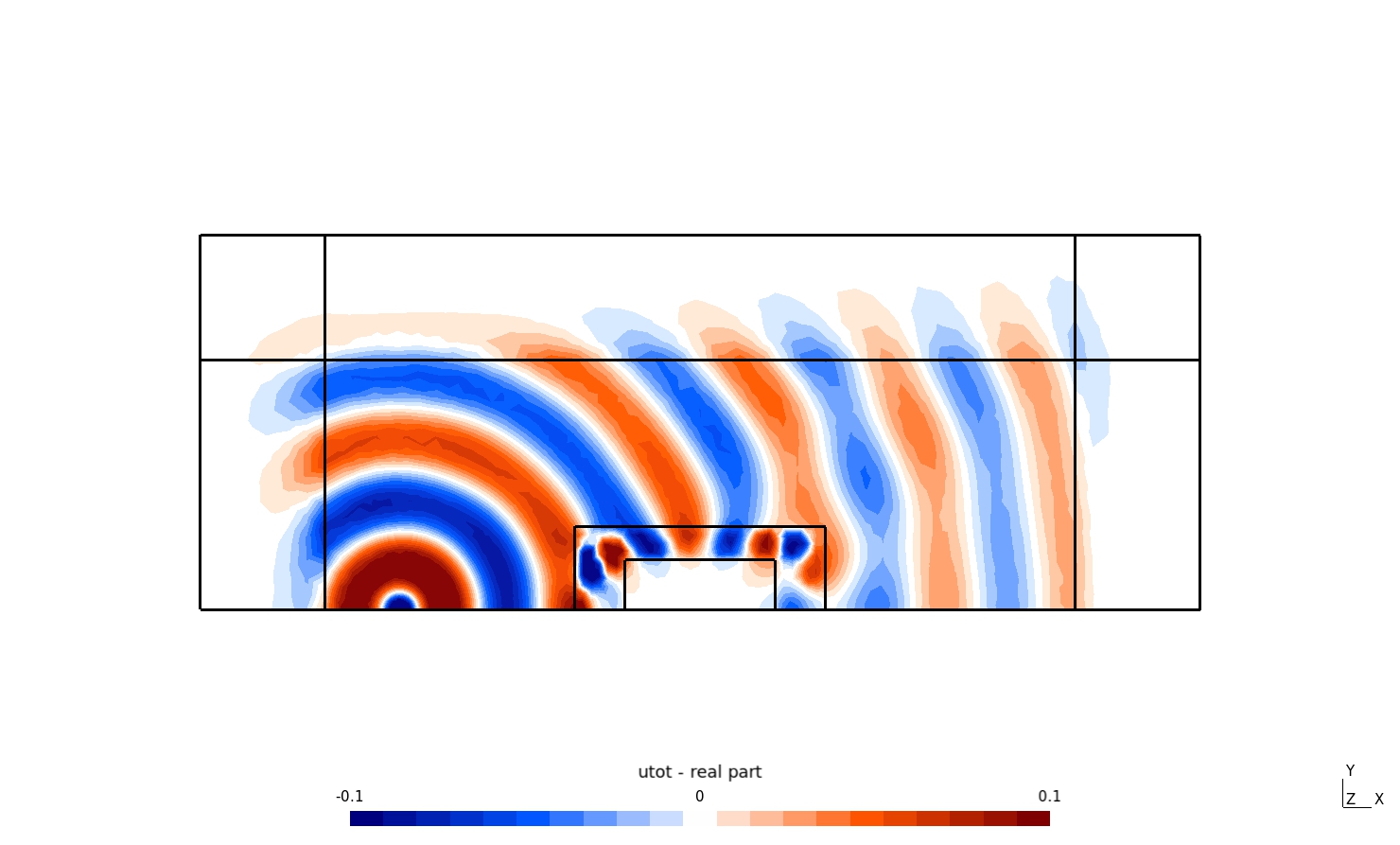}\\
         (b) & (e) \\
        \includegraphics[width=0.49\linewidth]{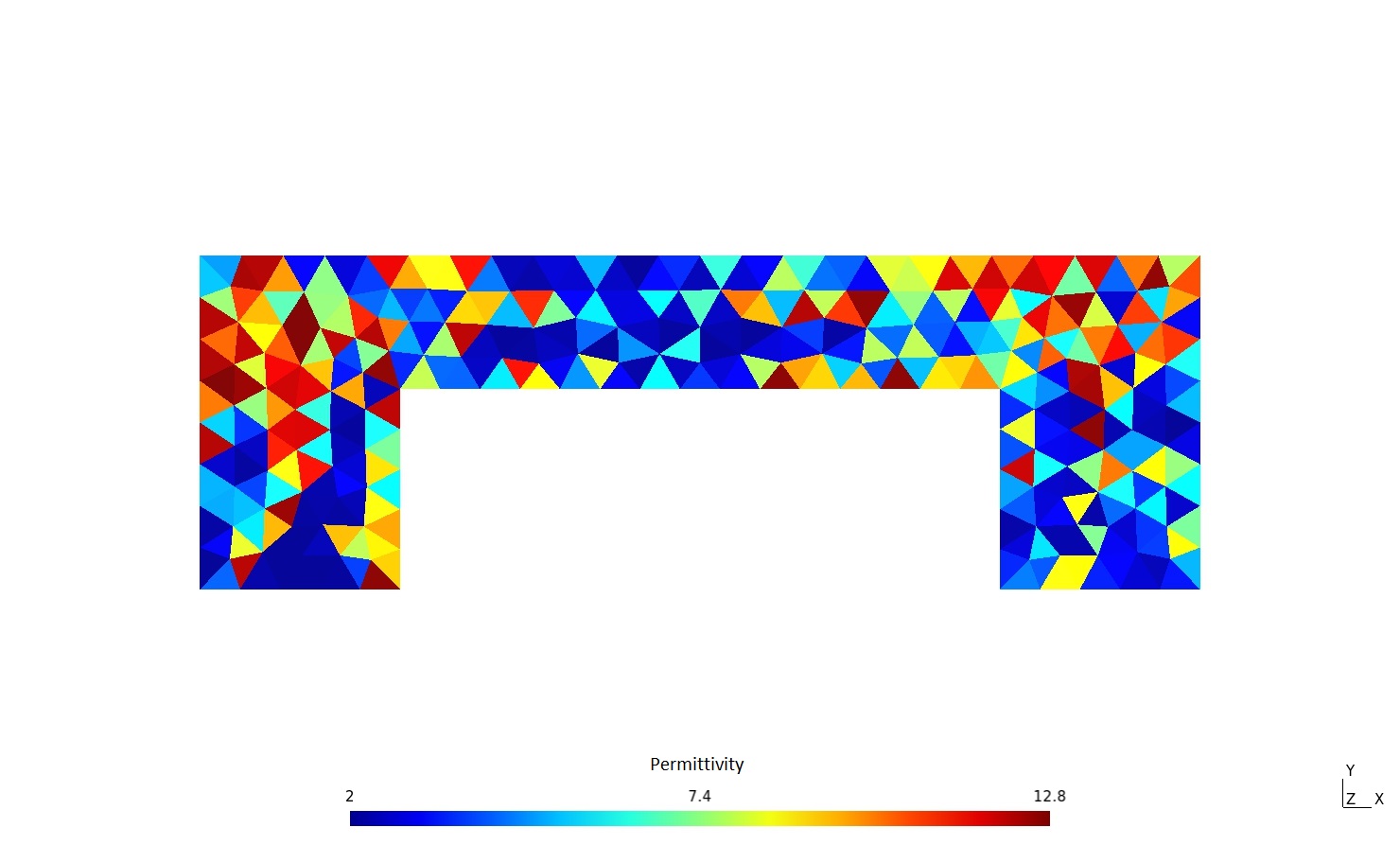}    &\includegraphics[width=0.49\linewidth]{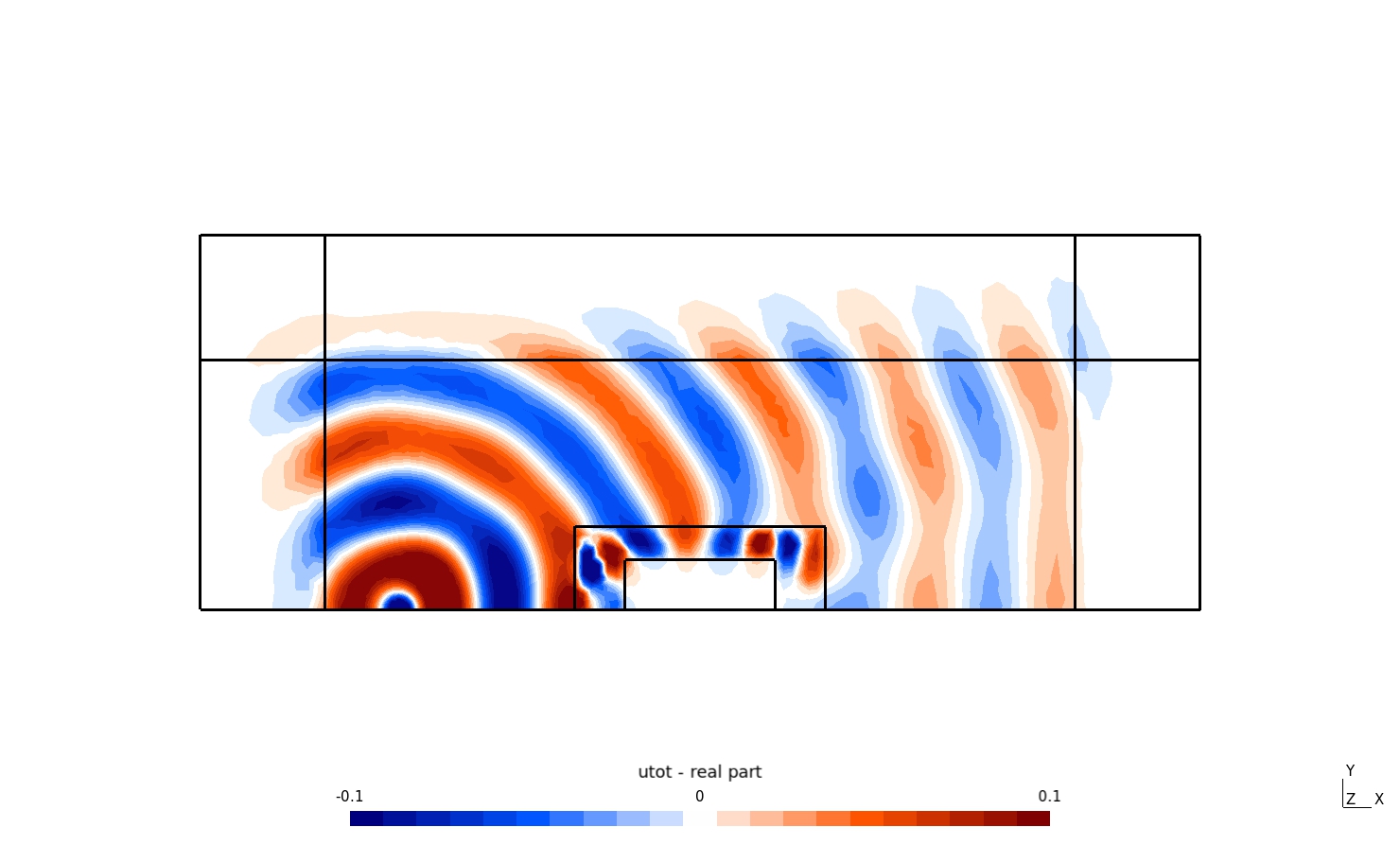}\\
         (c) & (f) \\
        \end{tabular}
        \caption{ Column 1: in (a) Best invisibility cloak, in (b) Best compromise between invisibility and protection cloak, and in (c), Best protection cloak; Column 2: in (d) Best invisibility result, in (e) Best compromise between invisibility and protection result, and in (f), Best protection result. \label{fig:f13}}}
\end{figure}

In this subsubsection, we display the results obtained by NSGA-II: the Pareto front in Fig. \ref{fig:continuousparetofront1000_200}; the protection and invisibility performances in Table \ref{Table:continuousNSGAII1000_200}; the cloak design and wave propagation field in Fig. \ref{fig:f13}. 

\begin{table*}[h!]
	\centering
	\resizebox{1.0\textwidth}{!}{
\begin{tabular}{c|ccc|c}
\hline
 & Protection $f_{1}$ & Invisibility $f_{2}$ & $\frac{f_{1}+f_{2}}{2}$&Links to figures\\
\hline
 Best Protection&\bf{\myround{7}{0.00006305848490381322}}&\myround{7}{0.00023408700901736073}&\myround{7}{0.0001485727469605870}&fig. \ref{fig:f13}(f)\\
 Best Invisibility&\myround{7}{0.0007782354112551082}&\bf{\myround{7}{0.000025917328991204385}}&\myround{7}{0.0004020763701231563}&fig. \ref{fig:f13}(d)\\
 Best Compromise&\myround{7}{0.0001529618980280156}&\myround{7}{0.00006740873459052437}&\bf{\myround{7}{0.0001101853163092700}}&fig. \ref{fig:f13}(e)\\
\hline
			\end{tabular}}
			\caption{Comparison of a NSGA-II bi-objective method in protection and invisibility for a half-rectangular cloak. ${\rm T_{max}}=1000$, $Q=200$ \label{Table:continuousNSGAII1000_200} }
	\end{table*}

\newpage

\newpage

\subsection{Spectral tolerance and comparison with random cloaks}\label{subsec:spectraltolerance}
To further verify the robustness of our approach to a change in frequency, we compute the invisibility and protection criteria for several frequency values around the central freespace wavelength $\lambda_{0}$. We compare the results obtained on the optimized cloaks to the results obtained with random cloaks obtained by filling each voxel of the design space by an arbitrary integer (resp. real) value in $\{7,10,12\}$ (resp. [7,12]). For the central freespace wavelength $\lambda_{0}$: for both C1 (protection) and C2 (invisibility) these criteria are 20 times smaller for the optimized cloaks: 0.00015 vs 0.0022 for $C_1$; and 0.00006 vs 0.0012 for $C_2$.

\begin{figure}[H]
	\centering{
        \includegraphics[width=0.9\textwidth]{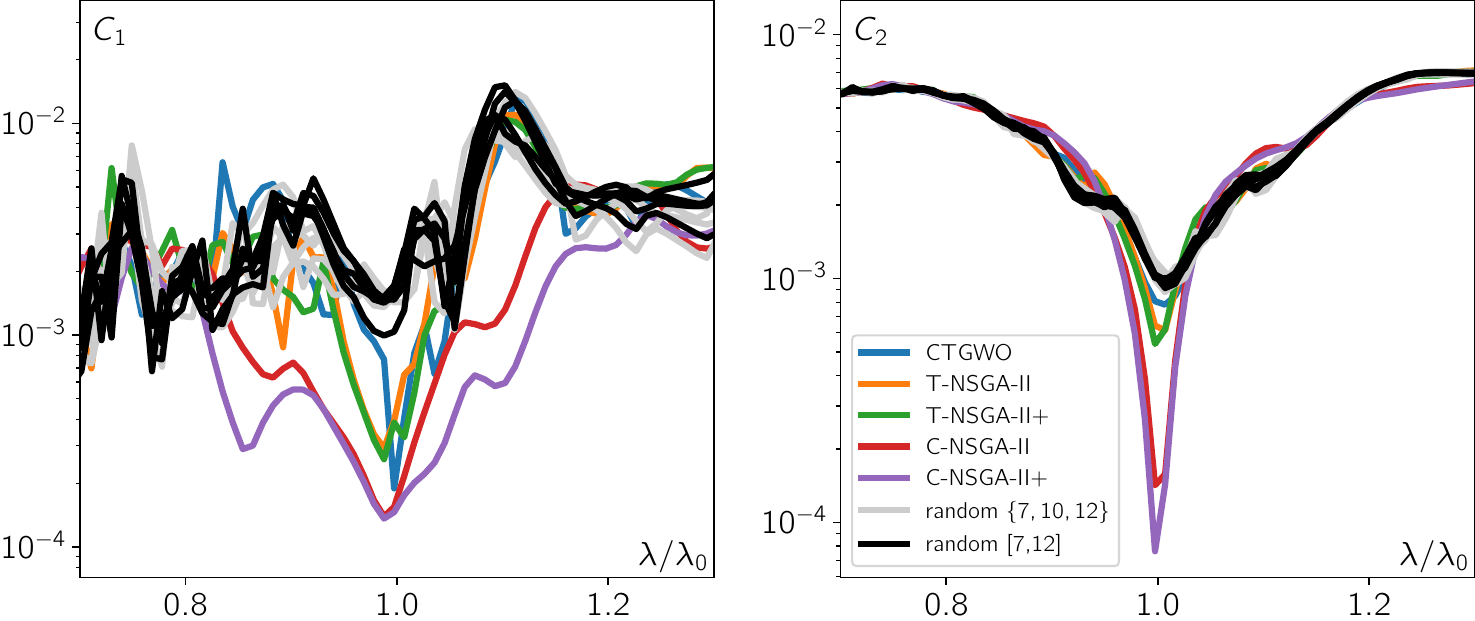}
    \caption{Spectral tolerance of the best candidates in terms of protection ($C_1$, left panel) and invisibility ($C_2$, right panel). The wavelength in abscissa is normalized by $\lambda_0$, the freespace wavelength targeted for the optimization process. The blue curves corresponds to the CTGWO method, the orange ones to T-NSGA-II, the green ones to T-NSGA-II+, the red ones to C-NSGA-II and the purple ones to C-NSGA-II+. The five grey (resp. black) curves in each plot represent the spectral behavior of $C_1$ and $C_2$ for random cloaks obtained by filling each voxel of the design space by an arbitrary integer (resp. real) value in $\{7,10,12\}$ (resp. [7,12]).  \label{fig:random}}}
\end{figure}

We infer from these results that it is worth optimizing the structure of the cloaks: this good behavior at the target frequency is obtained at the expense of slightly worse performances at other frequencies, in particular for the invisibility which appears to be quite resonant (optimized cloaks lead to a better invisibility than any random  realizations for $\lambda/\lambda_0\in[0.95,1.03]$), as opposed to the protection which turns out to be more broadband (the cloak optimized with C-NSGA-II+ lead to a better protection than any random  realizations for $\lambda/\lambda_0\in[0.82:1.22]$). We note that the optimization method operating over a continuous space lead to a more broadband response (see red and purple curves in Fig.~\ref{fig:random}).

\newpage 

\section{Discussion}\label{sec:DiscussionCloaks}

We distinguish two situations: the ternary case, where we afford three possible values for epsilon, and the continuous case, where we afford any real value between 7 and 12.\\
In the ternary case, when ${\rm T_{max}}=250$ and $Q=100$, the mono-objective approach implemented with the proposed CTGWO yields the best results in terms of protection, that is, $1.56082~10^{-4}$ and score, that is, $\frac{f_{1}+f_{2}}{2}=5.12828~10^{-4}$. Fig. \ref{fig:f5}(f) confirms this with a good protection behavior obtained with CTGWO. So in these conditions the proposed CTGWO algorithm yields the best trade-off between protection and invisibility.\\ 
NSGA-II yields the best result in terms of invisibility, that is, $5.36271~10^{-4}$, at the expense of a mean value $\frac{f_{1}+f_{2}}{2}=8.06398~10^{-4}$ which is larger than the score obtained by CTGWO. 
However the interest of NSGA-II is to enable an end-user to choose to favor one criterion (for instance invisibility) rather than the other. When ${\rm T_{max}}=1000$ and $Q=200$ affording therefore 8 times more trials of each objective function, the best protection reaches $1.23355~10^{-4}$, with an associated mean value $\frac{f_{1}+f_{2}}{2}=4.02089~10^{-4}$. 
In Figs. \ref{fig:f7} and \ref{fig:f9} we can check the coherence of the shape of the wavefront and the score values obtained.
A less realistic manner to improve the results is to enable the search for any real permittivity value between 7 and 12.\\ 
In the continuous case, still with $25000$ trials of both $f_{1}$ and $f_{2}$, we reach a best protection criterion $7.84514~10^{-5}$, a best invisibility criterion $3.27419~10^{-5}$, and a best mean value $1.41643~10^{-4}$. The results are improved but the corresponding physical constraints are very much strengthened, as we assume we can choose any material for any voxel in the cloak. In the continuous case, with ${\rm T_{max}}=1000$ and $Q=200$, we reach very low best protection ($6.30584~10^{-5}$), best invisibility ($2.59173~10^{-5}$), and best mean ($1.10185~10^{-4}$) values. The visual results in the continuous case (see Figs. \ref{fig:f11} and \ref{fig:f13}) are very convincing, and clearly illustrate the difference between the best protection, the best invisibility, the best compromise, and the best protection cases.
Finally, the tolerance with respect to the operating freespace incident wavelength showed a broadband behavior in terms of protection and a rather narrowband behavior for the invisibility criterion, as shown by the comparison to random cloaks in Fig.~\ref{fig:random}.

\section{Conclusion}\label{sec:ConclusionCTGWOCloakDesign}
In this work, we address the issue of an electromagnetic cloak's design in the transverse magnetic (TM) polarization (whereby the magnetic field is perpendicular to the $(xy)$-plane containing the computational domain). This polarization has been chosen as the model can then be adapted to plasmonics by assuming some Drude-like dispersion in the permittivity \cite{kadic2012transformation}. More precisely, this means that $\sigma$ in (\ref{eq:wave}) should be frequency dependent, and $\chi$ is set to $1$.
However, the transverse electric polarization would be also worth investigating, in that case $\sigma$ is set to $1$ in (\ref{eq:wave}), and $\chi$ plays the role of the permittivity.

Our objective is here to achieve the best compromise between protection and invisibility for TM waves. In other words, we are looking for the best trade-off between protection while considering the wave amplitude inside the cloak, and invisibility while considering the wave behavior outside of the cloak.
This is a large scale bi-objective optimization problem. We propose two approaches: in the first one we transform this problem into a mono-objective optimization problem and seek for the best mean value of protection and invisibility criteria. In the second one we look for the best protection the best invisibility, and the best trade-off with a bi-objective optimization algorithm. GWO is a well known mono-objective optimization algorithm which reaches the desired solution with a reduced number of iterations. We propose a novel mono-objective version of GWO, namely the chaotic ternary GWO, with three main innovations: ad hoc update rules to face ternary search spaces, evolving map and chaotic sequences to improve exploration abilities, and division of the pack into two groups to improve diversity. We apply this algorithm, and the comparative GWO and amixedGWO (both mono-objective) as well as the NSGA-II (bi-objective) to solve the considered cloak design problem. In the considered cloaking application, and with the help of 25000 evaluations of these criteria, the proposed CTGWO algorithm yields the best mean value, that is, $5.128~10^{-4}$, hence the best 'trade-off', between protection and invisibility. It surpasses GWO ($29.871~10^{-4}$), amixedGWO ($6.526~10^{-4}$); and the best trade-off provided by NSGA-II ($5.306~10^{-4}$). 
A possible prospect could consist in altering the ternary map in CTGWO to favor one particular material, among the three which are chosen to build the cloak. This could help in implementing the cloak with a preferred material.   

\newpage 

\section*{CRediT authorship contribution statement}
{\bf Ronald Aznavourian}: Conceptualization, Software, Validation, Writing. {\bf Guillaume Demesy}: Software, Validation, Formal analysis. {\bf Sebastien Guenneau}: Conceptualization, Validation, Supervision, Final preparing.
{\bf Julien Marot}: Conceptualization, Software, Editing, Writing, Supervision, Final preparing.

\section*{Declaration of Competing Interest}
No author associated with this paper has disclosed any potential or pertinent conflict which may be perceived to have impending conflict with this work.

\bibliographystyle{elsarticle-num}



\bibliography{./Bib/BibASC}

\newpage

\appendix


\end{document}